\documentclass[fleqn,usenatbib]{mnras}

\usepackage{newtxtext,newtxmath}
\usepackage[flushleft]{threeparttable}
 
\usepackage[T1]{fontenc}

\DeclareRobustCommand{\VAN}[3]{#2}
\let\VANthebibliography\thebibliography
\def\thebibliography{\DeclareRobustCommand{\VAN}[3]{##3}\VANthebibliography}

\usepackage{graphicx}	
\usepackage{amsmath}	
\usepackage{stackengine}
\usepackage{threeparttable}

\usepackage[normalem]{ulem}

\newcommand{\ergs}{\mbox{ erg s}^{-1}}

\newcommand{\kms}{\mbox{ km s}^{-1}}
\newcommand{\kpc}{\mbox{~kpc}}
\newcommand{\pc}{\mbox{~pc}}

\title[Observable signatures of jet-ISM interaction]{Modelling observable signatures of jet-ISM interaction: thermal emission and gas kinematics}

\author[Meenakshi et. al]{Moun Meenakshi$^{1}$,\thanks{E-mail:mounmeenakshi@iucaa.in}
Dipanjan Mukherjee$^{1}$,\thanks{E-mail:dipanjan@iucaa.in} Alexander Y. Wagner$^{2}$, Nicole P. H. Nesvadba$^{3}$,\newauthor Geoffrey V. Bicknell$^{4}$, Raffaella Morganti$^{5,6}$, Reinier M. J. Janssen$^{7,8}$, Ralph S. Sutherland$^{4}$ and \newauthor Ankush Mandal$^{1}$\\
$^{1}$Inter-University Centre for Astronomy and Astrophysics, Pune- 411007, India\\
$^{2}$University of Tsukuba, Center for Computational Sciences, Tennodai 1-1-1, 305-0006, Tsukuba, Ibaraki, Japan\\
$^3$Université de la Côte d'Azur, Observatoire de la Côte d'Azur, CNRS, Laboratoire Lagrange, Bd de l'Observatoire, CS 34229,06304 Nice cedex 4, France\\
$^4$Research School of Astronomy and Astrophysics, The Australian National University, Canberra, ACT 2611, Australia\\
$^5$ASTRON, the Netherlands Institute for Radio Astronomy, Oude Hoogeveensedijk 4, 7991 PD, Dwingeloo, The Netherlands\\
$^6$Kapteyn Astronomical Institute, University of Groningen, Postbus 800, 9700 AV Groningen, The Netherlands\\
$^7$Jet Propulsion Laboratory, California Institute of Technology, 4800 Oak Grove Dr., Pasadena, CA 91109, USA\\
$^8$Department of Astronomy, California Institute of Technology, 1216 E California Blvd., Pasadena, CA 91125, USA\\
}

\date{Accepted 2022 August 5. Received 2022 August 1; in original form 2022 March 17}

\pubyear{2022}

\begin{document}
\label{firstpage}
\pagerange{\pageref{firstpage}--\pageref{lastpage}}
\maketitle

\begin{abstract}
Relativistic jets are believed to have a substantial impact on the gas dynamics and evolution of the interstellar medium (ISM) of their host galaxies. In this paper, we aim to draw a link between the simulations and the observable signatures of jet-ISM interactions by analyzing the emission morphology and gas kinematics resulting from jet-induced shocks in simulated disc and spherical systems. We find that the jet-induced laterally expanding forward shock of the energy bubble sweeping through the ISM causes large-scale outflows, creating shocked emission and high-velocity dispersion in the entire nuclear regions ($\sim2$~kpcs) of their hosts. The jetted systems exhibit larger velocity widths (> 800 $\kms$), broader Position-Velocity maps and distorted symmetry in the disc's projected velocities than systems without a jet. We also investigate the above quantities at different inclination angles of the observer with respect to the galaxy. Jets inclined to the gas disc of its host are found to be confined for longer times, and consequently couple more strongly with the disc gas. This results in  prominent shocked emission and high-velocity widths, not only along the jet's path, but also  in the regions perpendicular to them. Strong interaction of the jet with a gas disc can also distort its morphology. However, after the jets escape their initial confinement, the jet-disc coupling is weakened, thereby lowering the shocked emission and velocity widths.

\end{abstract}




\section{Introduction}
Black-hole accretion can result in the release of a large amount of energy in the form of radio jets from an Active Galactic Nucleus (AGN) \cite[see][for a review]{blandford19}. \citet{sabater_2019} showed that radio activity is very common in massive galaxies ($>10^{11}M_\odot$). The study of the properties of compact and extended radio sources is thus an active field of research especially with the advent of many state-of-the-art observational facilities \citep{morganti_1997,chris_1998,montse_2017,hardcastle_2020,mooney_2021,suma_2022}. Simulations of jets interacting with the interstellar medium (ISM) of their host \citep{Saxton2005,Sutherland07,wagner11,dipanjan16} have shown that the jets are easily diverted and split, percolating through the porous ISM. Such a ``flood and channel" phase of the jet's evolution \citep{mukherjee2022} can transport momentum and energy isotropically to large distances \citep{Saxton2005, Sutherland07,dipanjan16}. 
Jets can potentially transfer up to $40\%$ of their kinetic energy into accelerating the diffuse and dense clumps to velocities $\sim 100-1000~\kms$ \citep{wagner12,collet_2016,dipanjan16}. Thus relativistic jets evolving through a dense ISM can be a potent source of feedback on the galaxy \citep{suma_2022}. 

Resolved observational studies of the kinematics and emission properties of multiple gas phases (neutral, ionized and molecular) often exhibit similar kinematic features across different gas phases, which points to the radio-jet being the driving mechanism of a genuine multi-phase outflow \citep{emonts,nicole07,raffaella15,oosterloo_2017,morganti_2021b,Vayner_2021b}. Jet driven outflows have been detected in systems with a wide variety of jet powers, from low power jets \citep[$\mathrm{kinetic\, jet\, power, P_J}\lesssim 10^{44}\ergs$ e.g.][]{tadhunter_2000,garc14,tadhunter_2014,dasyra_2015,harrison15,girdhar_2022} to more powerful jets  \citep[$\mathrm{P_J}\gtrsim 10^{45}\ergs$][]{nicole08,nesvadba_2017}. Understanding the complex interplay of jets with the ISM of their hosts requires spatially resolving the observed gas kinematics in such systems for multiple gas phases \citep[e.g.][]{oosterloo,solorzano03,nesvadba10,nesvadba_2017,zovaro_2019,zovaro_2019b,ruffa_2020a,Venturi21,girdhar_2022}.


The evolution of a jet through a galaxy's ISM can be broadly divided into four phases \citep{Sutherland07,morganti_2021a}: a flood and channel phase, a spherical energy-driven bubble phase, a jet break-out phase and finally a classical phase. Observational studies show that these different stages of jet-ISM interaction determine the evolution and morphology of gas kinematics in such systems \citep{schulz_2021,morganti_2017}, which can be traced by line emission observed with long-slit and imaging spectroscopic observations \citep[see e.g.][]{harrison15,morganti_2017,zovaro_2019}. 

Jet-induced shocks can sometimes dominate the ionization and emission in the extended emission line regions (EELR) of their hosts \citep{montse_1999,tadhunter_2002,shih_2013}, and the surface brightness morphology of the EELR depends on how the jet propagates through the ISM \citep{Falcke_1998,koss_2015,clive18,couto_2020}. Simulations also find that the jets can heat and ionize the ISM in their vicinity \citep{Meenakshi_2022a}, although the extent of shock-ionization in the large-scale galaxies is not known. However, very few theoretical studies have explored the impact of jet-ISM interaction on shock emission from the dense gas and the kinematics of such shocked gas \citep{Sutherland07,dipanjan5063,cecil_21}.

Recent high resolution simulations of jet-ISM interactions \citep{dipanjan16,dipanjan2018} are well suited to predict the distribution of shocked gas as the jet evolves through the ISM and the strength of the shocked emission. In this paper, we therefore create a direct link between hydrodynamic simulations of jet-ISM interactions and observations of jet-driven outflows through synthetic [\ion{O}{III}]~($\lambda$ = 500.7~nm) emission line data calculated by post-processing these simulations. Thus, we explore the observable signatures of jet-ISM interactions in the nuclear regions ($\sim2$~kpc) of both spherical and disc galaxies, which were the subject of the simulations.
We use different kinematic measures, commonly used in observational studies, to constrain the effect of jets on their host galaxies, as has been done in several highly resolved studies \citep[see e.g.][]{Venturi21,girdhar_2022,suma_2022}.

The paper is structured as follows. In Sec.~\ref{method}, we give a brief overview of our method for computing the emission and kinematic features. We present the results for spherical systems hosting a higher-power ($10^{45}\,\ergs$) jet and compare it with the results for a lower-power ($10^{44}\,\ergs$) jet in Sec.~\ref{results_spherical}. In Sec.~\ref{results_disc}, we discuss the evolution of the emission and gas kinematics in galaxies with kpc scale gaseous discs where relativistic jets with a kinetic power of $10^{45}\,\ergs$ are launched at different angles with respect to the minor axis (Sec.~\ref{edge-on}). In Sec.~\ref{pv-maps}, we explore different diagnostics to predict the impact of the jet on the observed velocity field of discs. Finally, we discuss the major implications of our study in Sec.~\ref{discussion}, and summarise the main findings in Sec.~\ref{summary}.

\section{Methodology}
\label{method}
In this section, we discuss our methodology for studying the thermal emission and gas kinematics in jet-harbouring systems. We use the results of jet-ISM simulations presented in \citet[][hereafter M2016]{dipanjan16} and \citet[][hereafter M2018b]{dipanjan2018} in postprocess to estimate the expected emission from shocked gas and its kinematics in the central nuclear region ($\sim 2$ kpc). Although both papers explore the interaction of jets evolving through a dense ISM , they differ in the morphology of the dense gas. While M2016 explores the evolution of a jet through a spherically distributed dense ISM of radius 2~kpcs, M2018b studies the interaction of jets inclined to gaseous discs at different angles with respect to the disc normal. These discs extend to 2~kpc in radius and have a thickness of 1~kpc in the central regions and 2~kpc outwards. Initially, the density distribution in both of these systems is set-up as fractal with a lognormal density distribution and a Kolmogorov spectrum, using the publicly available \textsc{pyfc}\footnote{\url{https://pypi.python.org/pypi/pyFC}} module. Thereafter, the fractal distribution is apodized with a warm gas density profile ($n_w(r,z)$) in the presence of the external gravitational potential, which for the spherical system in terms of cylindrical coordinates $r$ and $z$, is given as:
\begin{equation*}
   \frac{n_w(r,z)}{n_{w0}} = \mathrm{exp}\Bigg\{-\frac{1}{\sigma_{t}^2}[\phi(r,z)-\phi(0,0)]\Bigg\}
\end{equation*}
Here $\phi(r,z)$ is the external gravitational potential which includes contribution from the baryonic (stellar) and dark-matter components and $\epsilon=0.93$, $n_{w0}$ is the central density and $\sigma_t$ is the total velocity dispersion.
For the disc systems, the number density of the warm gas is given as:
\begin{equation*}
    \frac{n_w(r,z)}{n_{w0}} = \mathrm{exp}\Bigg\{-\frac{1}{\sigma_{t}^2}[\phi(r,z)-\epsilon^2\phi(r,0)-(1-\epsilon^2)\phi(0,0)]\Bigg\}
\end{equation*}

We refer the reader to the two papers for  detailed description of the simulation set up. The key parameters pertaining to the simulations are presented in Table~\ref{tab:sim_table}.

\begin{table}
	\centering
	\caption{Simulations of jet-ISM interaction in spherical \citep{dipanjan16} and  disc \citep{dipanjan2018} systems used in this study.}
	\label{tab:sim_table}
	\begin{threeparttable}
	
	\begin{tabular}{|c|l|l|l|l|l|} 
		\hline
		Simulation & Jet Power & $n_{w0}$ & \hspace{-1.2cm} ${\theta_\mathrm{J}}^b$ & \hspace{-0.8cm} $\Gamma^c$ & \hspace{-1.5cm} Gas Mass \\
		label & ($\mathrm{P_J}$$[\mathrm{erg\,s^{-1}}])$ & ($\mathrm{cm^{-3}})^a$ & &  ($10^{9}\,\mathrm{M_\odot}$)\\
		\hline
		    & & Spherical systems  &  \\
	    Bs & $10^{45}$ & 150 & \hspace{-1cm} - & \hspace{-0.8cm} - & \hspace{-1.5cm} 1.75$^d$ \\
		Cs & $10^{44}$ & 150 & \hspace{-1cm} - & \hspace{-0.8cm} - &\hspace{-1.5cm} 1.75$^d$ \\
		
		& & Discs &   \\
			no jet & \hspace{0.2cm}- & 200 & \hspace{-1cm} - & \hspace{-0.8cm} - &\hspace{-1.5cm} 5.71 \\
		B & 10$^{45}$ & 200 & \hspace{-1cm} $0^\circ$ & \hspace{-0.8cm} 5 &\hspace{-1.5cm} 5.71\\
		D & 10$^{45}$ & 200 & \hspace{-1cm} $45^\circ$ & \hspace{-0.8cm} 5 & \hspace{-1.4cm}5.71\\
		E & 10$^{45}$ & 200 & \hspace{-1cm} $70^\circ$ & \hspace{-0.8cm} 5 & \hspace{-1.4cm}5.71\\
		\hline
	\end{tabular}
	\begin{tablenotes}
	     \item[a] The density of the dense gas at the center of the disc.
         \item[b] Angle of inclination of the jet measured from the normal to the disc.
       \item[c] Jet Lorentz factor.
       \item[d] Mass in the half domain (upper) of the spherical system
       \item[] For the discs, we use similar nomenclature to that of M2018b. The names for the spherical systems given in M2016 are appended by `s' (for spherical) to differentiate them from the disc simulations.
       \item[] The times of the simulation mentioned in this work are lower than those in M2016 and M2018b by the jet injection times, which are 1.87~Myr and 0.17~Myr, respectively.
	\end{tablenotes}
	\end{threeparttable}
\end{table}

\subsection{Estimating the emission from shocked gas}
\label{method_emission}
The shocked ISM impacted by the jet is expected to emit at a variety of lines over a broad range of wavelengths. We have chosen to express emission from [\ion{O}{III}], peaking at a temperature $\sim 10^5$~K, as representative of the expected emission features from the shocked regions. Other phases of multi-phase ISM can also be expected to follow similar qualitative trends in morphology and kinematics \citep{raffaella15}. To verify this, we also explore different observed quantities from collisionally-ionized and non-ionized gas, which are discussed later in Sec.~\ref{diff_phase}. In this paper, we focus only on the emission from the collisionally-ionised gas, and neglect emission from other alternative processes, such as star formation or AGN photo-ionization. Thus, we are probing one important benchmark where we isolate one process (shocks here), and do a detailed analysis of its impact. Therefore, our work is particularly relevant to the sources where photoionization is not globally dominant (such as LINERS and LERGs). It can also be applied to AGN dominated by radiation, if the shocks and photoionization are co-spatial \citep{moy_2002,dagostino2019} in a system, as it has been commonly observed that the gas with larger line widths seems to be aligned with the radio emission \citep{raffaella07,santoro_2020}, although the line ratios are indicative of photoionization.

We calculate the emissivity from the collisionally excited atoms in the shocked gas using the results from \textsc{mappings v} \citep{mappings_2018}. The total cooling function, and cooling curves for [\ion{O}{III}] have been obtained as a function of gas temperature, using \textsc{mappings v}, as shown in Fig.~\ref{fig:cool-curve}.
These curves correspond to the non-equilibrium cooling of the solar-abundant shocked gas. The corresponding emission for a given computational cell is obtained by identifying its temperature at the given time of the snap-shot and using the look-up table obtained from \textsc{mappings v}. We have assumed optically thin emission throughout both the spherical and disc systems in our study. To verify the validity of this assumption we performed \textsc{cloudy} \citep[][]{cloudy17} computations using the AGN SED from \citet{Meenakshi_2022a} ($\log\,L=10^{45}\,\mathrm{\ergs}$) for the disc plane of Sim.~D at 2.31~Myr. We found a mean column density for the neutral gas till the last cell ionized to be $5.5\times 10^{21}\,\mathrm{cm^{-2}}$ and the mean optical depth for the absorption of [$\ion{O}{III}$] radiation is $4.79\times 10^{-5}$. We scaled this optical depth for the highest column density in our simulation which is $\sim 2.9\times 10^{24}\,\mathrm{cm^{-2}}$. This gives a small value of the total optical depth ($\sim$0.025) and thus, has a negligible effect on the total integrated emission along the Line of Sight (LOS). Therefore, under this assumption, the interpolated emission is multiplied with density squared to obtain the corresponding emissivity of the cell, which is then integrated along the LOS to estimate the total observed flux on the image plane.

[\ion{O}{III}] emission in our simulations results from regions shocked by the jet flows as well as mixing of hot gas with the cold clouds. However, due to the limited resolution of the simulations, the post-shock cooling zones are likely to be unresolved, which we discuss in detail in Appendix~\ref{cooling-length}. An approximate estimate of the shock cooling length, obtained by comparing the pre-shock and post-shock conditions, gives a value $0.014-1~\pc$, which is smaller than the resolution length for inner regions of the simulation box (i.e. 6~pc). Thus, one must note that the observed flux obtained from our analysis are likely upper limits, as one may over-estimate the volume of the emitting region due to the lack of resolution. Adaptive mesh refinement (AMR) simulations of such jet-ISM interactions are required to properly resolve the cooling layers, and more accurately estimate the luminosity of shocked gas.

 However, what is pertinent for this analysis is the qualitative trends in the morphology of the shocked emission and the expected kinematics, which depends on the transfer of kinetic energy and local turbulence. Furthermore, these observed characteristics are also shaped by the line of sight of the observer, which we explore in this study. Thus, although the luminosity of shocked emission may have uncertainties due to constraints of spatial resolution, the qualitative trends of observable features of jet-ISM interactions explored in this study provides valuable insights on the physics of jets evolving in gas-rich environments and serves as a reference for better understanding similar spatially resolved observations.

\subsection{Diagnostics to probe kinematics of the shocked gas}
Observational studies investigating the signatures of ionized outflows in galaxies often use velocity width measures to probe the kinematics and physical nature of the gas \citep[see e.g.][]{nicole08,harrison15,coatman,ferguilo20,speranza_2021,Venturi21}. To characterise the velocity kinematics of the shocked gas in our simulations, we also calculate such widths, particularly W80 widths of the [\ion{O}{III}] emission. This measure corresponds to the velocity range covering $\sim80\%$ of the total emitted luminosity in a spectral line. Here we closely follow the approach used by \citet{whittle_1985,nadia,coatman} for estimating the W80 widths. However, we have verified that other approaches, as described in \citet{heckman_1981}, give an overall similar morphology of velocity widths in the image plane (see Appendix~\ref{widths-compare}). We ignore the thermal width of the line, which being a few tens of km/s at $\sim10^5$ K, is negligible compared to the broadening \citep[for e.g. 1000~$\kms$, see][]{nicole08} due to bulk motions (such as turbulence and outflows) in the jet-disturbed gas. 

The steps to create a 2D image of [\ion{O}{III}] surface brightness, and consequently compute the W80 widths from the simulations are as follows:
\begin{itemize}
    \item First, an image plane orientation is chosen by specifying the polar and azimuthal coordinates of the observer denoted by $\theta_\mathrm{I}$ and $\phi_\mathrm{I}$ (`I' for inclination angle of the observer), respectively.
    \item Then we define a 2D image plane with a uniform resolution of 6~pc in the central 3~kpc, and stretched elsewhere to create an image of  [\ion{O}{III}] surface brightness and projected LOS velocities.
    \item 
    For the computation of W80 widths, we divide the image plane into broader spatial domains of resolution of $100\times 100~\pc^2$.
    \item We then estimate the luminosity and the line of sight velocity ($\mathrm{V_{LOS}}$) for all the cells in the galaxy. The luminosity of all cells in a domain of the image plane is binned in a histogram with a bin width ($\Delta v$) of 30~$\kms$. The chosen bin width corresponds to a spectral resolution of $R = \lambda/ \Delta \lambda \approx c/\Delta v = 10000$.
    The total luminosity in the domain is then calculated by integrating over all the bins of the histogram. 
    \item To find the W80 width, we integrate from the right and left sides of the histogram to estimate the velocity containing $10\%$ of the total luminosity from both of the sides. The spectra can be over-sampled due to the finite bin size, and so we divide the bins at the edges of the spectra into finer bins for more accurate W80 integral.

\end{itemize}
In our study, we fix the azimuthal angle ($\phi_\mathrm{I}$) at $270^\circ$, which corresponds to the X-Z plane with X changing from negative to positive values while going from left to right side on the image plane. Henceforth, for all instances of imaging the galaxy and computing subsequent diagnostics such as W80, we mention only $\theta_\mathrm{I}$ in the image plane of different observed quantities, where  $\theta_\mathrm{I}=90^\circ$ corresponds to an edge-on view.

To probe the variation of the kinematic features along a given spatial direction in the image plane, we also construct the synthetic Position-Velocity (PV) maps using the [\ion{O}{III}] flux and line of sight velocities ($\mathrm{V_{LOS}}$) in the disc. Such diagnostics are useful in highlighting the change in the velocity structure of the gas as a result of motions along the jet or inferring changes in the rotation curve. Similar to \citet{dipanjan5063}, the PV maps are constructed by considering a synthetic slit of width 500~pc in the image plane and binning the integrated flux along the slit into 130 spatial bins for a length interval of $[-2,2]~\kpc$ and $\mathrm{V_{LOS}}$ range of $[-800,800]~\kms$. The results of the analysis using PV maps are discussed in Sec.~\ref{disc_dynamics}.
\begin{figure*}
\centerline{
\def\arraystretch{1.0}
\setlength{\tabcolsep}{0.0pt}
\begin{tabular}{lcr}
     \hspace{-1cm}
    \includegraphics[width=\linewidth]{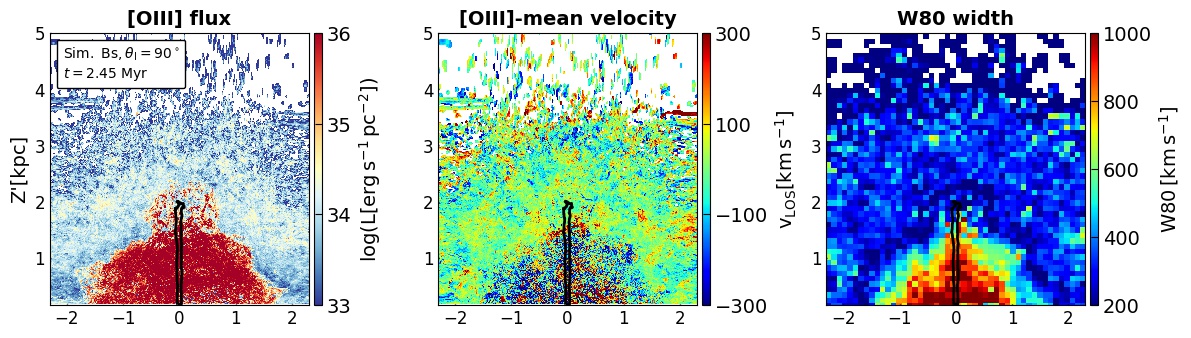}
     \end{tabular}}
 \centerline{
\def\arraystretch{1.0}
\setlength{\tabcolsep}{0.0pt}
\begin{tabular}{lcr}
     \hspace{-1cm}
    \includegraphics[width=\linewidth]{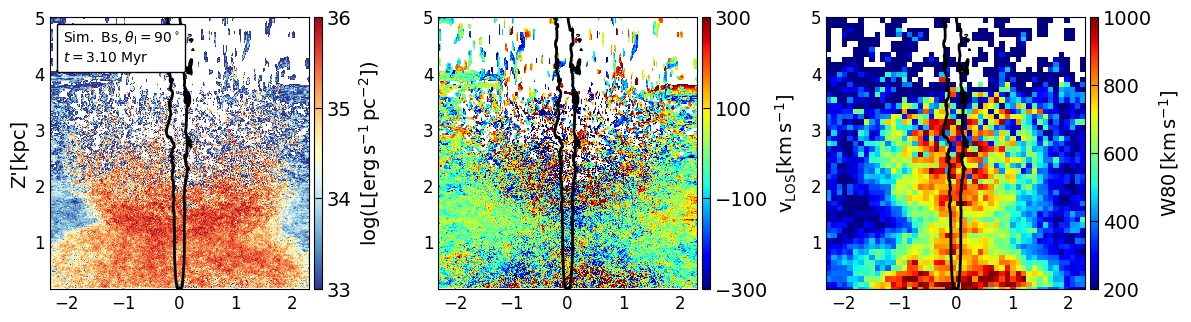}
     \end{tabular}}
\centerline{
\def\arraystretch{1.0}
\setlength{\tabcolsep}{0.0pt}
    \begin{tabular}{lcr}
     \hspace{-1cm}
   \includegraphics[width=\linewidth]{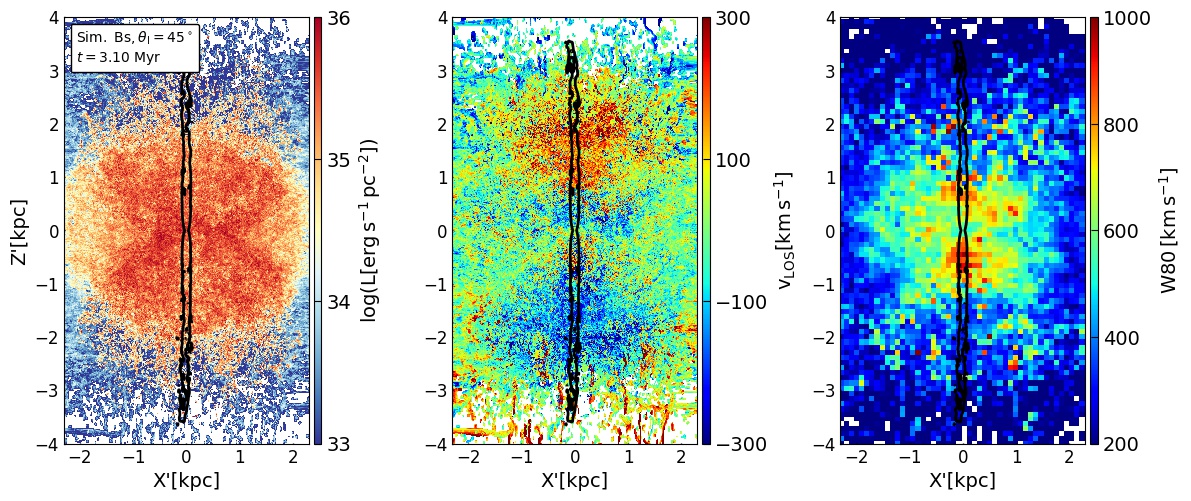} 
     \end{tabular}}
       \caption{\textbf{Left to right:} Integrated [\ion{O}{III}] flux, [\ion{O}{III}] emissivity weighted mean LOS velocities, and W80 widths at different viewing angles ($\theta_\mathrm{I}$) for Sim.~Bs ($\mathrm{P_J=10^{45}}\ergs$) at the mentioned times. The edge-on view maps in the top and middle panel are shown for the upper half of the spherical system, and in the bottom panel the quantities in the upper half are reflected in the bottom region to produce the final projected images}. The azimuthal angle $\phi_\mathrm{I}$ is fixed at $270^\circ$ for all the plots. The black contour shows the projected jet tracer at a value of 0.5 (maximal value is 1).
\label{fig:O_SimBs}
\end{figure*}

\begin{figure}
 \centerline{
\def\arraystretch{1.0}
\setlength{\tabcolsep}{0.0pt}
\begin{tabular}{lcr}
  \hspace{-0.5cm}
      {\includegraphics[width=\linewidth]{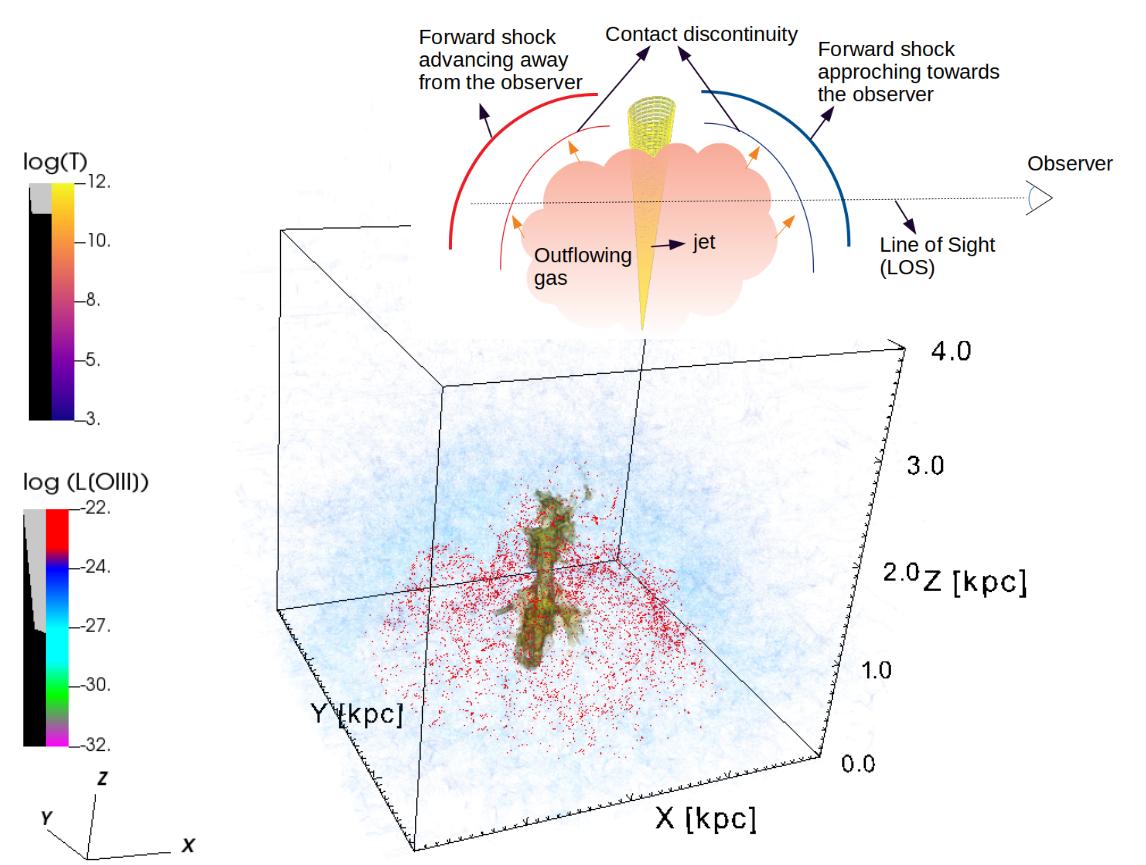}}
     \end{tabular}}
\caption{3D volume rendering of the [\ion{O}{III}] emissivity (i.e. $\log (L[\ion{O}{III}])$) where $L$ is in the units of $\ergs\,\mathrm{cm^{-3}}$ and temperature ($\log (T))$ of Sim.~Bs at 2.45~Myr, prepared using \textsc{visit} visualization software. The high [\ion{O}{III}] emission (in red) is limited to the regions in close vicinity of the jet (hot plasma in yellow), as also shown in Fig.\ref{fig:O_SimBs}. The grey-black bar shows the opacity level for the corresponding emissivity value (black means highly obscured). On the top right is a pictorial depiction showing LOS of an observer passing through the lateral outflows driven by the jet in the upper domain of the spherical systems.}
  \label{vol_render}
\end{figure}

\begin{figure*}
 \centerline{
\def\arraystretch{1.0}
\setlength{\tabcolsep}{0.0pt}

\begin{tabular}{lcr}
     \hspace{-1cm}
    \includegraphics[width=0.34\linewidth]{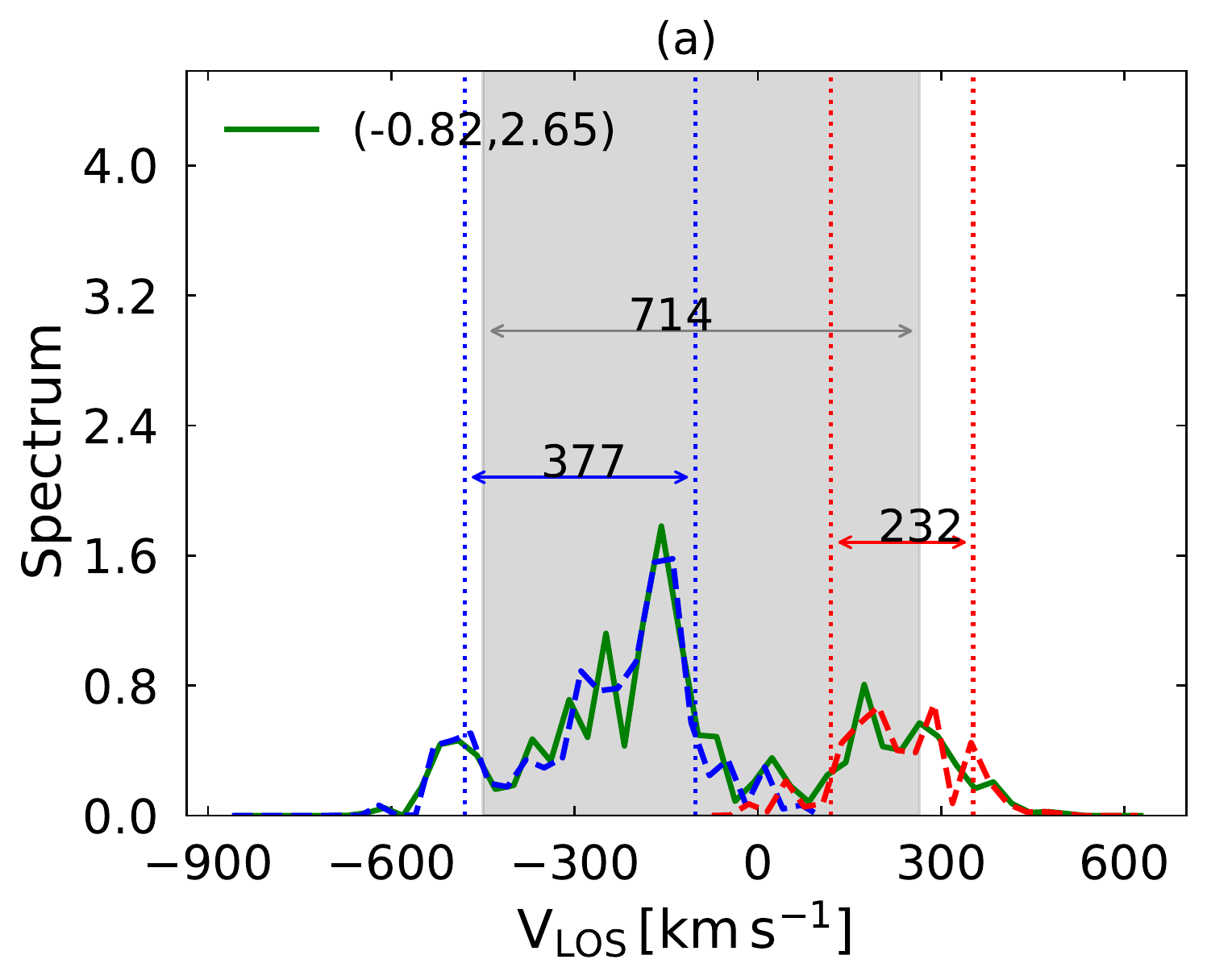} &
    \includegraphics[width=0.34\linewidth]{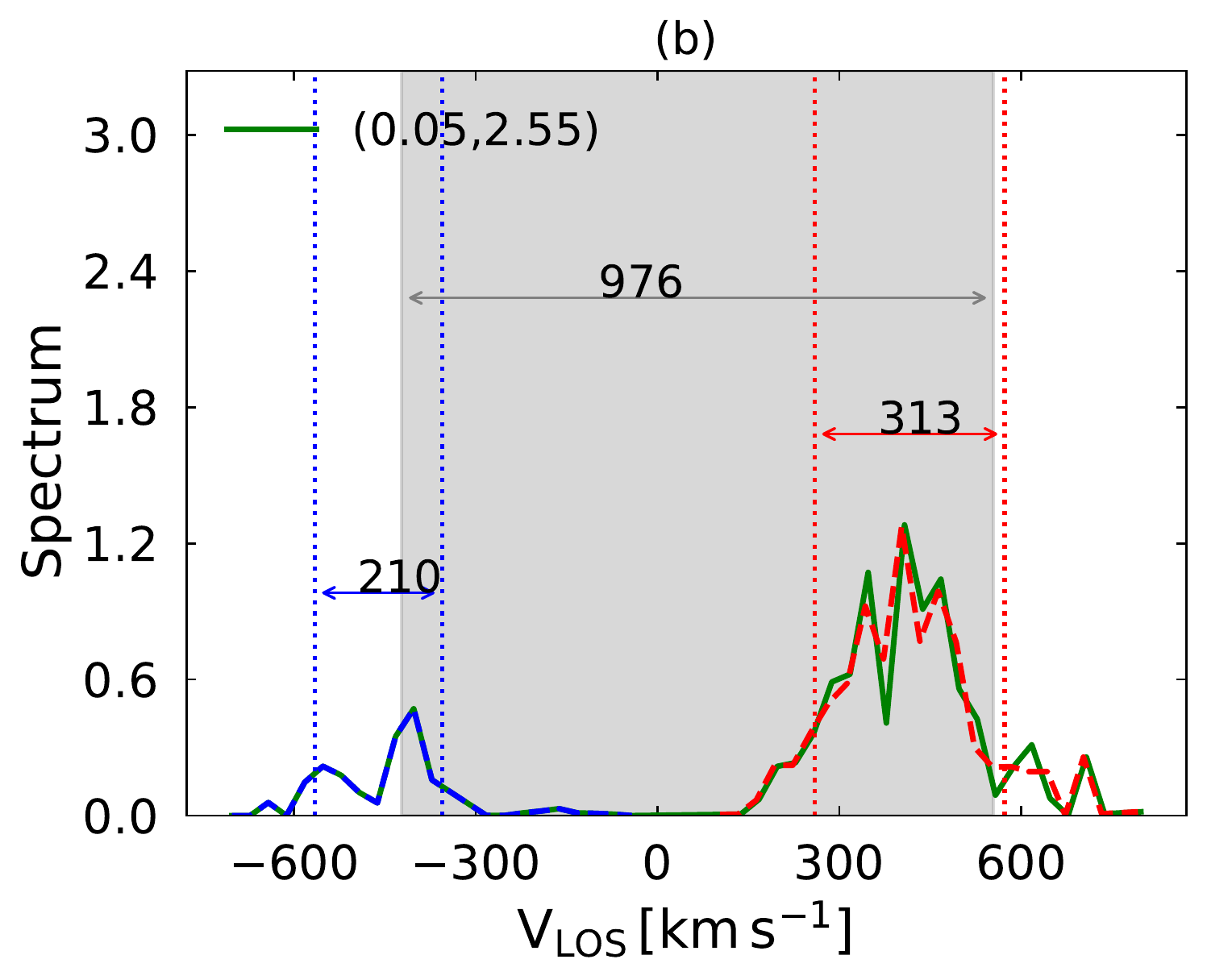} & 
    \includegraphics[width=0.345\linewidth]{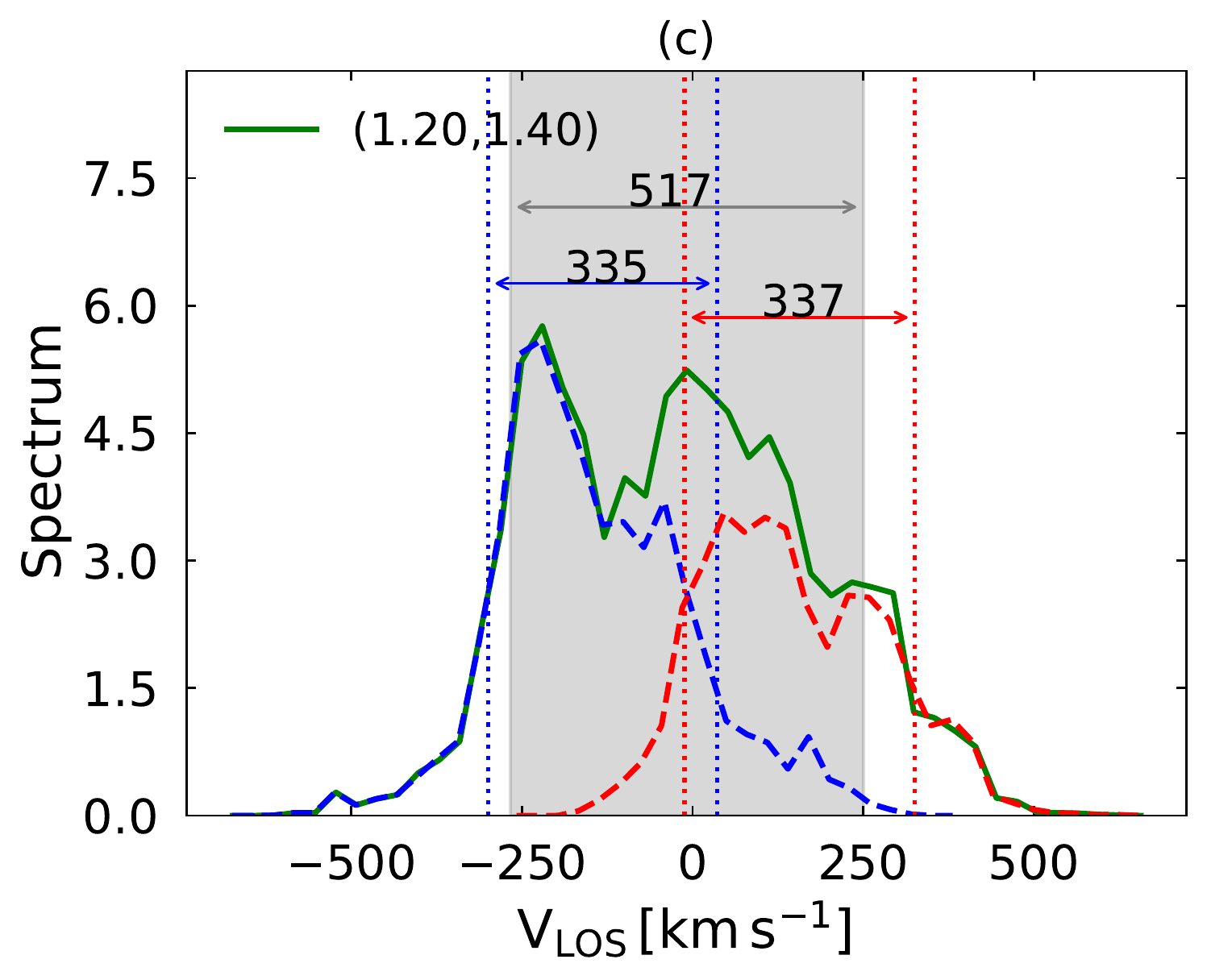} 
     \end{tabular}}

\centerline{
\def\arraystretch{1.0}
\setlength{\tabcolsep}{0.0pt}
\begin{tabular}{lcr}
     \hspace{-1cm}
    \includegraphics[width=0.34\linewidth]{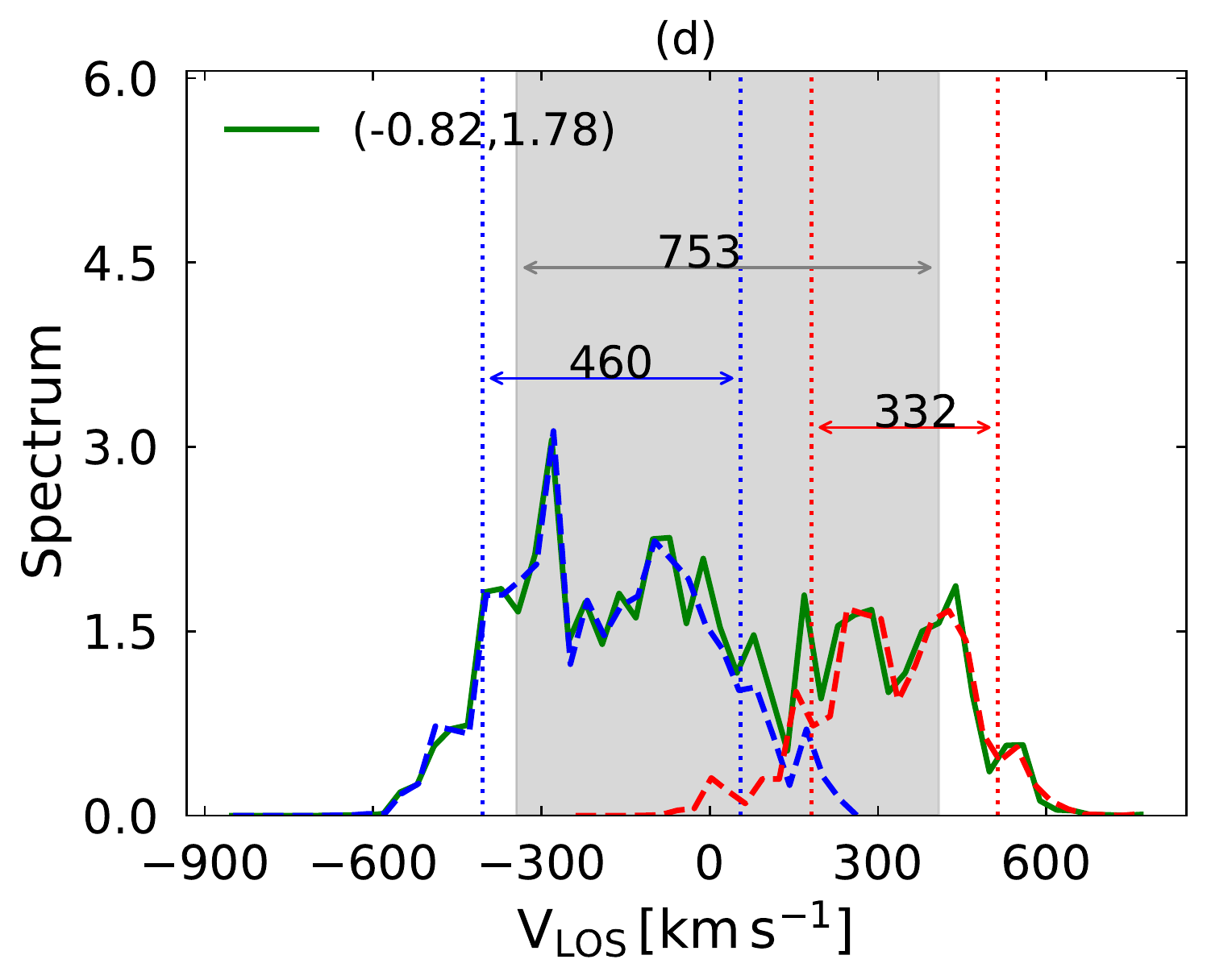} &
    \includegraphics[width=0.34\linewidth]{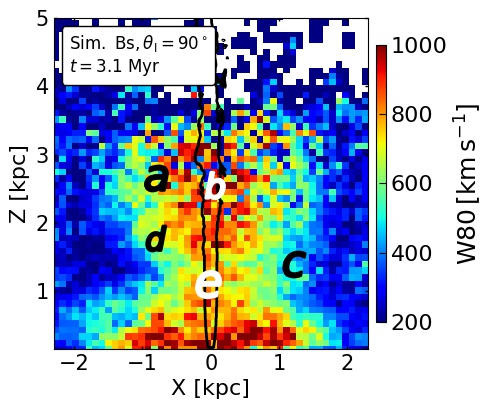}  &
    \includegraphics[width=0.345\linewidth]{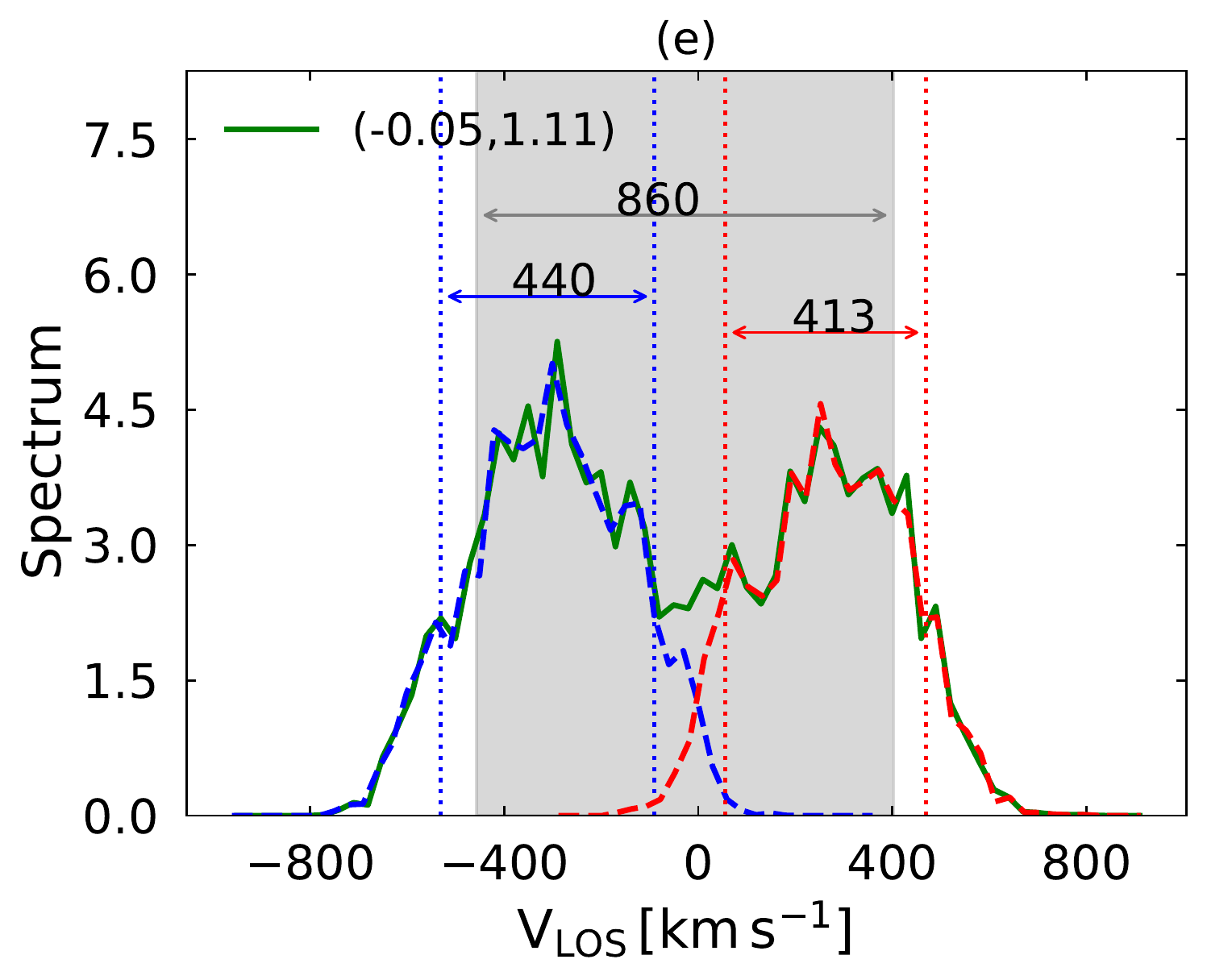}
     \end{tabular}}
  
       \caption{[\ion{O}{III}] ($\lambda$ = 500.7 nm) emission spectrum (normalised to $10^{36}\, \mathrm{erg\,s^{-1}\,(km\,s^{-1})^{-1}}$) for calculating W80 widths at various domains (represented using letters `a' to `e') in the edge-on view image plane for Sim.~Bs ($\mathrm{P_J=10^{45}}\ergs$) at 3.1~Myr.  The black contour shows the projected jet tracer at a value of 0.5  (maximal value is 1). The spectrum plot for the domains are labelled with the same letter which is used in the W80 map and the coordinates (in kpc) of the centre of the corresponding domain are also shown. The green curve shows the total spectrum for the full domain (100 $\pc^2$), and the gray shaded regions represent the W80 width for the total spectrum. The individual spectra for the regions in -Y and +Y direction are plotted as blue and red dashed curves, respectively, where the former axis is closer to the observer and vice-versa. The region between dotted blue and red vertical lines shows the W80 widths for the spectrum in regions at -Y and +Y axis (shown with blue and red arrows, respectively).}

\label{fig:w80_Sim_Bs}
\end{figure*}

\begin{figure}
    \centering
    \includegraphics[width=0.8\linewidth]{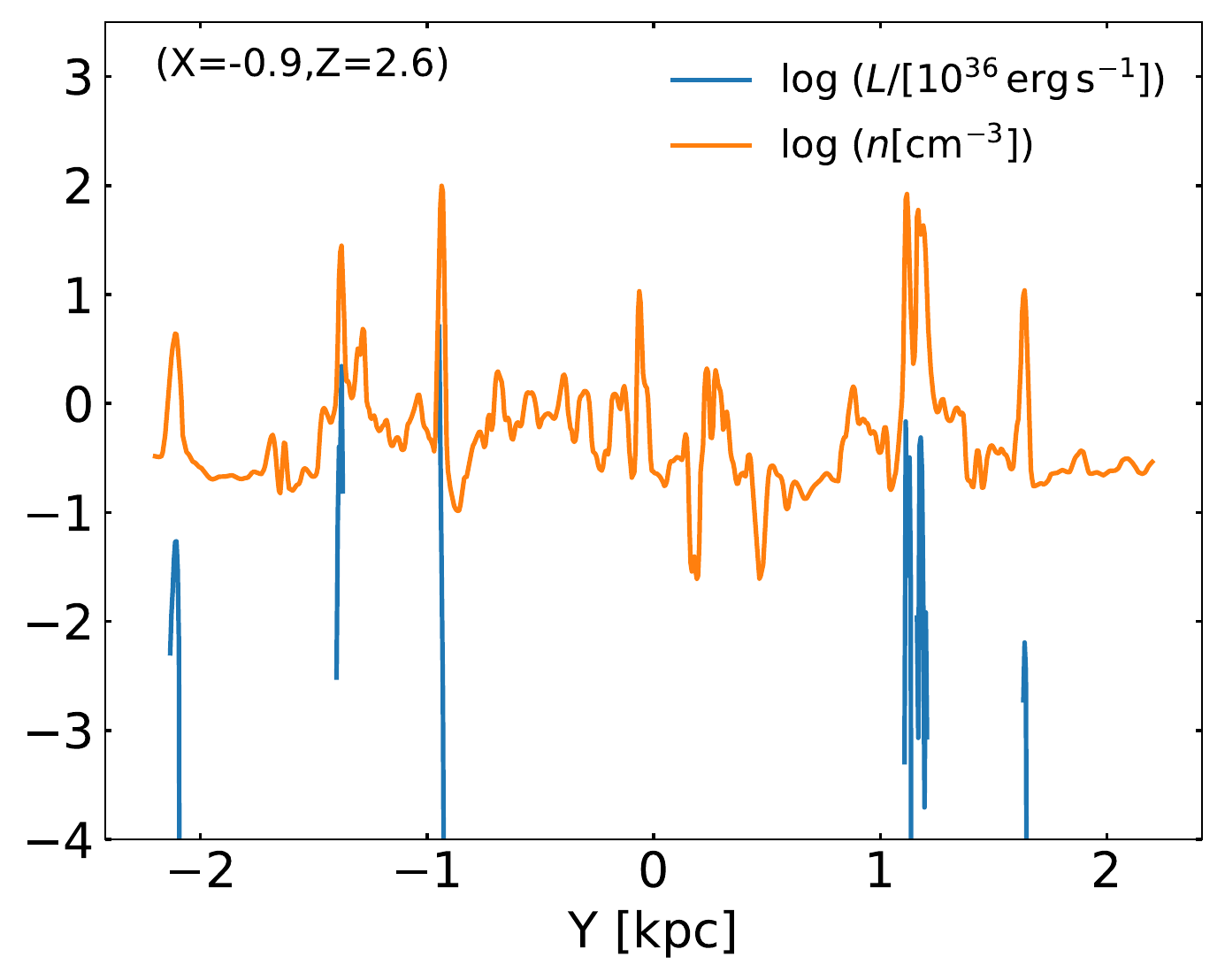}
    \caption{[\ion{O}{III}] luminosity (scaled to $10^{36}\ergs$) and Density [$\mathrm{cm^{-3}}$] along a LOS for Sim.~Bs at 3.1~Myr. The coordinates of the LOS on the image plane (X,Z) are mentioned at the top-left. The shocked gas clumps are the dominant source of shocked emission along the LOS (see Fig.~\ref{fig:cool-curve} for [\ion{O}{III}] cooling curve), as other regions are either heated to much higher temperatures or left unhindered by the shocks.}
    \label{fig:den_flux}
\end{figure}

\begin{figure*}
\centerline{
\def\arraystretch{1.0}
\setlength{\tabcolsep}{0.0pt}
\begin{tabular}{lcr}
     \hspace{-1cm}
    \includegraphics[width=\linewidth]{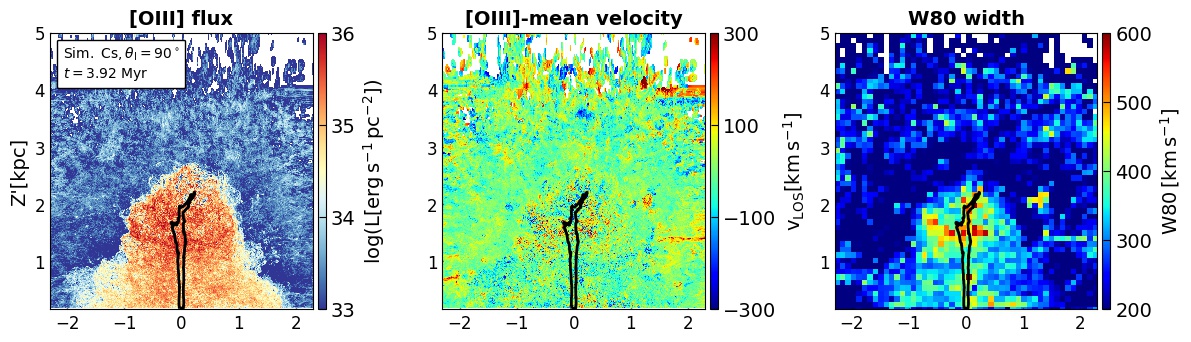}
     \end{tabular}}
       
 \centerline{
\def\arraystretch{1.0}
\setlength{\tabcolsep}{0.0pt}
\begin{tabular}{lcr}
     \hspace{-1cm}
    \includegraphics[width=\linewidth]{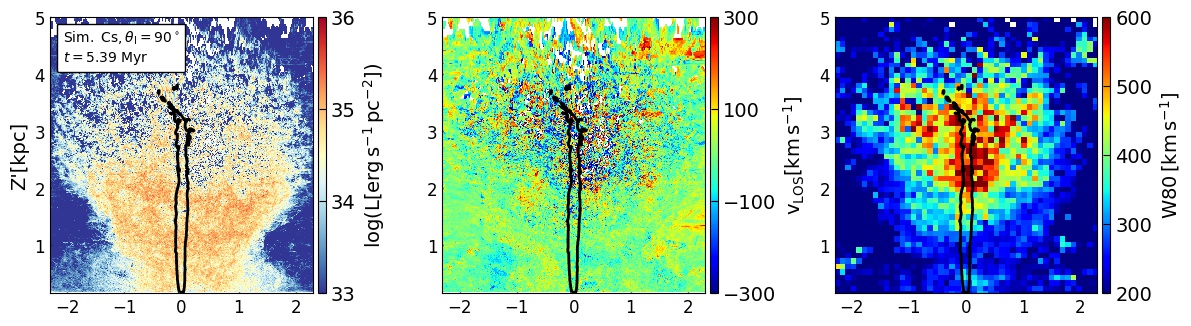}
     \end{tabular}}
\centerline{
\def\arraystretch{1.0}
\setlength{\tabcolsep}{0.0pt}
    \begin{tabular}{lcr}
     \hspace{-1cm}
   \includegraphics[width=\linewidth]{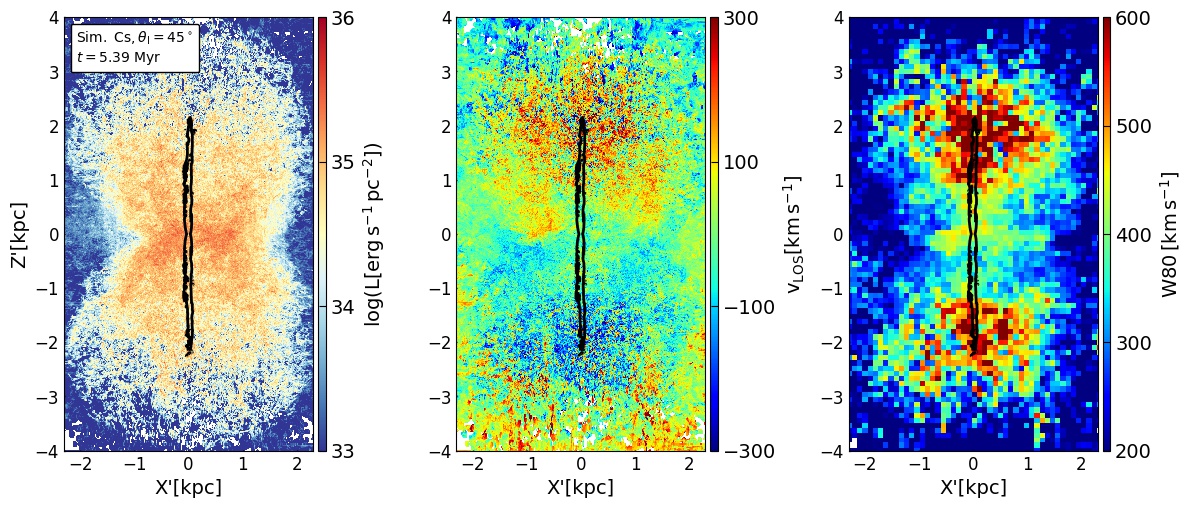} 
     \end{tabular}}
       \caption{\textbf{Left to right:} Same as Fig.~\ref{fig:O_SimBs} for Sim.~Cs ($\mathrm{P_J=10^{44}}\ergs$) at the mentioned times. The black contour shows the projected jet tracer at a value of 0.5  (maximal value is 1).}
\label{fig:O_SimCs}
\end{figure*}

\begin{figure*}
 \centerline{
\def\arraystretch{1.0}
\setlength{\tabcolsep}{0.0pt}
\begin{tabular}{lcr}
     \hspace{-1cm}
    \includegraphics[width=\linewidth]{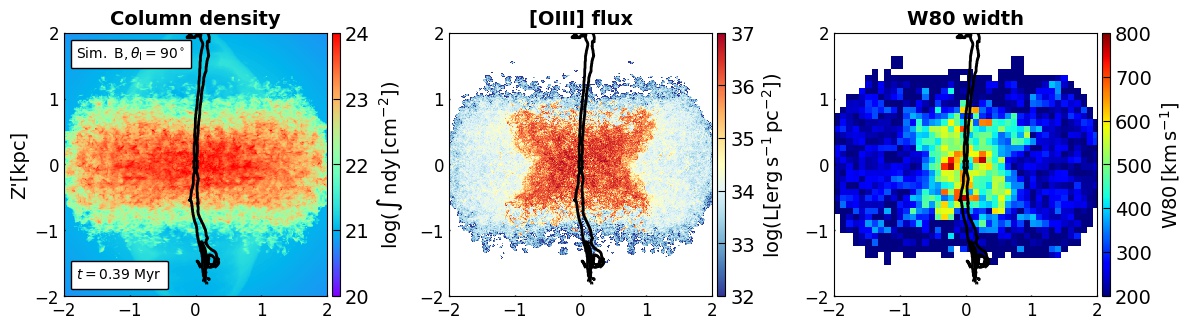}
     \end{tabular}}
\centerline{
\def\arraystretch{1.0}
\setlength{\tabcolsep}{0.0pt}
    \begin{tabular}{lcr}
     \hspace{-1cm}
   \includegraphics[width=\linewidth]{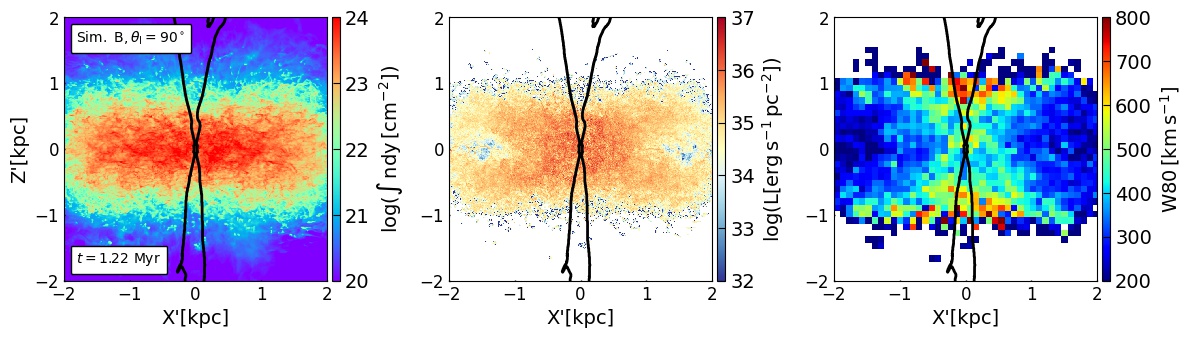} 
     \end{tabular}}
       \caption{\textbf{Left to right:} Column Density, integrated [\ion{O}{III}] flux, W80 widths at the edge-on view ($\theta_\mathrm{I}=90^{\circ},\phi_\mathrm{I}=270^{\circ}$) for Sim.~B ($\theta_\mathrm{J}=0^\circ$,$\mathrm{P_J=10^{45}}\ergs$) at different snapshots of time. The black contour shows the projected jet tracer at a value of 0.5 (maximum value is 1).}
\label{fig:OIII-simB}
\end{figure*}

\section{Results for the thermal emission in spherical systems}\label{results_spherical}
In this section we discuss the observed properties of shocked gas in simulations of relativistic jets emerging from a spherically distributed ISM, viz. simulations Bs and Cs in Table~\ref{tab:sim_table}. We first present the results of the evolution of the morphology of [\ion{O}{III}] emission and the nature of the gas kinematics for simulation Bs, with a kinetic jet power of $\mathrm{P_J} = 10^{45}~\ergs$. The different maps at the edge-on view are shown for the upper-half of the spherical system, as the values in the lower-half are expected to be a reflected image of the former. However for the inclined view, we reflect the physical quantities from the upper to the lower half and show the total integrated values on the image plane. We subsequently compare the results with that in Sim.~Cs, which has a jet of power lower by an order of magnitude ($\mathrm{P_J} = 10^{44}~\ergs)$.

\subsection{Evolution of different observed quantities}

We perform the post-processing analysis to study the evolution and morphology of various observed quantities for the higher-power jetted system, Sim.~Bs ($\mathrm{P_J}=10^{45}\,\ergs$), at 2.45~Myr and 3.1~Myr. At the former time, the jet is confined\footnote{This implies that the jet is restricted in the galaxy potential and is yet to break out.} in the galaxy, whereas the jet has escaped from the system in the latter, which is also the last snapshot of this simulation.
We show the plots for integrated [\ion{O}{III}] flux, LOS velocities, and W80 widths for these snapshots at different image planes in Fig.~\ref{fig:O_SimBs}. The projected jet tracer\footnote{The jet-tracer is a passive scalar quantity in \textsc{pluto} which gets advected with fluid flow. It represents the regions filled with jet plasma inside the computational domain. More details can be found in Chapter 2 of the \textsc{pluto} user-guide, which can be downloaded at:
\url{http://plutocode.ph.unito.it/userguide.pdf}} at a value of 0.5 is shown with black contours. We describe our findings in the sections below.

\subsubsection{\textbf{\texorpdfstring{Morphology of [\ion{O}{III}]}{OIII} surface brightness}}
\label{spherical_emiss}

As has been pointed out in M2016, the jet driven bubble heats-up the gas and produces strong shocks as it progresses through the system. One can see that the confined jet at 2.45~Myr in Fig.~\ref{fig:O_SimBs} produces enhanced shocked emission from the regions in the vicinity of the jet, also shown in the 3D volume rendered image in Fig.~\ref{vol_render}, which is prepared using \textsc{visit} \footnote{\url{https://wci.llnl.gov/simulation/computer-codes/visit}} visualization software. The vertical push and local outflows from the escaping jet create the bubble-shaped emission region at the jet head. However, after the jet-breakout from its spherical system, the intensity of emission is reduced due to decreased jet-ISM interaction. The jet no longer actively interacts with the gas now (M2016), which leads to a decrease in the integrated shocked emission from the spherical system at 3.1~Myr. The jet, while escaping from its spherical system, pushes and disperses the gas fragments in the vertical direction, which distorts the gas distribution and produces a hour glass shaped emission morphology when viewed edge-on. At lower inclination ($\theta_\mathrm{I}=45^\circ$), the projection of emitting regions appears ``biconic(al)" as one sees more of the laterally extended shocked gas in the projected image plane.

\subsubsection{\textbf{Spatial variation of gas kinematics}}
\label{spherical_kinematics}
\begin{itemize}
    \item \textbf{Morphology of W80 widths on the image plane:}\\  The confined jets in the galaxy are known to interact vigorously with the gas, causing very high velocities and broad line widths \citep{harrison15,dipanjan16}. As can be seen in Fig.~\ref{fig:O_SimBs}, the jet at 2.45~Myr is still actively interacting with the ISM, giving rise to much higher W80 widths (up to 1000~$\kms$) in the central 1~kpc of the spherical system. This is caused due to the local outflows and the jet-driven bubble, which expands with the jet evolution into the ISM. Even after the jet has broken out at 3.1~Myr, and evolved to large scales, high widths are spread along the whole jet path and also up to $\pm 1$~kpc perpendicular to the jet. This occurs due to the jet-injected spherical energy bubble which is sweeping through the dense gas (M2016), inducing local turbulence and causing strong outflows. As can be seen in the velocity map at 3.1~Myr ($\theta_\mathrm{I}=45^\circ$), the presence of jet induced laterally expanding forward shock of the energy bubble sweeping through the ISM causes bulk motions of the ISM gas towards and away from the observer (see cartoon image in Fig.~\ref{vol_render}). This causes large W80 widths in the regions affected by the jet, which we discuss in detail below. At $\theta_\mathrm{I}=45^\circ$ image plane, W80 widths larger than $600~\kms$ values are extended in direction perpendicular to the jet, following the high emission regions at this orientation (see left panel).
   
\item \textbf{Nature of synthetic [\ion{O}{III}] spectra:}
In Fig.~\ref{fig:w80_Sim_Bs}, we show the [\ion{O}{III}] spectra for some domains (marked using letters `a' to `e') in the edge-on image plane for Sim.~Bs at 3.1~Myr. These domains are chosen to explore the spectral variation in the regions closer and farther away from the jet, extending to $\pm~1$~kpc perpendicular to the jet and $3$~kpc along the vertical direction.
The green curve in the spectra represents the total spectrum for each domain ($100~\pc^2$), and the shaded gray regions show the total W80 width. In order to investigate the impact of the outflows driven by the radial expansion of the energy bubble, we compute the spectra separately for the two halves of the Y axis, along which a spectrum is computed. They are represented by the blue and red dashed curves. which represents the spectrum along the negative and positives halves of the Y axis, respectively. The W80 widths for each half is represented by the dashed blue and red vertical lines. This is done to capture the advancing and receding parts of the energy bubble respectively. The extent of W80 widths for the two individual (blue and red curves) and the total spectrum (green) is further marked with blue, red and gray colored double-arrows, respectively. 

We firstly note that the spectra appear spiky with small scale features. This occurs as the sampling of the [\ion{O}{III}] emitting regions along the LOS is not continuous. A given LOS crosses discrete clouds and filaments as it probes the computational volume, of which only some of regions are shocked to emit in [\ion{O}{III}] (see Fig.~\ref{fig:den_flux}). This can cause spikes at some places when the spectra from various LOS are overlapped to construct the final spectra for the domain. However, convolving the spectra over broader spatial regions will average the effects of individual clouds and produce smoother distributions.
We find that most of the domains exhibit double-peaked spectra which occurs due to the laterally expanding forward shock of the energy bubble sweeping through the ISM (see Fig.~\ref{vol_render}). One peak in these spectra denotes the receding half and the other shows the advancing half region, as shown in the velocity map in the lower panel of Fig.~\ref{fig:O_SimBs}. However, each of the peaks is also broad with widths of 200-500$~\kms$, which is solely due to internal turbulence and local outflows in each of the expanding half regions. We obtain a varied nature of the spectra at different spatial locations which depends on the nature of the underlying gas distribution and the strength of the jet-ISM interaction. At some domains the spectra have a broader distribution, as they originate from regions that have high column depth, being inside the galaxy's core radius. The other regions are in the outskirts and sample a few clouds and hence well separated in peaks. Nonetheless, all of these spectra clearly illustrate that the large-scale outflows along the counter axes give rise to double peaked spectrum, which leads to higher W80 widths (up to 1000~$\kms$) when integrated along the LOS.

\subsubsection{\textbf{Effect of the lower-power jet on observed quantities}}

For the lower-power jetted system Sim.~Cs ($\mathrm{P_J}=10^{44}~\ergs$), we study the snapshots at 3.92~Myr and 5.39~Myr respectively. We show different observed measures at different image planes for these snapshots in Fig.~\ref{fig:O_SimCs}. The jet in Sim.~Cs reaches a height of 2.5~kpc at 3.92~Myr, which the higher power jet reaches at a much earlier time (~2.45~Myr, see Fig.~\ref{fig:O_SimBs}). This is because the lower-power jet has to work more to clear the gas due to its lower available momentum. However, the extent of the [\ion{O}{III}] emitting region is similar to that in Sim.~Bs. This results due to the longer confinement time for the lower-power jet in Sim.~Cs, enabling it to spread out in a large volume and shock the ISM, as pointed out in M2016. The surface brightness of the [\ion{O}{III}] emission, however, is lower here when compared to Sim.~Bs. This is because the shocks from the jet in Cs are weaker than its higher-power counterpart, and hence the dense regions are less likely to be heated to high temperatures ($T\sim10^5$~K), leading to a lower integrated shocked [\ion{O}{III}] emission. As the confined jet in Fig.~\ref{fig:O_SimCs} slowly pushes against the ISM and evolves in size, it produces a fountain-shaped emission when viewed edge-on at 5.39~Myr. This morphology appears nearly ``biconical-shaped" at lower inclinations.

The right panel in Fig.~\ref{fig:O_SimCs} shows that the confined lower-power jet produces widths of up to 400-600~$\kms$ near the jet-head. As compared to the confined higher-power jet, which produces high W80 widths in wide regions (see Fig.~\ref{fig:O_SimBs}), the high-widths here are limited to regions close to the jet working surface. The lower-power jet finds it difficult to clear its path (see M2016), and consequently, loss in momentum of such jets can result in lower values of W80 relative to a higher-power jet. At 5.39~Myr, when the jet in Sim.~Cs has broken out from the ISM, only the regions near the jet-head are found to have high W80 widths ($>500~\kms$). The turbulence in the regions below the jet-head decay, and result in lowering of the W80 values along the jet-base. As can be seen in the velocity map at 5.39~Myr ($\theta_\mathrm{I}=45^\circ$), the lower-power jet, similar to Sim.~Bs, also drive large scale outflows along its axis; however,the kinetic impact on the gas is less at the jet base, causing low velocities and settling and in-fall of the gas in the presence of external gravitational potential (found from the velocity maps of upper half plane only). This is contrary to Sim.~Bs, where we observed high-widths near the jet base as the bulk gas outflows in these regions are maintained even after the jet was no longer actively interacting with the ISM.

\end{itemize}

\section{Results for the thermal emission in gas discs}
\label{results_disc}

In the subsequent sections, we present the expected morphology and kinematics of the shocked emission in kpc scale gas discs. We discuss the dependence of various observed quantities, such as column density, flux and W80 widths on the orientation of the observer in different jetted systems in Sec.~\ref{edge-on}. In Sec.~\ref{pv-maps}, we discuss how the presence of jet-ISM interaction can shape the velocity field and thus, affect the observed rotation profile of the disc.

\begin{figure*}
 \centerline{
\def\arraystretch{1.0}
\setlength{\tabcolsep}{0.0pt}
\begin{tabular}{lcr}
     \hspace{-1cm}
    \includegraphics[width=\linewidth]{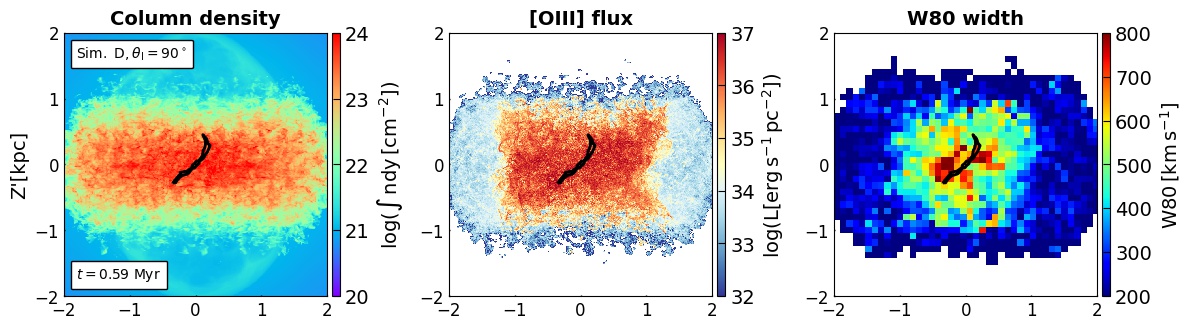}
     \end{tabular}}
\centerline{
\def\arraystretch{1.0}
\setlength{\tabcolsep}{0.0pt}
    \begin{tabular}{lcr}
     \hspace{-1cm}
   \includegraphics[width=\linewidth]{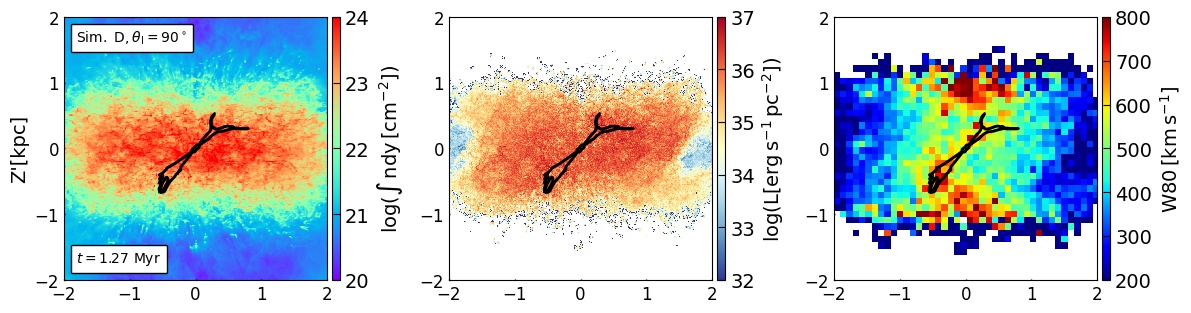} 
     \end{tabular}}
     
     \centerline{
\def\arraystretch{1.0}
\setlength{\tabcolsep}{0.0pt}
    \begin{tabular}{lcr}
     \hspace{-1cm}
   \includegraphics[width=\linewidth]{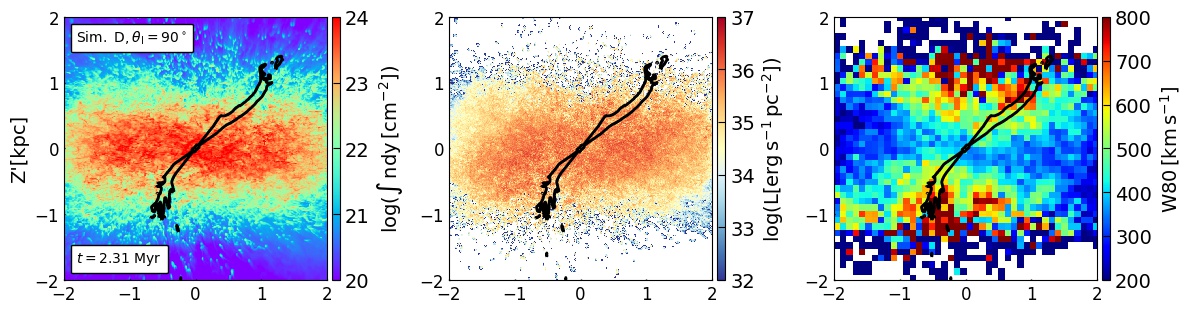}
     \end{tabular}}
     \centerline{
\def\arraystretch{1.0}
\setlength{\tabcolsep}{0.0pt}
    \begin{tabular}{lcr}
     \hspace{-1cm}
   \includegraphics[width=\linewidth]{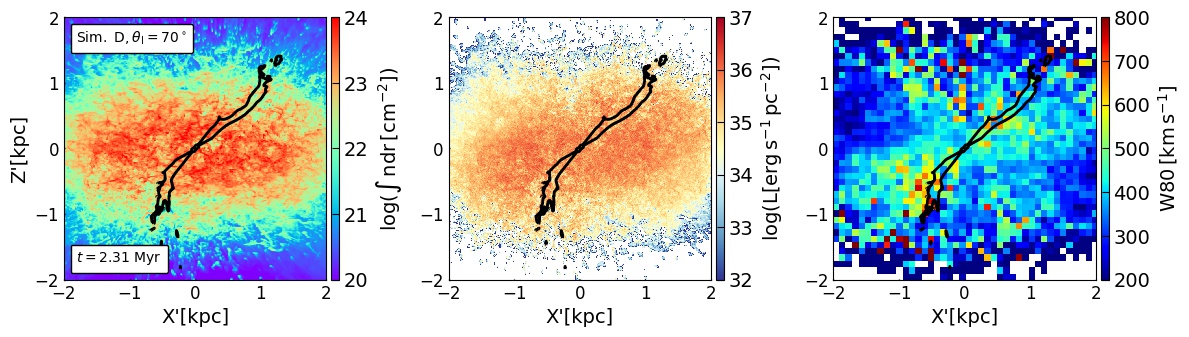} 
     \end{tabular}}
      \caption{\textbf{Left to right:} Column Density, integrated [\ion{O}{III}] flux, W80 widths on the mentioned image planes for Sim.~D ($\theta_\mathrm{J}=45^\circ$,$\mathrm{P_J=10^{45}}\ergs$) at different snapshots of time. The black contour shows the projected jet tracer at a value of 0.5 (maximum value is 1).}
\label{fig:OIII-simD}
\end{figure*}

\begin{figure*}
 \centerline{
\def\arraystretch{1.0}
\setlength{\tabcolsep}{0.0pt}
\begin{tabular}{lcr}
     \hspace{-1cm}
    \includegraphics[width=\linewidth]{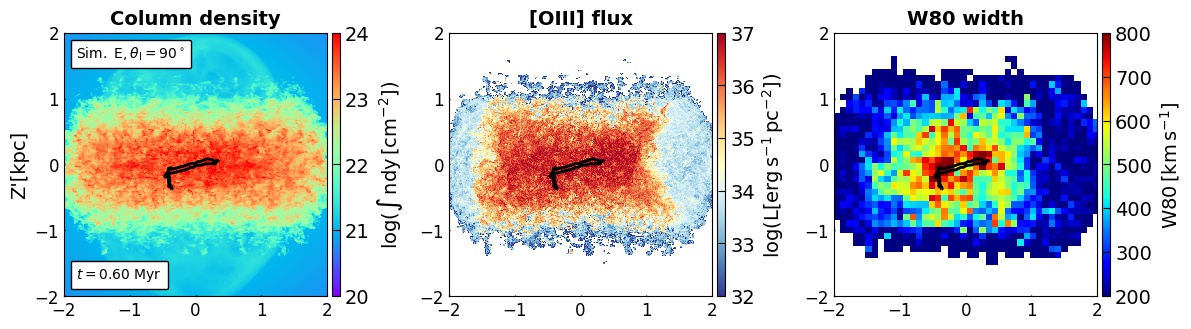}
     \end{tabular}}
\centerline{
\def\arraystretch{1.0}
\setlength{\tabcolsep}{0.0pt}
    \begin{tabular}{lcr}
     \hspace{-1cm}
   \includegraphics[width=\linewidth]{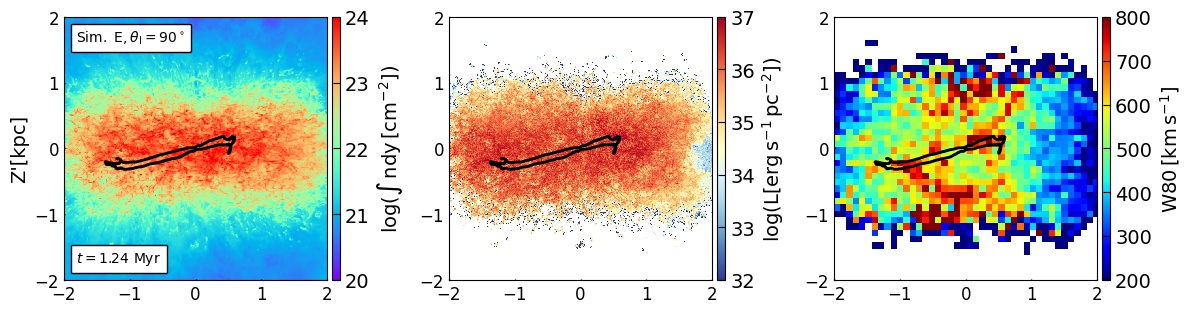} 
     \end{tabular}}
      \centerline{
\def\arraystretch{1.0}
\setlength{\tabcolsep}{0.0pt}
    \begin{tabular}{lcr}
     \hspace{-1cm}
    \includegraphics[width=\linewidth]{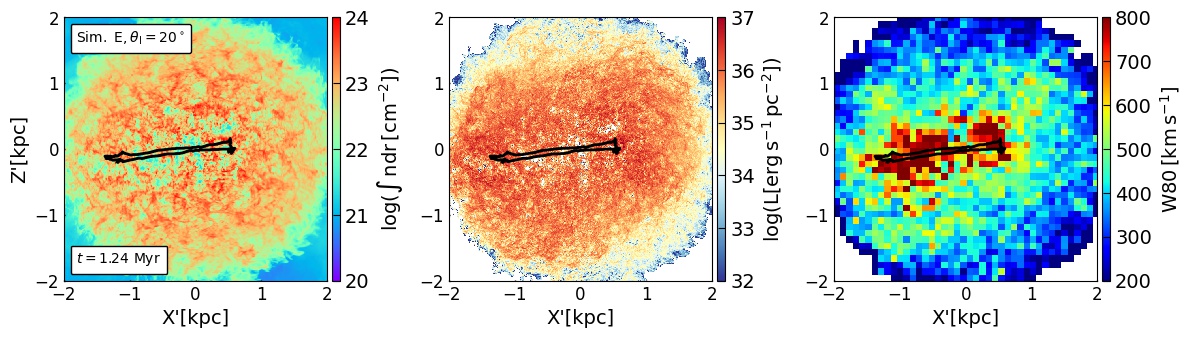}
     \end{tabular}}
       \caption{\textbf{Left to right:} Column Density, integrated [\ion{O}{III}] flux, W80 widths on the mentioned image planes for Sim.~E  ($\theta_\mathrm{J}=70^\circ$,$\mathrm{P_J=10^{45}}\ergs$) at different snapshots of time. The black contour shows the projected jet tracer at a value of 0.5 (maximum value is 1).}
\label{fig:OIII-simE}
\end{figure*}

\begin{figure*}
 \centerline{
\def\arraystretch{1.0}
\setlength{\tabcolsep}{0.0pt}
\begin{tabular}{lcr}
     \hspace{-1cm}
     {\includegraphics[width=0.88\linewidth]{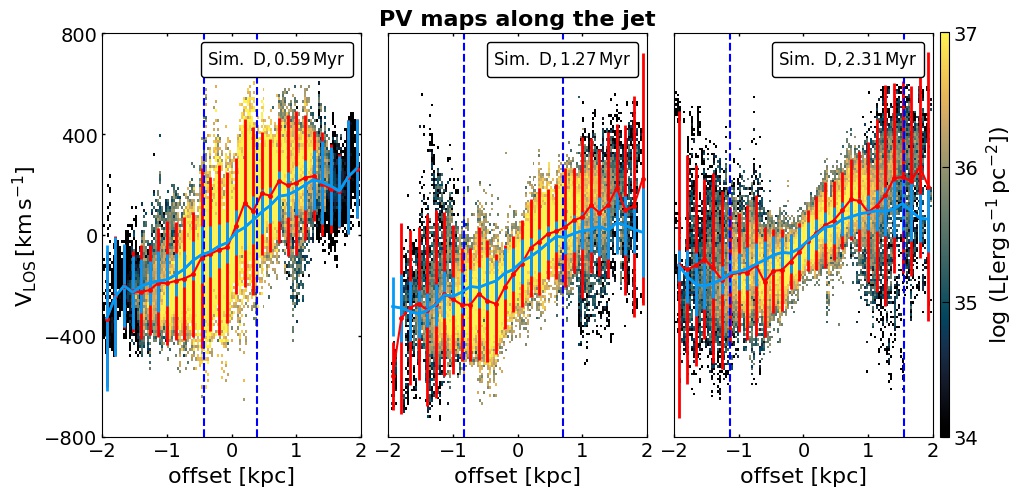}}
     \end{tabular}}
\caption{Time evolution of PV-diagrams for the slit placed along the jet ($S_0$) for Sim.~D ($\theta_\mathrm{J}=45^\circ, \mathrm{P_J}=10^{45}\ergs$) at the edge-on view ($\theta_\mathrm{I}=90^{\circ},\phi_\mathrm{I}=270^{\circ}$) of disc. The column density, W80 maps and jet tracer at these snapshots are shown in Fig.~\ref{fig:OIII-simD}. The red curve shows the mean velocity curve with $\pm2\sigma$ deviation from the mean values} for the Sim.~D (jetted system in general) at the mentioned times. From left to right, the light-blue curve depicts the mean velocity with $\pm2\sigma$ deviation for the no-jet simulation at 0.59 Myr, 1.23 Myr, and 1.37 Myr, respectively. The blue vertical dashed lines represent the extent of the jet (at a value of 0.5) tracer along the slits.
 \label{fig:pv_edge}
\end{figure*}

\begin{figure*}
 \centerline{
\def\arraystretch{1.0}
\setlength{\tabcolsep}{0.0pt}
\begin{tabular}{lcr}
     \hspace{-1cm}
    \includegraphics[width=0.33\linewidth]{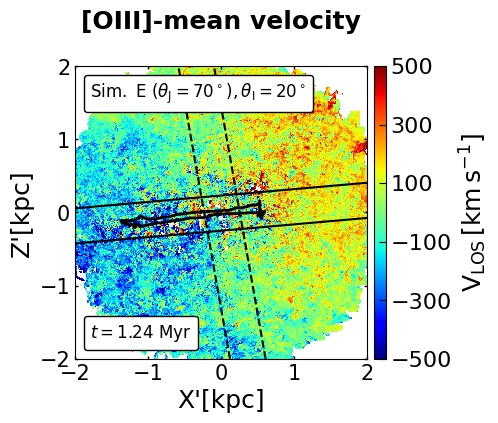}&
     \includegraphics[width=0.57\linewidth]{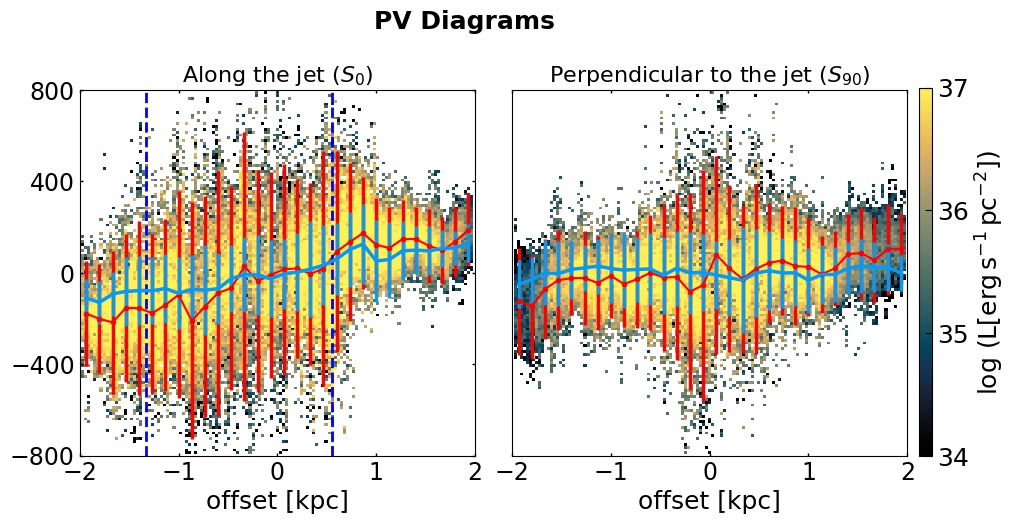}
     \end{tabular}}
     \caption{\textbf{Left:} [\ion{O}{III}]-weighted mean velocity observed at $\theta_\mathrm{I}=20^\circ,\phi_\mathrm{I}=270^\circ$ image plane for Sim.~E ($\theta_\mathrm{J}=70^\circ, \mathrm{P_J}=10^{45}\ergs$) at 1.24~Myr. Two slits of 500~pc widths are placed, where slit along the jet ($S_0$) is shown by bold black lines and dashed lines mark the slit placed perpendicular to the jet ($S_{90}$). The projected jet tracer at a value of 0.5 is shown by black contour.
     \textbf{Middle and Right:} PV diagram for the slits $S_0$ and $S_{90}$, respectively.
     The color convention for the mean velocity curves for the jetted and non-jetted system (at 1.23~Myr) is similar to that used in Fig.~\ref{fig:pv_edge}. The blue vertical dashed lines represent the extent of the jet tracer along the slit $S_0$.}
\label{fig:pv-simE}
\end{figure*}

\begin{figure*}
\centerline{
\def\arraystretch{1.0}
\setlength{\tabcolsep}{0.0pt}
\begin{tabular}{lcr}
     \hspace{-.5cm}
     {\includegraphics[width=0.88\linewidth]{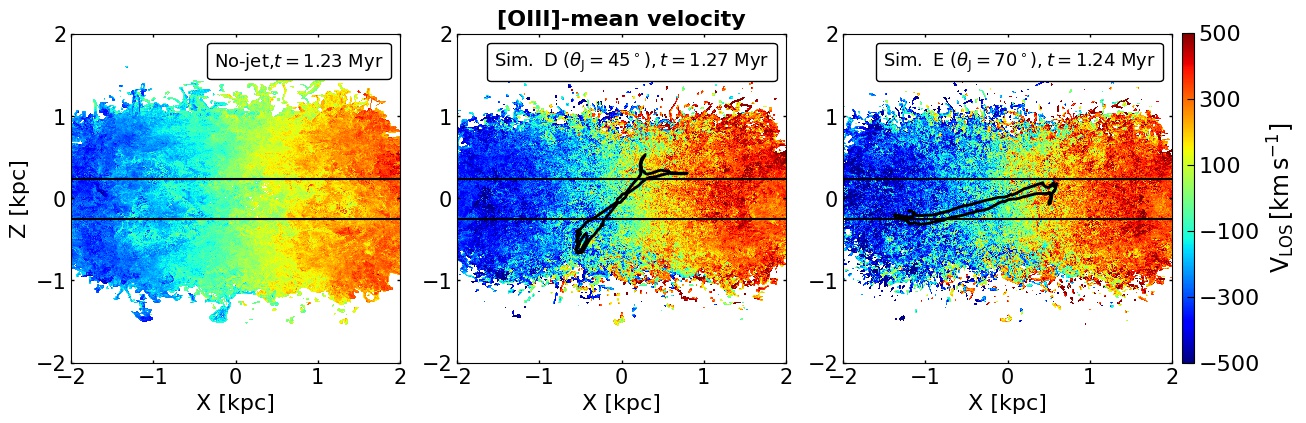}}
     \end{tabular}}
 \centerline{
\def\arraystretch{1.0}
\setlength{\tabcolsep}{0.0pt}
\begin{tabular}{lcr}
     \hspace{-1cm}
     {\includegraphics[width=0.88\linewidth]{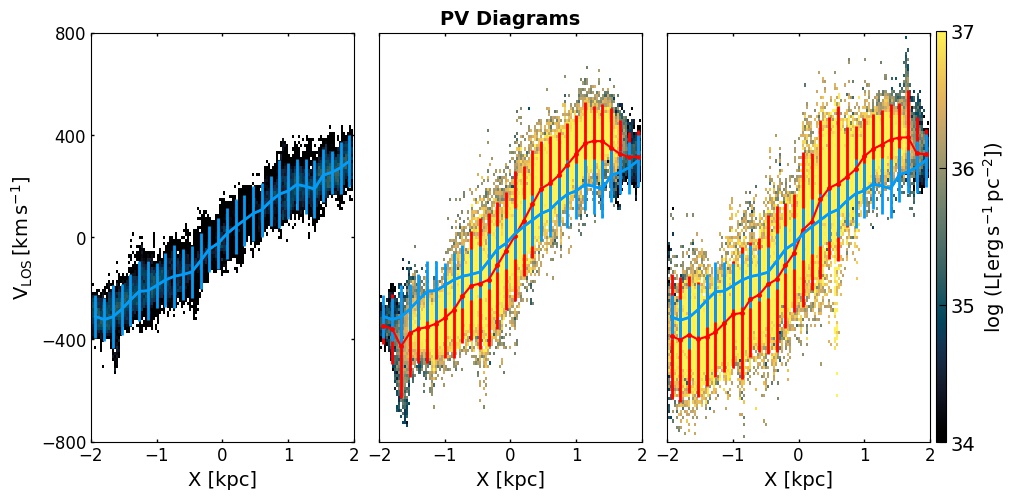}}
     \end{tabular}}
\caption{\textbf{Top:} [\ion{O}{III}] weighted mean velocity maps at the edge-on view ($\theta_\mathrm{I}=90^{\circ},\phi_\mathrm{I}=270^{\circ}$) for `no-jet' simulation (Left), Sim.~D ($\theta_\mathrm{J}=45^\circ$) (Middle) and Sim.~E ($\theta_\mathrm{J}=70^\circ$) at (Right) at similar times. The projected jet tracer at a value of 0.5 (maximum value is 1) is shown with black contour. \textbf{Bottom:} PV-diagrams for a slit of width 500~pc placed along the major axis of the disc, as shown in top row. The color convention for the mean velocity curves for the jetted (at mentioned times) and non-jetted case is similar to that used in Fig.~\ref{fig:pv_edge}. The PV plot shows the deviation of the disc's velocity field under the influence of the jets as compared to the rotation curve of settled disc.} 
  \label{fig:rot_curve}
\end{figure*}

\begin{figure}
\centering
   \includegraphics[width=\linewidth]{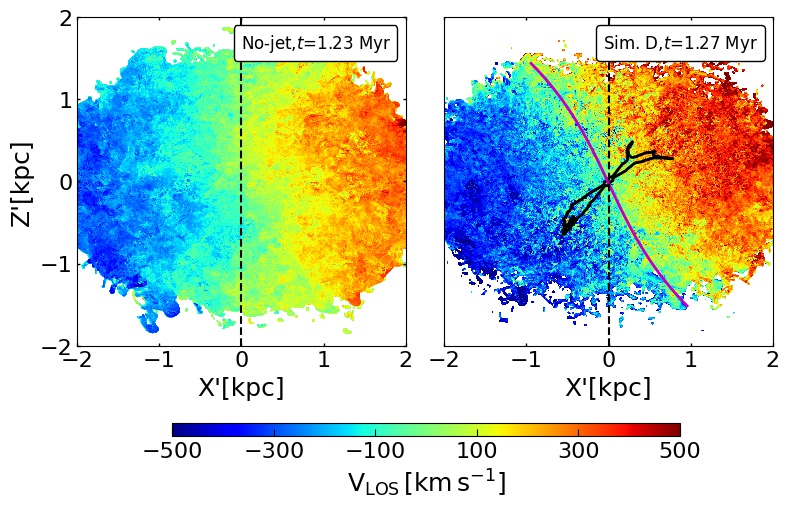}
    \caption{[\ion{O}{III}] weighted mean velocity map for `no-jet' simulation (\textbf{Left}) and Sim.~D ($\theta_\mathrm{J}=45^\circ$) (\textbf{Right}) at an orientation of $\theta_\mathrm{I}=70^\circ, \phi_\mathrm{I}=270^\circ$. The magenta curve traces the mean of the zero-velocity regions (kinematic centres) with $\mathrm{V_{LOS}}<\pm 10\,\kms$. The black vertical dashed line passes through the minor-axis of the disc (X' = 0). The black contour in the bottom panel shows the projected jet tracer at a value of 0.5 (maximal value is 1).}
    \label{fig:vel_70_deg}
\end{figure}

\subsection{\textbf{Evolution of different observed quantities}}
\label{edge-on}
In this section, we present the kinematics and morphology of [\ion{O}{III}] emission in kpc-scale gas discs interacting with a pair of relativistic jets ($\mathrm{P_J}=10^{45}\ergs$) inclined at different angles with respect to the minor axis, which are shown in Table~\ref{tab:sim_table}. 
For Sim.~B ($\theta_\mathrm{J}=0^\circ$), we focus on the snapshots at 0.39~Myr and 1.22~Myr, where at the former time the vertical jet just escapes out from the disc, and the other is the last snapshot of the simulation. We analyze three snapshots for Sim.~D ($\theta_\mathrm{J}=45^\circ$), i.e., 0.59~Myr, 1.27~Myr, and 2.31~Myr, where at the first two, the jet is confined inside the disc and in the third the jet has broken out of the disc and is evolved to large scales. The jet breakout time for this simulation is at around 1.37~Myr. For Sim.~E ($\theta_\mathrm{J}=70^\circ$), we present the results at 0.6~Myr and 1.24~Myr. In this case, the jet remains confined in the disc until the end of the simulation (i.e. 1.24~Myr).

We show the column density, integrated [\ion{O}{III}] flux, and W80 widths at the above mentioned times at different viewing angles for different simulations in Fig.~\ref{fig:OIII-simB},~\ref{fig:OIII-simD} and~\ref{fig:OIII-simE}. The projected jet tracer at a value of 0.5 is plotted with the black contour.

\subsubsection{\textbf{\texorpdfstring{Morphology of [\ion{O}{III}]}{OIII} surface brightness}}
\label{evol_emission_disc}
We find that the [\ion{O}{III}] emitting regions in all the disc simulations exhibit almost similar behavior with the jet evolution in their host's disc (see Fig.~\ref{fig:OIII-simB}, Fig.~\ref{fig:OIII-simD} and Fig.~\ref{fig:OIII-simE} for Sim.~B ($\theta_\mathrm{J}=0^\circ$), Sim.~D ($\theta_\mathrm{J}=45^\circ$) and Sim.~E ($\theta_\mathrm{J}=70^\circ$), respectively). We see strongly enhanced emission in the regions that are shocked by the expanding bubble or the jet streams. In Sim.~B, at the earlier stages of jet evolution (0.39~Myr), the high flux emitting region is confined to the central regions of the disc and appears to have an `X-shaped' morphology. This is because the outflows from the jet can easily expand in the lower densities perpendicular to the disc plane, whereas they are hindered by high column depths in the central parts, where the jet-driven bubble expands at a slower pace. In Sim.~D (0.59~Myr) and Sim.~E (0.6~Myr), this region exhibits a `rectangular-shaped' morphology, as the jets here are confined in the disc, and both the lobes are interacting with the disc gas on either side.

At later stages, the jet-driven bubble penetrates further into the ISM, driving outflows across the whole disc (M2018b), as indicated by the disrupted gas in the column density maps. This results both from the lateral expansion of the bubble into the disc starting from the central location, as well as the backflows impinging on the outer edges of the disc, driving shocks. This causes thermal shocked emission from a larger region of the disc, as compared to earlier times.
For the snap-shots in Sim.~B and Sim.~D  at 1.22~Myr and 2.31~Myr respectively, the jet has evolved to scales larger than the disc size. After the jet breaks out from the disc, the coupling of the jet with the disc gas weakens, resulting in a reduction of the intensity of shocked emission (see Fig.~\ref{fig:OIII-simB} and Fig.~\ref{fig:OIII-simD}). Contrarily, in Sim.~E (see Fig.~\ref{fig:OIII-simE}), the jet being launched into the disc remains confined and strongly coupled to the disc until the end of the simulation. Thus, the jet induces strong shocks without a significant decrease in the [\ion{O}{III}] flux, which shows emission elongated along the edge-on viewed disc. However, the apparent morphology of emission when viewed at $\theta_\mathrm{I}=20^\circ$ i.e. closer to the face-on view, appears nearly circular, as the whole nuclear disc is shocked from the spread of the jet plasma and backflows.

 In Sim.~D, the [\ion{O}{III}] emitting regions at 2.31~Myr exhibits an `S-shaped' morphology at the edge-on view and at $\theta_\mathrm{I}=70^\circ$ (see Fig.~\ref{fig:OIII-simD}). This is possibly caused by the inward push of the deflected jet onto the gas disc, and  the [\ion{O}{III}] arms tend to appear below the deflected jet, on either side of the disc plane.

\subsubsection{\textbf{Spatial variation of gas movements}}
\label{disc_dynamics}
\begin{itemize}
    \item \textbf{Enhanced W80 widths in confined jets:}\\
The right panels of Figs.~\ref{fig:OIII-simB},~\ref{fig:OIII-simD} and~\ref{fig:OIII-simE} show the evolution of W80 widths in the edge-on image plane, as the jet progresses into the ISM for different simulations.
Jets at the earlier stages of evolution in these systems (i.e. at 0.39~Myr in Sim.~B ($\theta_\mathrm{J}=0^\circ$), at 0.59~Myr in Sim.~D ($\theta_\mathrm{J}=45^\circ$), and 0.6~Myr in Sim.~E ($\theta_\mathrm{J}=70^\circ$)), causes greater widths up to 800~$\kms$, primarily in the central $\sim500$ pc, both along the jet axis and perpendicular to it. Such high
velocity widths are also observed for the higher-power jet confined in
a gas-rich spherical system (see Fig.~\ref{fig:O_SimBs}). This indicates that powerful young confined jets vigorously interacting with the ISM launch a near-spherical energy bubble, irrespective of the nature of the large scale distribution of the gas. At 1.27~Myr in Sim.~D, when the jet is about to break-out, the high velocity dispersion can be seen to follow the jet trail as it progresses in the disc, and high W80 values have already started accumulating at the upper and lower edges of the disc.\\

 \item \textbf{Morphology of W80 widths for extended jets:}\\
As the jet breaks out and evolves outside the disc in Sim.~B and Sim.~D, it drives outflows along the minor axis. This results in high-velocity widths at the outer edges of the disc along the direction perpendicular to the disc plane, showing a biconical shape aligned around X'=0, from where the outflowing gas exits the disc. However, the velocity dispersion also simultaneously decreases in the core regions of the disc as the jet has decoupled from the central parts of the disc. This leads to lowering of the W80 values in the central regions (see  Fig.~\ref{fig:OIII-simB} and~\ref{fig:OIII-simD}). 
This becomes more prominent in Sim.~D at 2.31~Myr, where one can clearly see two high-velocity width cones with values larger than 800 km/s, indicating stronger outflows and turbulence towards the lobes of the jet. However at $\theta_\mathrm{I}=70^\circ$, the W80 widths show an `X-shaped' morphology. The regions along the jet, as well as perpendicular to it, have higher velocity dispersion (see Fig.~\ref{fig:OIII-simD}). The observer at this orientations sees the jet-impacted kinematics of the regions both along and across it.

The feature of high W80 values perpendicular to the disc plane can be seen clearly in the later stages of Sim.~E (see the middle panel in Fig.~\ref{fig:OIII-simE}). This results from the deflected jet plasma outflows along the minor axis, which are launched by a highly inclined jet \citep{dipanjan5063,dipanjan2018}. 
However, unlike Sim.~B and Sim.~D, we do not see a reduction in the velocity dispersion in the central disc regions of Sim.~E as the jet is still vigorously interacting with the disc gas, injecting turbulence all across the disc. One can notice that, due to the longer confinement time and the resulting stronger jet-disc coupling (as shown in M2018b), the inclined jets produce higher W80 widths in the central regions of the disc, when compared to the vertical jet at almost similar times (see Fig.~\ref{fig:OIII-simB},~\ref{fig:OIII-simD} and~\ref{fig:OIII-simE} at 1.22~Myr, 1.27~Myr and 1.24~Myr, respectively).
At closer to the face-on view ($\theta_\mathrm{I}=20^\circ$) for Sim.~E, the whole of the disc appears to have high-velocity dispersion, with the highest values attained in the regions in the vicinity of the jet (see bottom panel in Fig.~\ref{fig:OIII-simE}). This also depicts that the jet affects both the regions along its path, as well as the gas at large scales perpendicular to it, as discussed further in detail below.

\item \textbf{Gas kinematics along and perpendicular to the jet:}\\
 In order to explicitly evaluate the effect of the jet on the gas kinematics in directions along and perpendicular to it, we construct mock Position-Velocity (PV) maps of the [\ion{O}{III}] emitting gas. The slits are chosen to be of width $\sim500$~pc, and oriented along the jet (hereafter $S_0$) for Sim.~D ($\theta_\mathrm{J}=45^\circ$) for different times (viz. 0.59~Myr, 1.27~Myr, and 2.31~Myr) in Fig.~\ref{fig:pv_edge}.
We also over-plot the mean-velocity curve with $\pm2\sigma$ deviation (in red) at the same times. The slit following the jet is inclined to the disc normal by $45^\circ$, and the blue dashed line marks the extent of the jet tracer along the slit (see Fig.~\ref{fig:OIII-simD} for the extent of jet tracer). For comparison, we have over-plotted the corresponding mean velocity with 2$\sigma$ deviation along the slit for the `no-jet' case (in light-blue), at almost similar times for the first two PVs, and we use the last snapshot (i.e. 1.37~Myr) for the third. This curve shows the expected mean velocity along the slit for the disc settling in the presence of the gravitational potential, without being disrupted by any external driving agent like a jet. One can clearly see large deviations in the velocities in Sim.~D as compared to the settled disc. This can have a significant effect on the apparent disc's velocity field, which we discuss later in Sec.~\ref{pv-maps}. 
We find that the PV curves show significant changes with time as the jet progresses into the galaxy. At 0.59~Myr, the PV map shows a general broadening (up to 700~$\kms$) in the central regions of the disc ($r \lesssim 1$~kpc) where the jet is confined and decreases farther out.
At later times, the PV map shows a significant increase in slope in the central region from that of the `no-jet' simulation, along with the enhanced dispersion and spiky features at the outer radii of the slit. At 2.31~Myr, the jet has escaped out from the disc, and the outflows along the minor axis spread the turbulence in the outer edges of the disc (M2018b), which enhances the velocity dispersion at the two ends of the slit. This gives rise to high W80 bi-cones, as shown in Fig.~\ref{fig:OIII-simD} and results in the `dumbbell shaped' structure of the PV maps at late times in Fig.~\ref{fig:pv_edge}. One can notice low flux emitting regions at the edges of the PV maps in Fig~\ref{fig:pv_edge}. At 0.59~Myr, it happens because several of these low-density regions at the upper and lower part of the disc are unshocked in the absence of jet interaction (see Fig.~\ref{fig:OIII-simD} for the emission and W80 maps). At 2.31~Myr these regions are heated to much high temperatures due to the shocks from the jet (see Fig.~\ref{fig:cool-curve}), and produce low integrated flux in [$\ion{O}{III}$].

 To demonstrate that the jet also affects gas away from main axis of flow, we compare the PV maps constructed for both along the jet ($S_0$) and perpendicular (hereafter $S_{90}$) to it for Sim.~E ($\theta_\mathrm{J}=70^\circ$) at 1.24~Myr in Fig.~\ref{fig:pv-simE}. The slits are marked in the velocity map (in the left panel), for an image plane at $\theta_\mathrm{I}=20^\circ$.
One can clearly see that the presence of the jet causes strong deviations in the jetted galaxy from the mean velocity curve, as compared to the `no-jet' case. Along $S_0$ slit, there are deviations $\gtrsim800~\kms$ from the mean curve in the regions lying between the blue vertical lines, which may indicate bulk motion along with turbulent velocities in the regions directly in contact with the jet. The regions beyond the jet extent have line widths up to $400~\kms$, and decreases farther out. For the $S_{90}$ slit, the broadening in regions farther from the jet reach $\sim400~\kms$ and are enhanced to 700 $\kms$ near the jet, indicating that jets can have a wide-spread affect on the kinematics of gas in the disc. One can also notice some large prominent deviations from the mean velocity curve at some places along both the slits, resulting in spiky structures. Other studies  \citep[e.g.][]{dipanjan5063,suma_2022} have also observed such spikiness in the PV maps of simulated and observed systems, which is caused due to the acceleration of individual clumps by the local outflows.
\end{itemize}
\subsection{\textbf{How do jets affect the disc's velocity field?}}
\label{pv-maps}
In this section, we discuss how the jet-ISM interaction can affect the projected observed velocities and hence, the observed disc rotation profile. For this, we compare the projected velocity maps for the `no-jet' and the jetted simulations at similar times.
We show the edge-on viewed [\ion{O}{III}]-weighted mean velocity for the `no-jet' simulation, Sim.~D ($\theta_\mathrm{J}=45^\circ$) and  Sim.~E ($\theta_\mathrm{J}=70^\circ$) at 1.23~Myr, 1.27~Myr and 1.24~Myr, respectively, in the top panel of Fig.~\ref{fig:rot_curve}. The velocity maps clearly show that the observed LOS velocities of the disc in the jetted systems are significantly disturbed due to outflows, especially at the outer edges of disc along the direction perpendicular to the disc plane. The corresponding W80 widths and emission maps for Sim.~D and Sim.~E are shown in Fig.~\ref{fig:OIII-simD} and Fig.~\ref{fig:OIII-simE}, respectively.

Using the velocity maps in Fig.~\ref{fig:rot_curve}, we also construct the PV maps for the slit of width $500~\pc$ placed along the major axis of the disc, which are shown in the bottom panel. The slit is marked with black horizontal lines in the velocity maps.
For different jet inclination angles, these snapshots assist in comparing the effect of jet-driven outflows on the observed rotation profile of the disc at almost similar times. 
One can see that the light-blue curve, i.e. the no-jet system, clearly shows lower velocities and smaller deviations as compared to the jet-harbouring systems in Fig.~\ref{fig:rot_curve}. Moreover, the PV curves for the jetted simulations show anti-symmetric deviations from the mean curve of the `no-jet' case, and we find an enhancement of the mean velocity's slope in the central regions of the slit, i.e. [-0.5,0.5]~kpc, to be $\sim$1.7 times its value in the `no-jet' case. This implies that the jet is steepening the rotation profile, i.e., providing boost in angular momentum, compared to the loss of rotational support due to dissipation in the `no-jet' case. Larger deviation in the jetted mean curve indicates that the apparent rotational motions of the disc are disturbed by the jet-driven outflows, as they act on the rotating disc, as compared to settled disc. Thus we find that jets can indeed interact with a rotating disc and affect the rotation profile of the gas. Other studies \citep[e.g.][]{raffaella15,dipanjan5063,suma_2022} have also found such signatures of jets influencing the rotation profile in the central regions of a gas disc. In Fig.~\ref{fig:rot_curve}, the flux is reduced at the edges of the PV maps for Sim.~D and Sim.~E, as was also observed in Fig.~\ref{fig:pv_edge}. This lowering occurs because the jet-driven shocks have not yet reached the regions at ends of the slit which leads to low integrated emission, as can be inferred from the evolution of the [$\ion{O}{III}$] maps in Fig.~\ref{fig:OIII-simD} and~\ref{fig:OIII-simE}.

 In Fig.~\ref{fig:vel_70_deg}, we show how the jet-ISM interaction can shape the velocity field of the disc, when compared to the `no-jet' case at similar times. We show the [\ion{O}{III}] weighted mean velocity for the `no-jet' and Sim.~D at 1.23~Myr and 1.27~Myr respectively, at the similar disc orientation ($\theta_\mathrm{I}=70^\circ$). The magenta curve represents the mean zero-velocity curve (kinematic centres) at the given orientation of the disc, fitted using a third-order polynomial.
 For reference we mark the projected minor axis i.e X'=0 using black vertical dashed line.
 The distortion in the projected velocity field of Sim.~D is highly deviated from the projected minor axis, and diverges far from it in the upper and lower regions of the disc. To investigate the cause of such realignments, we build a toy model that closely mimics the scenarios in these systems, which is presented in Appendix~\ref{vel_map}. Our model demonstrates how the outflows from the jet, which impart an outward push on the disc gas, can disrupt the symmetry of velocity field in the rotating disc.

\section{Discussion}
\label{discussion}
In this study, we have explored the impact on the emission morphology and kinematics of gas shocked by an evolving jet in the central few kpcs of their host galaxies, by tracing the simulated [\ion{O}{III}]~($\lambda$ = 500.7~nm) emission. We find that the total [\ion{O}{III}] luminosity in the jetted systems can reach up to $\sim10^{43}\ergs$, which is up to two orders of magnitude less than the jet kinetic power. Since we are unable to resolve the shocks in our simulations (see Appendix~\ref{cooling-length}), this gives an upper estimate for the expected shocked [\ion{O}{III}] emission. However, the qualitative inferences shown in our study for the total emission are valid. In the following sections we discuss the main findings and some implications regarding jet-ISM interactions from our study.
\subsection{Nature of jet-ISM interactions and their spatial extent}
Our analysis shows that jets can have a wide-spread impact on the ISM, and can cause shocked emission and extreme gas velocities and line widths in the whole nuclear regions (see Sec.~\ref{results_spherical} and~\ref{results_disc}). Such signatures of large-scale jet-ISM interaction have been observed in a number of jet-harbouring systems \citep{nicole07,holt_2008,johnston_2010,holt_2011,santoro_2018}. 
Using various diagnostics (i.e. W80 widths and PV maps at different observer's orientation), we find that the jets not only have a substantial impact on the regions along their axis; but can also cause turbulent motions and outflows in regions far away ($\sim2~\kpc$). Such stronger ionized outflows, with FWHM broader than $500~\mathrm{\kms}$ along the jet axis, have been detected in 3C~293 \citep{emonts,mahony16} using long-slit observations. Further, \citet{mahony16} found outflows with line-widths broader than $300~\mathrm{\kms}$ in a direction perpendicular to the jet (3.5~kpc), analogous to the high PV widths (upto $400~\kms$) observed in our study along slit $S_{90}$ (see Fig.~\ref{fig:pv-simE}). It is believed that the jet-driven energy bubble causing a widespread effect on the disc, as shown in jet simulations \citep{Sutherland07,wagner11,wagner12,dipanjan16,dipanjan2018}, is responsible for enhanced velocity dispersion on large scales in the galaxy.

We verify that the W80 widths for the `no-jet' simulation at 1.37~Myr (last snapshot) are confined to $<300\,\mathrm{km\,s^{-1}}$, which is in agreement with the values expected due to the rotation profile of the settled disc (peak rotation velocity $\sim300~\kms$). Also, the widths are uniformly distributed across the whole disc in the absence of any outflow driving agent. Our study does not incorporate any noise in the emission estimated from the analyzed systems, which is a very important element in the observations. Nonetheless, we show in Appendix~\ref{noise} that the noise can reduce the observed strength of apparent gas kinematics (W80 widths) in the galaxy, but does not alter qualitative conclusions. Below we discuss some main results from our analysis with relevant observational evidences.

\begin{itemize}
  \item{\textbf{Observed evolution of the gas properties during different stages of the jet-gas interaction:}} We find that the gas morphologies change over time as the jets interact with their ISM. At earlier times, the shocked emission and the extreme gas kinematics is limited to the central few pcs, which later extends to 2~kpcs following the jet breakout phases. This behaviour is in agreement with the various phases of jet evolution \citep{Sutherland07,wagner12,dipanjan16}, where the scale of jet-gas interaction grows with the expansion of jet-driven cocoon, resulting in large-scale outflows \citep[see][]{schulz_2021,morganti_2021b}.
  
   \item \textbf{Jet-gas interaction shaping the emission morphology of the systems:}
  When seen edge-on in discs, the shocked emission at earlier times generates a nearly `X-shaped' morphology because the jet driven outflows can easily grow in the lower densities above and below the disc plane, but are hampered in the central regions by high column depths (see Fig.~\ref{fig:OIII-simB}, ~\ref{fig:OIII-simD} and~\ref{fig:OIII-simE}).
The emission spreads across the whole nuclear disc as the jet-driven bubble covers the entire disc in subsequent phases. Similarly in spherical systems (Fig.~\ref{fig:O_SimBs} and~\ref{fig:O_SimCs}), the emission is limited to the vicinity of the jet at the earlier stages, while later the breaking out of the jets, via pushing and clearing out matter in their path, create an hour-glass or fountain shaped emission morphology.
   
    \item \textbf{Distortions in the observed emission morphology: } 
    In Sim.~D ($\theta_\mathrm{J}=45^\circ,\mathrm{P_J}=10^{45}\ergs$), we notice a distortion in the emission morphology of the disc, when viewed at high inclinations (see Fig.~\ref{fig:OIII-simD}).
This `S-shaped' emission morphology is expected to be caused by the jet creating denser shocked regions by forcing the gas along its leading edges, resulting in a distorted morphology in the emission as it gets closer to the edge-on view of the disc. Such distortions in the optical emission have been observed in various sources, such as Mrk~34, Mrk~573, and ESO~428-G14 \citep{Falcke_1998,falcke_1996}. 

    \item \textbf{Laterally expanding forward shocks in jetted systems: } We predict that the jet-driven large-scale laterally expanding forward shock of the energy bubble sweeping through the ISM leads to double-peaked spectra, when integrated along the LOS passing through such regions (see Fig.~\ref{fig:w80_Sim_Bs}). Such double-peaks have been attributed to several different sources, such as  the presence of dual AGN \citep{rubinur_2019,maschmann_2020}, AGN-driven outflows \citep{nevin_2018,rakshit_2018,kharb_2021} or rotating discs \citep{michael_2009}. In our analysis, we find such double-peaks in the non-rotating spherical system, which are caused solely due to the presence of lateral outflows caused by the jet (see Fig.~\ref{vol_render}).
   
    \item \textbf{Jets causing outflows along the minor axis: }
    In the disc systems, the jets are expected to cause strong outflows along the minor axis, from where the jet plasma escapes out easily \citep{dipanjan5063,dipanjan2018}. This leads to high W80 biconical structures at the upper and lower regions of the disc when viewed edge-on (see Fig.~\ref{fig:OIII-simB}, ~\ref{fig:OIII-simD} and~\ref{fig:OIII-simE}).
However, the jet launched at $70^\circ$ to the disc normal (Sim.~E) causes prominently high dispersion along the direction perpendicular to the disc plane, i.e. extending perpendicular to the jet (see Fig.~\ref{fig:OIII-simE}). Such enhanced velocity dispersion perpendicular to the jet has been recently observed in spatially resolved studies of gas kinematics in a few systems such as IC~5603 \citep{Venturi21}, where the jet is ejected very close to the disc plane. 
    
   \item \textbf{Realigned symmetry of the velocity field in jetted discs : }
   We find that the presence of a jet can provide the rotational support by enhancing the angular momentum of the disc (see Fig.~\ref{fig:rot_curve}), and also disrupt the symmetry of the projected velocity field (Fig.~\ref{fig:vel_70_deg}). Such distortions/realignments have been observed in the velocity field of several systems, and are believed to be caused due to disc warps or non-circular motions \citep{ruffa_2019b,ruffa_2020a} or the presence of a nuclear bar \citep{belete_2021} as found using \textsc{diskfit} \citep{naray_2012}. However, by using an analytical toy model, we show that such realignments in our study are caused by the jet-driven bubble pushing the gas radially outwards from the centre of the disc (see Appendix~\ref{vel_map}).
  
\end{itemize}

\begin{figure}
 \centerline{
\def\arraystretch{1.0}
\setlength{\tabcolsep}{0.0pt}
\begin{tabular}{lcr}
     {\includegraphics[width=0.7\linewidth]{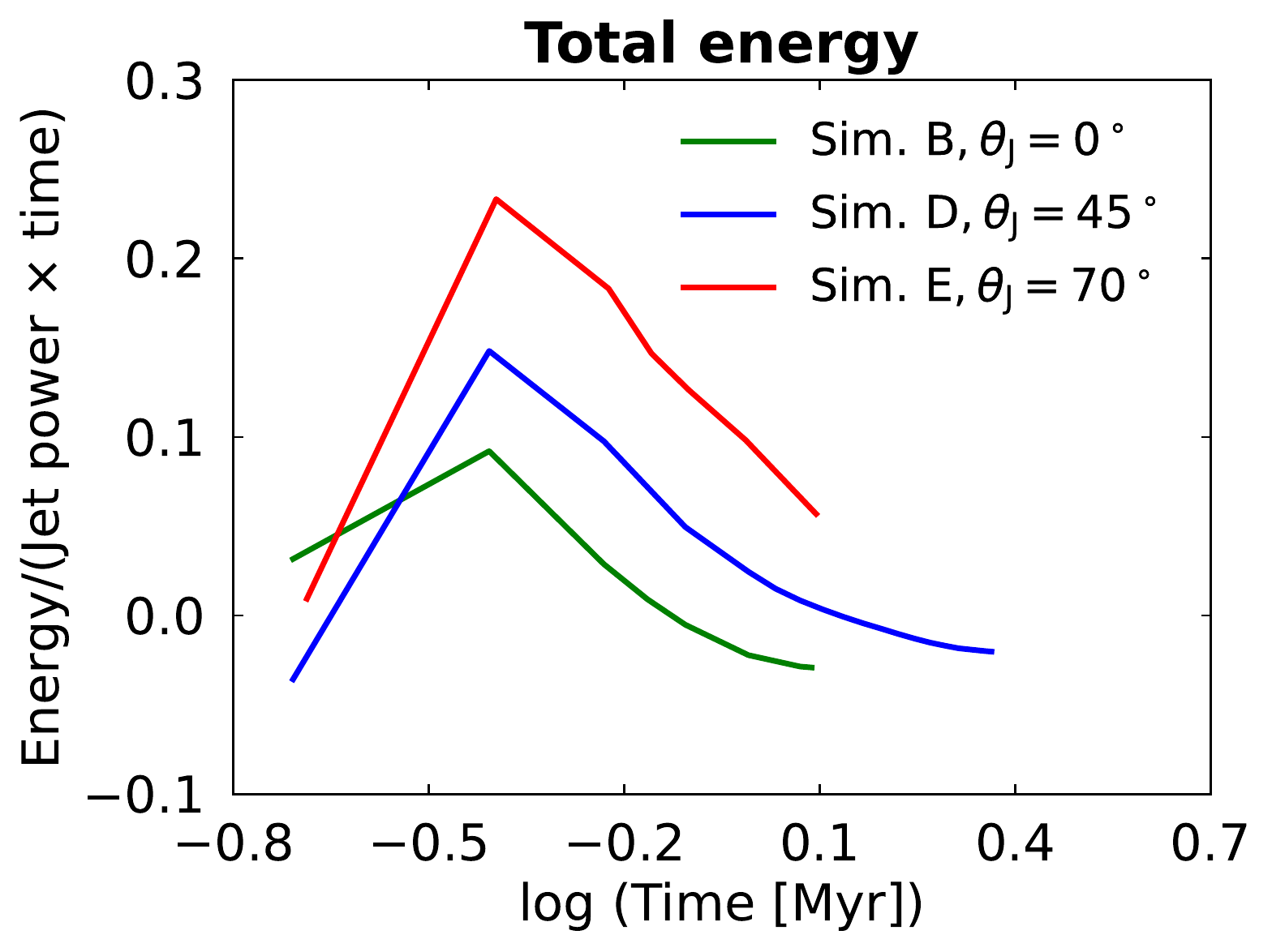}} \\
     {\includegraphics[width=0.7\linewidth]{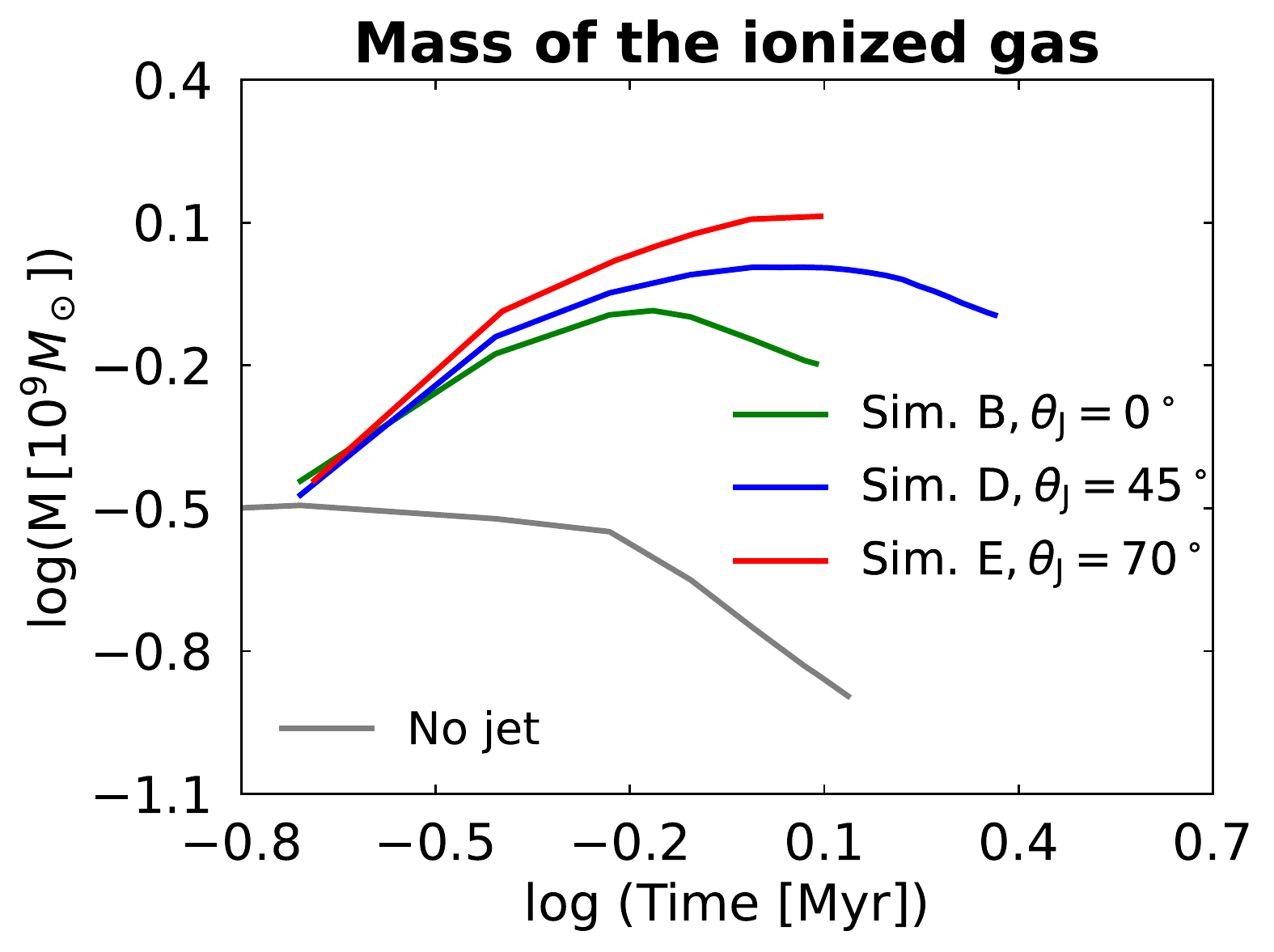}}
     \end{tabular}}
\caption{Ratio of total energy of the simulation domain and the energy injected by the jet (\textbf{Top}) and mass of the collisionally-ionized gas (\textbf{Bottom}) estimated using Saha Equation as a function of time. Only the cells with $n>0.5\,\mathrm{cm^{-3}}$ are used for the estimates of energy and ionized gas mass.}
  \label{fig:energy-sim}
\end{figure}

\begin{figure*}
 \centerline{
\def\arraystretch{1.0}
\setlength{\tabcolsep}{0.0pt}
\begin{tabular}{lcr}
     \hspace{-1cm}
      {\includegraphics[width=0.35\linewidth]{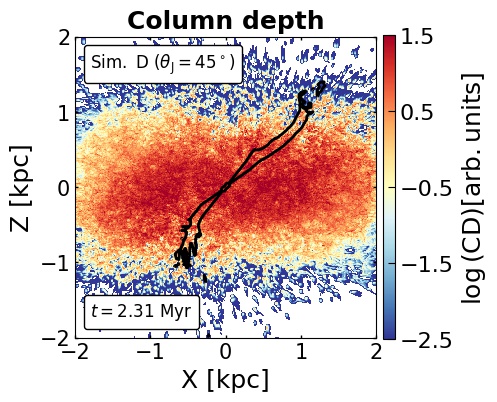}}&
       {\includegraphics[width=0.335\linewidth]{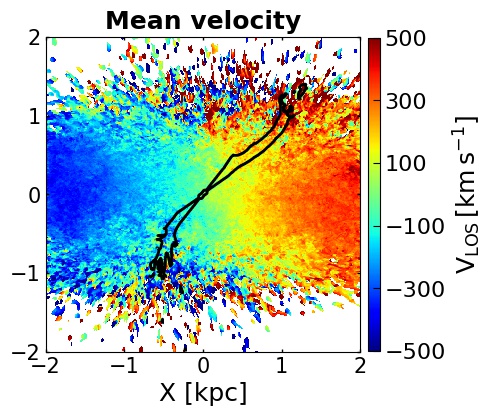}}&
     {\includegraphics[width=0.33\linewidth]{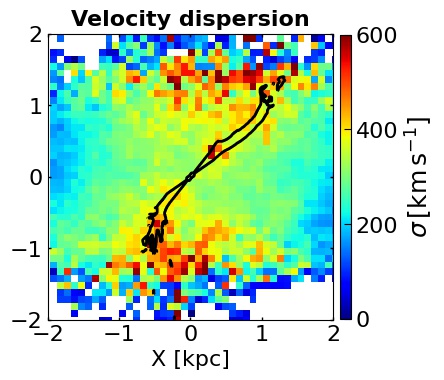}}
     \end{tabular}}
     \vspace{-0.3cm}
      \centerline{
\def\arraystretch{1.0}
\setlength{\tabcolsep}{0.0pt}
\begin{tabular}{lcr}
     \hspace{-1cm}
      {\includegraphics[width=0.35\linewidth]{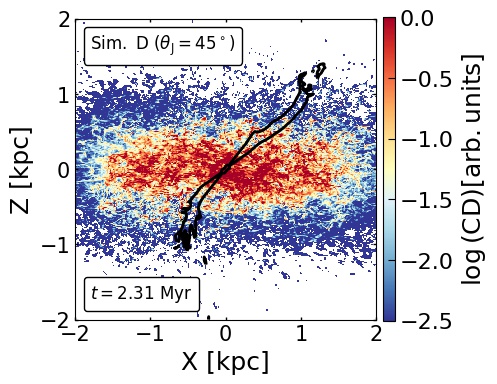}}&
       {\includegraphics[width=0.335\linewidth]{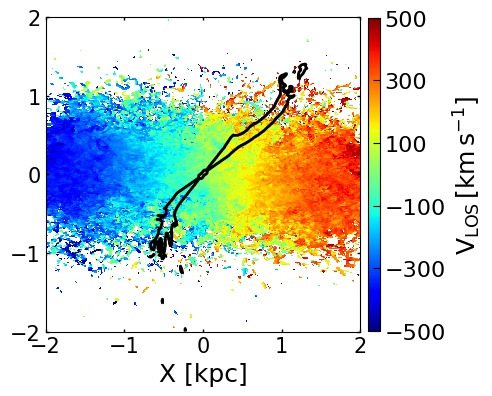}}&
     {\includegraphics[width=0.334\linewidth]{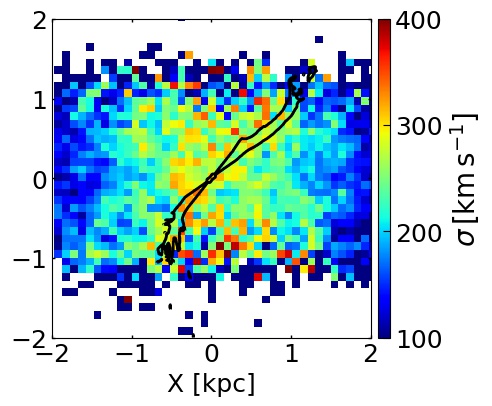}}
     \end{tabular}}
\caption{\textbf{Top:} Edge-on view ($\theta_\mathrm{I}=90^\circ,\phi_\mathrm{I}=270^\circ$) maps of column-depth (CD), mass-weighted mean LOS velocity, mass-weighted mean velocity dispersion from the collisionally-ionized gas in Sim.~D ($\theta_\mathrm{J}=45^\circ,\mathrm{P_J}=10^{45}\ergs$) at 2.31~Myr. \textbf{Bottom:} Same as top panel but for the collisionally-non-ionized (or neutral) gas in Sim.~D. A cut-off density value of $0.5\,\mathrm{cm^{-3}}$ is used for the estimates. The black contour shows the projected jet tracer at a value of 0.5 (maximal value is 1).}
  \label{fig:saha-flux}
\end{figure*}

\subsection{{Time evolution of energy and ionized gas mass in the jetted systems}}
\label{energy-mass}
In this section, we discuss how the total energy in the simulation domain as a fraction of energy injected by jet, and the collisionally-ionized gas mass, which we estimate using the Saha equation \citep{paddy_2006}, evolve with time. We show the plots of energy and ionized gas mass in disc systems from Table~\ref{tab:sim_table} in Fig.~\ref{fig:energy-sim}. In the top panel, one can see that as the jet evolves in the ISM, the energy input increases to a maximum and begins to decrease at later stages. Jets with higher inclination to the disc inject more energy into the system, and the maximum fraction is 0.23 for Sim.~E ($\theta_\mathrm{J}=70^\circ$), similar to the values from \citet{wagner12,dipanjan16}. The bottom panel shows that the gas in jetted systems is shock-heated leading to much higher shock-ionized gas mass, than the `no-jet' disc. In Sim.~B ($\theta_\mathrm{J}=0^\circ$) and Sim.~D ($\theta_\mathrm{J}=45^\circ$), the ionized gas mass decreases at later stages when the jet has broken out from the disc, as the gas begins to cool in the absence of direct jet-ISM interaction. This also causes lowering of [\ion{O}{III}] emission flux after the jets escape out from their disc in these simulations in Fig.~\ref{fig:OIII-simB} and~\ref{fig:OIII-simD}. However, for Sim.~E ($\theta_\mathrm{J}=70^\circ$), the jet remains confined in the disc, and thus, the ionized gas mass keeps on increasing and attains a plateau at later times. For the `no-jet' simulation, the hot gas mass in the disc decreases as the gas settles and cools down under the gravitational potential.

\subsection{\textbf{Observed quantities for different phases of gas in the ISM}}
\label{diff_phase}
In the main study we focus on the kinematics and morphology of [\ion{O}{III}] as a tracer of shocked gas; however, the qualitative features are a proxy for the general velocity kinematics of the multiple gas phases in the ISM. To verify this, we compare the [$\ion{O}{III}$] results obtained earlier with the observations of gas in different phases in the sections below.
\begin{itemize}
\item \textbf{Comparing the morphology of collisionally ionized and non-ionized gas with [\ion{O}{III}]:}
    To check if [\ion{O}{III}] emitting shocked gas closely follows the emission and kinematics of the underlying gas distribution in the ISM, we compare their observed properties with those of collisionally ionized and collisionally-non-ionized gas. We use the Saha ionization equation \citep{paddy_2006} to estimate the collisionally ionized gas density in the cells \footnote{For simplicity, we assume the cells to be filled with only hydrogen atoms.} \citep[as done in][]{Meenakshi_2022a}. We use a value of $C\geq10^7$ \citep[see eq. 3 in][]{Meenakshi_2022a} to label the cell as fully ionized i.e. $x=1$, where $x$ is the  ionization fraction. We limit our analysis to regions with density ($n$) greater than $>0.5\,\mathrm{cm^{-3}}$ to focus on the dense gas clouds and exclude the regions filled with jet plasma. The emission strength, i.e. $x^2n^2$, and $(1-x)^2n^2$ for the ionized and non-ionized gas, respectively\footnote{Assuming optical thin limit, emission strength is proportional to the density squared}, is integrated along the LOS to estimate the column depth (a tracer for the total integrated emission).
    
   In Fig.~\ref{fig:saha-flux}, we show the column-depth, mass-weighted mean velocity, and mass-weighted velocity dispersion ($\sigma$) \citep{mandal_2021} for the collisionally ionized and non-ionized gas at the edge-on view image plane of Sim.~D ($\theta_\mathrm{J}=45^\circ,\mathrm{P_J}=10^{45}\ergs$) at 2.31~Myr. As one can see in the top-left panel, the column depth of the ionized gas extends to $\pm1~\kpc$ along the vertical direction, which results from the gas clumps being shred and ionized by the escaping jet. Contrarily, the non-ionized gas is confined in $\pm0.5~\kpc$ from the disc plane. However, the high column depth regions from both phases exhibit the `S-shaped' morphology swirling around the jet, although more prominent for the ionized phase, which also closely matches the total [\ion{O}{III}] emission morphology at 2.31~Myr in Fig.~\ref{fig:OIII-simD}. The projected velocities from the ionized phase show turbulent motions in the upper and lower regions of the disc, which leads to the high-velocity dispersion and biconical structures (see right panel). This velocity dispersion is contributed by various phases, and therefore the high widths at the cones are comparatively lower ($<600~\kms$) than what we observe for the [\ion{O}{III}] emitting gas in Fig.~\ref{fig:OIII-simD} (up to $1000~\kms$).
The non-ionized gas phase, on the other hand, does not show such strong turbulent motions, and the resulting velocity dispersion is limited to $\sim350~\kms$. However, the high-dispersion shows an ``hourglass'' shape with enhancing widths in upper and lower half of the disc.

We verify that for the `no-jet' case ($t=$1.37~Myr), the velocity dispersion for the collisionally ionized and non-ionized phases across the whole disc attain values up to $<300~\kms$. The widths in the settled disc are almost uniformly distributed, except some enhancements are observed closer to the projected minor axis (up to $350~\kms$), caused due to the settling and collapse of the gas in the disc due to external potential.
These estimates are lower than the dispersion in the ionized gas in Sim.~D, but comparable to the values for the neutral gas in Fig.~\ref{fig:saha-flux}. This suggests that the jet-driven outflows in the disc does not strongly perturb the dynamics of neutral gas, as it does for the ionized gas, as has
also been observed in the multi-phase study of IC 5063 \citep{raffaella07}. Such weak outflows in the neutral as well as warm molecular phase can be from the post-shock cooling gas or secondly, these can be the embedded dense clouds that are not ionized all the way through. Similar kinetic behaviour is also found for the molecular (CO) and ionized phase in PKS~0023-26 \citep{shih_2013,santoro_2020,morganti_2021b}. Nonetheless, the qualitative features of emission and gas kinematics remain similar for different gas phases, suggesting jet interaction with multiple phases in the ISM. 

\subsection{Prevalence of extended shocked gas at wide range of phases:}
In Sec.~\ref{diff_phase}, we find predictable observable signatures in emission and jet disturbed gas linked to different phases. Our analysis indicate that the jet interaction with different phases of gas can cause similar emission morphology across different shock-ionized gas phases. Such similar morphologies across [\ion{O}{III}], $H\alpha$, [\ion{S}{II}] and X-rays has been observed by \citet{Maksym_2017} in the resolved study of NGC~3393. Also, spectroscopic studies of several systems \citep{martin_2010,hardcastle_2012,wang_2011,Maksym_2019} indicate co-spatial emission for X-rays and [\ion{O}{III}], which is aligned with the radio emission in these sources. This marks an interplay between the different gas components, indicating how they are all being governed by jet-gas-interaction in these sources. Moreover, this also suggests that different shocked gas phases are also expected to exhibit similar qualitative distribution in kinematic behaviours, as we have shown in Sec.~\ref{diff_phase}.
Indeed, such features in kinematics across different gas phases, i.e., neutral \citep{oosterloo,morganti_2005,mahony13}, ionized ($[\ion{O}{III}]$) \citep{emonts,shih_2013,raffaella07} and molecular gas \citep{dasyra_2015,raffaella15,oosterloo_2017,oosterloo_2019,morganti_2021b} are observed in several galaxies harbouring radio jets.

\subsection{Strength of shocked emission as compared to other sources:}

Due to the lack of a radiative central source in our simulations, our analysis focuses on the shocked emission due to ionization from the jet-induced shocks. Other ionizing mechanism, such as AGN radiation and star activity, can, nevertheless, make substantial contributions to the surface brightness of shocked gas in observed galaxies.
 Photo-ionization due to AGN radiation is found to have a dominant contribution in the emission from several sources \citep{montse_1997,raffaella07,humphrey2008,santoro_2020}; even though the gas kinematics shows clear signatures of jet-induced outflows. On the contrary, excitation from hot gas in optical/X-rays is also found to be dominated by the shocks in several studies \citep{montse_1999,hardcastle_2010,hardcastle_2012,shih_2013,lanz_2015,murthy_2019,travascio_2021}. This suggests that shocks and photoionization may co-exist in many cases \citep[e.g.,][]{dagostino2019}, although the signatures of shocks are not always easy to distinguish, in particular from optical diagnostics alone \citep[][]{humphrey2008}.
Although the post-process investigation in the simulations used here found central AGN to be a weak photo-ionizing source, when compared to the jet of similar kinetic power \citep{Meenakshi_2022a}, it does indicate that jets can assist the radiation to spread over large distances by producing lines of sight with low column densities. This may also allow the ionizing radiation to ionize the post shock-cooling gas, potentially masking the signatures of shock ionization. Ideally, this should be examined more closely with a real time jet and radiation analysis, as was done for the AGN-driven winds in \citet{bieri17}.

\end{itemize}
\section{Conclusions}
\label{summary}
In this study, we have investigated the thermal emission and kinematics from shocked gas in the simulations from M2016 and M2018b, where we explored [\ion{O}{III}]~($\lambda$ = 500.7~nm) emission as a tracer of shocked gas. The emission maps were constructed using the results of \textsc{mappings v} \citep{mappings_2018} to estimate the emissivity from collisional ionization in the nuclear regions of host galaxies ($\sim$kpc scales). Below we summarize the main findings of our study.

\begin{enumerate}

  \item We find that the jets, as they evolve with time, shape the emission morphology of the host's ISM. In the spherical system, the jets push and shred the gas along their path and produce an hourglass or fountain shape distorted morphology at the edge-on view (see Fig.~\ref{fig:O_SimBs} and~\ref{fig:O_SimCs}). In discs, we find that the shocked emission exhibits a nearly `X- shaped' morphology in the central regions at the earlier times when seen edge-on, and extend to whole nuclear disc as the jets evolve to large scales (see Sec.~\ref{evol_emission_disc}).
  \item We observe an `S-shaped' distortion in the emission morphology from the shocked gas in the disc when the jet is inclined towards it at $45^\circ$ (see Fig.~\ref{fig:OIII-simD}). We believe that such distortions result from the pressurization and shock-heating of the gas by the leading edges of the inclined jets.

    \item We find that the higher power jet ($10^{45}\ergs$) in the spherical system causes high velocity dispersion across the whole nuclear region; however, a lower power jet ($10^{44}\ergs$) can cause outflows only in the regions closer to the jet head and has negligible affect in the farther regions (see Sec.~\ref{results_spherical}). We find that the jet-driven large-scale laterally expanding forward shock of the energy bubble sweeping through the ISM result in high W80 widths (up to 1000~$\kms$) when integrated along the LOS (see Fig.~\ref{fig:w80_Sim_Bs}).
    
    \item In discs, the high W80 widths (up to 800$~\kms$) are confined to the central regions during the earlier stages of jet evolution. When the jet escapes out from the disc, the jet-disc coupling begins to decrease in the central regions of the disc, leading to decay in the apparent W80 widths in these regions. Meanwhile, outflows along the minor axis cause the high-width bicones to form at the disc's outer edges along the direction perpendicular to the disc plane (see Sec.~\ref{disc_dynamics}).
     
    \item When the jets are young and confined, the central region of the disc along the jet axis show high PV widths (up to $700\,\kms$). Later as the jet evolves, the dispersion rises in the upper and lower regions of the disc (when viewed edge-on), giving rise to a `dumbbell shaped' PV diagram (as inferred from Fig.~\ref{fig:pv_edge}).
    
    \item The jets not only affect their immediate vicinity, but also cause high PV widths in the regions perpendicular to it ($\sim 400 \,\kms$). As compared to the `no-jet' case, the stronger anti-symmetric deviations in the PV maps may indicate bulk motion caused due to outflows from the jet (see Fig.~\ref{fig:pv-simE}).
    
    \item The mean rotation curve shows that the jet-ISM interaction provides rotational support and angular momentum boost to the disc (see Sec.~\ref{pv-maps}), thereby increasing the slope of the PV maps. However, we also observe realignments in the symmetry of the disc's velocity field of the jetted simulations, which results from the jet-driven outflows pushing the gas outwards in the disc (see Appendix~\ref{vel_map}).
    
    \item As compared to the jet vertical to the disc, the inclined jets have a longer confinement time in the host system and consequently cause stronger jet-disc coupling. We find that this leads to more energy being injected into the disc, resulting in larger gas mass being collisionally-ionized by the shocks (see Sec.~\ref{energy-mass} ). 
    
    \item The similar features in emission and gas kinematics observed across different gas phases in the ISM (see Sec.~\ref{diff_phase}) support the finding \citep{raffaella15,morganti_2021b} that jets have a widespread effect on the ISM and leave a similar kinematic imprint on different gas phases being affected by the jet-induced shocks.

\end{enumerate}

\section*{Acknowledgements}
We thank the anonymous referee for his/her constructive comments that helped in improving the clarity of the manuscript. MM also thanks Tek P. Adhikari for useful discussions.
We acknowledge the use of the supercomputing facility at IUCAA\footnote{\url{http://hpc.iucaa.in}}, Pune
for the computations in this study. The simulations used in this study were carried out using the resources of the National Computational Infrastructure (NCI Australia), an NCRIS enabled capability supported by the Australian Government. AYW is supported by JSPS KAKENHI Grant Number 19K03862.  R.M.J. Janssen is supported by an appointment to the NASA Postdoctoral Program at the NASA Jet Propulsion Laboratory, administered by Oak Ridge Associated Universities under contract with NASA.

\section*{DATA AVAILABILITY}
No new data were generated in support of this research. The simulations used for this study will be shared on a reasonable request to the corresponding authors.




\bibliographystyle{mnras}
\bibliography{manuscript} 

\begin{thebibliography}{}
\makeatletter
\relax
\def\mn@urlcharsother{\let\do\@makeother \do\$\do\&\do\#\do\^\do\_\do\%\do\~}
\def\mn@doi{\begingroup\mn@urlcharsother \@ifnextchar [ {\mn@doi@}
  {\mn@doi@[]}}
\def\mn@doi@[#1]#2{\def\@tempa{#1}\ifx\@tempa\@empty \href
  {http://dx.doi.org/#2} {doi:#2}\else \href {http://dx.doi.org/#2} {#1}\fi
  \endgroup}
\def\mn@eprint#1#2{\mn@eprint@#1:#2::\@nil}
\def\mn@eprint@arXiv#1{\href {http://arxiv.org/abs/#1} {{\tt arXiv:#1}}}
\def\mn@eprint@dblp#1{\href {http://dblp.uni-trier.de/rec/bibtex/#1.xml}
  {dblp:#1}}
\def\mn@eprint@#1:#2:#3:#4\@nil{\def\@tempa {#1}\def\@tempb {#2}\def\@tempc
  {#3}\ifx \@tempc \@empty \let \@tempc \@tempb \let \@tempb \@tempa \fi \ifx
  \@tempb \@empty \def\@tempb {arXiv}\fi \@ifundefined
  {mn@eprint@\@tempb}{\@tempb:\@tempc}{\expandafter \expandafter \csname
  mn@eprint@\@tempb\endcsname \expandafter{\@tempc}}}

\bibitem[\protect\citeauthoryear{{Bewketu Belete} et~al.,}{{Bewketu Belete}
  et~al.}{2021}]{belete_2021}
{Bewketu Belete} A.,  et~al., 2021, \mn@doi [\aap]
  {10.1051/0004-6361/202140492}, \href
  {https://ui.adsabs.harvard.edu/abs/2021A&A...654A..24B} {654, A24}

\bibitem[\protect\citeauthoryear{{Bieri}, {Dubois}, {Rosdahl}, {Wagner}, {Silk}
   \& {Mamon}}{{Bieri} et~al.}{2017}]{bieri17}
{Bieri} R.,  {Dubois} Y.,  {Rosdahl} J.,  {Wagner} A.,  {Silk} J.,   {Mamon}
  G.~A.,  2017, \mn@doi [\mnras] {10.1093/mnras/stw2380}, \href
  {https://ui.adsabs.harvard.edu/abs/2017MNRAS.464.1854B} {464, 1854}

\bibitem[\protect\citeauthoryear{{Blandford}, {Meier}  \&
  {Readhead}}{{Blandford} et~al.}{2019}]{blandford19}
{Blandford} R.,  {Meier} D.,   {Readhead} A.,  2019, \mn@doi [\araa]
  {10.1146/annurev-astro-081817-051948}, \href
  {https://ui.adsabs.harvard.edu/abs/2019ARA&A..57..467B} {57, 467}

\bibitem[\protect\citeauthoryear{{Cappellari} \& {Copin}}{{Cappellari} \&
  {Copin}}{2003}]{michele_2003}
{Cappellari} M.,  {Copin} Y.,  2003, \mn@doi [\mnras]
  {10.1046/j.1365-8711.2003.06541.x}, \href
  {https://ui.adsabs.harvard.edu/abs/2003MNRAS.342..345C} {342, 345}

\bibitem[\protect\citeauthoryear{Cecil, Wagner, Bland-Hawthorn, Bicknell  \&
  Mukherjee}{Cecil et~al.}{2021}]{cecil_21}
Cecil G.,  Wagner A.~Y.,  Bland-Hawthorn J.,  Bicknell G.~V.,   Mukherjee D.,
  2021, \mn@doi [ApJ] {10.3847/1538-4357/ac224f}, 922, 254

\bibitem[\protect\citeauthoryear{{Coatman}, {Hewett}, {Banerji}, {Richards},
  {Hennawi}  \& {Prochaska}}{{Coatman} et~al.}{2019}]{coatman}
{Coatman} L.,  {Hewett} P.~C.,  {Banerji} M.,  {Richards} G.~T.,  {Hennawi}
  J.~F.,   {Prochaska} J.~X.,  2019, \mn@doi [\mnras] {10.1093/mnras/stz1167},
  \href {https://ui.adsabs.harvard.edu/abs/2019MNRAS.486.5335C} {486, 5335}

\bibitem[\protect\citeauthoryear{{Collet} et~al.,}{{Collet}
  et~al.}{2016}]{collet_2016}
{Collet} C.,  et~al., 2016, \mn@doi [\aap] {10.1051/0004-6361/201526872}, \href
  {https://ui.adsabs.harvard.edu/abs/2016A&A...586A.152C} {586, A152}

\bibitem[\protect\citeauthoryear{{Couto}, {Storchi-Bergmann}, {Siemiginowska},
  {Riffel}  \& {Morganti}}{{Couto} et~al.}{2020}]{couto_2020}
{Couto} G.~S.,  {Storchi-Bergmann} T.,  {Siemiginowska} A.,  {Riffel} R.~A.,
  {Morganti} R.,  2020, \mn@doi [\mnras] {10.1093/mnras/staa2268}, \href
  {https://ui.adsabs.harvard.edu/abs/2020MNRAS.497.5103C} {497, 5103}

\bibitem[\protect\citeauthoryear{{D'Agostino} et~al.,}{{D'Agostino}
  et~al.}{2019}]{dagostino2019}
{D'Agostino} J.~J.,  et~al., 2019, \mn@doi [\mnras] {10.1093/mnras/stz1611},
  \href {https://ui.adsabs.harvard.edu/abs/2019MNRAS.487.4153D} {487, 4153}

\bibitem[\protect\citeauthoryear{{Dasyra}, {Bostrom}, {Combes}  \&
  {Vlahakis}}{{Dasyra} et~al.}{2015}]{dasyra_2015}
{Dasyra} K.~M.,  {Bostrom} A.~C.,  {Combes} F.,   {Vlahakis} N.,  2015, \mn@doi
  [\apj] {10.1088/0004-637X/815/1/34}, \href
  {https://ui.adsabs.harvard.edu/abs/2015ApJ...815...34D} {815, 34}

\bibitem[\protect\citeauthoryear{{Emonts}, {Morganti}, {Tadhunter},
  {Oosterloo}, {Holt}  \& {van der Hulst}}{{Emonts} et~al.}{2005}]{emonts}
{Emonts} B.~H.~C.,  {Morganti} R.,  {Tadhunter} C.~N.,  {Oosterloo} T.~A.,
  {Holt} J.,   {van der Hulst} J.~M.,  2005, \mn@doi [\mnras]
  {10.1111/j.1365-2966.2005.09354.x}, \href
  {https://ui.adsabs.harvard.edu/abs/2005MNRAS.362..931E} {362, 931}

\bibitem[\protect\citeauthoryear{{Eracleous}, {Lewis}  \& {Flohic}}{{Eracleous}
  et~al.}{2009}]{michael_2009}
{Eracleous} M.,  {Lewis} K.~T.,   {Flohic} H. M.~L.~G.,  2009, \mn@doi [\nar]
  {10.1016/j.newar.2009.07.005}, \href
  {https://ui.adsabs.harvard.edu/abs/2009NewAR..53..133E} {53, 133}

\bibitem[\protect\citeauthoryear{{Falcke}, {Wilson}, {Simpson}  \&
  {Bower}}{{Falcke} et~al.}{1996}]{falcke_1996}
{Falcke} H.,  {Wilson} A.~S.,  {Simpson} C.,   {Bower} G.~A.,  1996, \mn@doi
  [\apjl] {10.1086/310299}, \href
  {https://ui.adsabs.harvard.edu/abs/1996ApJ...470L..31F} {470, L31}

\bibitem[\protect\citeauthoryear{{Falcke}, {Wilson}  \& {Simpson}}{{Falcke}
  et~al.}{1998}]{Falcke_1998}
{Falcke} H.,  {Wilson} A.~S.,   {Simpson} C.,  1998, \mn@doi [\apj]
  {10.1086/305886}, \href
  {https://ui.adsabs.harvard.edu/abs/1998ApJ...502..199F} {502, 199}

\bibitem[\protect\citeauthoryear{{Ferland} et~al.,}{{Ferland}
  et~al.}{2017}]{cloudy17}
{Ferland} G.~J.,  et~al., 2017, \rmxaa, \href
  {https://ui.adsabs.harvard.edu/abs/2017RMxAA..53..385F} {53, 385}

\bibitem[\protect\citeauthoryear{{Feruglio}, {Fabbiano}, {Bischetti}, {Elvis},
  {Travascio}  \& {Fiore}}{{Feruglio} et~al.}{2020}]{ferguilo20}
{Feruglio} C.,  {Fabbiano} G.,  {Bischetti} M.,  {Elvis} M.,  {Travascio} A.,
  {Fiore} F.,  2020, \mn@doi [\apj] {10.3847/1538-4357/ab67bd}, \href
  {https://ui.adsabs.harvard.edu/abs/2020ApJ...890...29F} {890, 29}

\bibitem[\protect\citeauthoryear{{Garc{\'\i}a-Burillo}
  et~al.,}{{Garc{\'\i}a-Burillo} et~al.}{2014}]{garc14}
{Garc{\'\i}a-Burillo} S.,  et~al., 2014, \mn@doi [\aap]
  {10.1051/0004-6361/201423843}, \href
  {https://ui.adsabs.harvard.edu/abs/2014A&A...567A.125G} {567, A125}

\bibitem[\protect\citeauthoryear{{Girdhar} et~al.,}{{Girdhar}
  et~al.}{2022}]{girdhar_2022}
{Girdhar} A.,  et~al., 2022, \mn@doi [\mnras] {10.1093/mnras/stac073}, \href
  {https://ui.adsabs.harvard.edu/abs/2022MNRAS.512.1608G} {512, 1608}

\bibitem[\protect\citeauthoryear{{Gonzalez-Martin}, {Acosta-Pulido}, {Perez
  Garcia}  \& {Ramos Almeida}}{{Gonzalez-Martin} et~al.}{2010}]{martin_2010}
{Gonzalez-Martin} O.,  {Acosta-Pulido} J.~A.,  {Perez Garcia} A.~M.,   {Ramos
  Almeida} C.,  2010, \mn@doi [\apj] {10.1088/0004-637X/723/2/1748}, \href
  {https://ui.adsabs.harvard.edu/abs/2010ApJ...723.1748G} {723, 1748}

\bibitem[\protect\citeauthoryear{{Hardcastle} \& {Croston}}{{Hardcastle} \&
  {Croston}}{2020}]{hardcastle_2020}
{Hardcastle} M.~J.,  {Croston} J.~H.,  2020, \mn@doi [\nar]
  {10.1016/j.newar.2020.101539}, \href
  {https://ui.adsabs.harvard.edu/abs/2020NewAR..8801539H} {88, 101539}

\bibitem[\protect\citeauthoryear{{Hardcastle}, {Massaro}  \&
  {Harris}}{{Hardcastle} et~al.}{2010}]{hardcastle_2010}
{Hardcastle} M.~J.,  {Massaro} F.,   {Harris} D.~E.,  2010, \mn@doi [\mnras]
  {10.1111/j.1365-2966.2009.15855.x}, \href
  {https://ui.adsabs.harvard.edu/abs/2010MNRAS.401.2697H} {401, 2697}

\bibitem[\protect\citeauthoryear{{Hardcastle} et~al.,}{{Hardcastle}
  et~al.}{2012}]{hardcastle_2012}
{Hardcastle} M.~J.,  et~al., 2012, \mn@doi [\mnras]
  {10.1111/j.1365-2966.2012.21247.x}, \href
  {https://ui.adsabs.harvard.edu/abs/2012MNRAS.424.1774H} {424, 1774}

\bibitem[\protect\citeauthoryear{{Harrison}, {Thomson}, {Alexander}, {Bauer},
  {Edge}, {Hogan}, {Mullaney}  \& {Swinbank}}{{Harrison}
  et~al.}{2015}]{harrison15}
{Harrison} C.~M.,  {Thomson} A.~P.,  {Alexander} D.~M.,  {Bauer} F.~E.,  {Edge}
  A.~C.,  {Hogan} M.~T.,  {Mullaney} J.~R.,   {Swinbank} A.~M.,  2015, \mn@doi
  [\apj] {10.1088/0004-637X/800/1/45}, \href
  {https://ui.adsabs.harvard.edu/abs/2015ApJ...800...45H} {800, 45}

\bibitem[\protect\citeauthoryear{{Heckman}, {Miley}, {van Breugel}  \&
  {Butcher}}{{Heckman} et~al.}{1981}]{heckman_1981}
{Heckman} T.~M.,  {Miley} G.~K.,  {van Breugel} W.~J.~M.,   {Butcher} H.~R.,
  1981, \mn@doi [\apj] {10.1086/159050}, \href
  {https://ui.adsabs.harvard.edu/abs/1981ApJ...247..403H} {247, 403}

\bibitem[\protect\citeauthoryear{{Holt}, {Tadhunter}  \& {Morganti}}{{Holt}
  et~al.}{2008}]{holt_2008}
{Holt} J.,  {Tadhunter} C.~N.,   {Morganti} R.,  2008, \mn@doi [\mnras]
  {10.1111/j.1365-2966.2008.13089.x}, \href
  {https://ui.adsabs.harvard.edu/abs/2008MNRAS.387..639H} {387, 639}

\bibitem[\protect\citeauthoryear{{Holt}, {Tadhunter}, {Morganti}  \&
  {Emonts}}{{Holt} et~al.}{2011}]{holt_2011}
{Holt} J.,  {Tadhunter} C.~N.,  {Morganti} R.,   {Emonts} B.~H.~C.,  2011,
  \mn@doi [\mnras] {10.1111/j.1365-2966.2010.17535.x}, \href
  {https://ui.adsabs.harvard.edu/abs/2011MNRAS.410.1527H} {410, 1527}

\bibitem[\protect\citeauthoryear{{Humphrey}, {Villar-Mart{\'\i}n}, {Vernet},
  {Fosbury}, {di Serego Alighieri}  \& {Binette}}{{Humphrey}
  et~al.}{2008}]{humphrey2008}
{Humphrey} A.,  {Villar-Mart{\'\i}n} M.,  {Vernet} J.,  {Fosbury} R.,  {di
  Serego Alighieri} S.,   {Binette} L.,  2008, \mn@doi [\mnras]
  {10.1111/j.1365-2966.2007.12506.x}, \href
  {https://ui.adsabs.harvard.edu/abs/2008MNRAS.383...11H} {383, 11}

\bibitem[\protect\citeauthoryear{{Johnston}, {Broderick}, {Cotter}, {Morganti}
  \& {Hunstead}}{{Johnston} et~al.}{2010}]{johnston_2010}
{Johnston} H.~M.,  {Broderick} J.~W.,  {Cotter} G.,  {Morganti} R.,
  {Hunstead} R.~W.,  2010, \mn@doi [\mnras] {10.1111/j.1365-2966.2010.16950.x},
  \href {https://ui.adsabs.harvard.edu/abs/2010MNRAS.407..721J} {407, 721}

\bibitem[\protect\citeauthoryear{{Kharb}, {Subramanian}, {Das}, {Vaddi}  \&
  {Paragi}}{{Kharb} et~al.}{2021}]{kharb_2021}
{Kharb} P.,  {Subramanian} S.,  {Das} M.,  {Vaddi} S.,   {Paragi} Z.,  2021,
  \mn@doi [\apj] {10.3847/1538-4357/ac0c82}, \href
  {https://ui.adsabs.harvard.edu/abs/2021ApJ...919..108K} {919, 108}

\bibitem[\protect\citeauthoryear{{Koss} et~al.,}{{Koss}
  et~al.}{2015}]{koss_2015}
{Koss} M.~J.,  et~al., 2015, \mn@doi [\apj] {10.1088/0004-637X/807/2/149},
  \href {https://ui.adsabs.harvard.edu/abs/2015ApJ...807..149K} {807, 149}

\bibitem[\protect\citeauthoryear{{Kuzio de Naray}, {Arsenault}, {Spekkens},
  {Sellwood}, {McDonald}, {Simon}  \& {Teuben}}{{Kuzio de Naray}
  et~al.}{2012}]{naray_2012}
{Kuzio de Naray} R.,  {Arsenault} C.~A.,  {Spekkens} K.,  {Sellwood} J.~A.,
  {McDonald} M.,  {Simon} J.~D.,   {Teuben} P.,  2012, \mn@doi [\mnras]
  {10.1111/j.1365-2966.2012.22126.x}, \href
  {https://ui.adsabs.harvard.edu/abs/2012MNRAS.427.2523K} {427, 2523}

\bibitem[\protect\citeauthoryear{{Lanz}, {Ogle}, {Evans}, {Appleton},
  {Guillard}  \& {Emonts}}{{Lanz} et~al.}{2015}]{lanz_2015}
{Lanz} L.,  {Ogle} P.~M.,  {Evans} D.,  {Appleton} P.~N.,  {Guillard} P.,
  {Emonts} B.,  2015, \mn@doi [\apj] {10.1088/0004-637X/801/1/17}, \href
  {https://ui.adsabs.harvard.edu/abs/2015ApJ...801...17L} {801, 17}

\bibitem[\protect\citeauthoryear{{Mahony}, {Morganti}, {Emonts}, {Oosterloo}
  \& {Tadhunter}}{{Mahony} et~al.}{2013}]{mahony13}
{Mahony} E.~K.,  {Morganti} R.,  {Emonts} B.~H.~C.,  {Oosterloo} T.~A.,
  {Tadhunter} C.,  2013, \mn@doi [\mnras] {10.1093/mnrasl/slt094}, \href
  {https://ui.adsabs.harvard.edu/abs/2013MNRAS.435L..58M} {435, L58}

\bibitem[\protect\citeauthoryear{{Mahony}, {Oonk}, {Morganti}, {Tadhunter},
  {Bessiere}, {Short}, {Emonts}  \& {Oosterloo}}{{Mahony}
  et~al.}{2016}]{mahony16}
{Mahony} E.~K.,  {Oonk} J.~B.~R.,  {Morganti} R.,  {Tadhunter} C.,  {Bessiere}
  P.,  {Short} P.,  {Emonts} B.~H.~C.,   {Oosterloo} T.~A.,  2016, \mn@doi
  [\mnras] {10.1093/mnras/stv2456}, \href
  {https://ui.adsabs.harvard.edu/abs/2016MNRAS.455.2453M} {455, 2453}

\bibitem[\protect\citeauthoryear{{Maksym}, {Fabbiano}, {Elvis}, {Karovska},
  {Paggi}, {Raymond}, {Wang}  \& {Storchi-Bergmann}}{{Maksym}
  et~al.}{2017}]{Maksym_2017}
{Maksym} W.~P.,  {Fabbiano} G.,  {Elvis} M.,  {Karovska} M.,  {Paggi} A.,
  {Raymond} J.,  {Wang} J.,   {Storchi-Bergmann} T.,  2017, \mn@doi [\apj]
  {10.3847/1538-4357/aa78a4}, \href
  {https://ui.adsabs.harvard.edu/abs/2017ApJ...844...69M} {844, 69}

\bibitem[\protect\citeauthoryear{{Maksym} et~al.,}{{Maksym}
  et~al.}{2019}]{Maksym_2019}
{Maksym} W.~P.,  et~al., 2019, \mn@doi [\apj] {10.3847/1538-4357/aaf4f5}, \href
  {https://ui.adsabs.harvard.edu/abs/2019ApJ...872...94M} {872, 94}

\bibitem[\protect\citeauthoryear{{Mandal}, {Mukherjee}, {Federrath},
  {Nesvadba}, {Bicknell}, {Wagner}  \& {Meenakshi}}{{Mandal}
  et~al.}{2021}]{mandal_2021}
{Mandal} A.,  {Mukherjee} D.,  {Federrath} C.,  {Nesvadba} N. P.~H.,
  {Bicknell} G.~V.,  {Wagner} A.~Y.,   {Meenakshi} M.,  2021, \mn@doi [\mnras]
  {10.1093/mnras/stab2822}, \href
  {https://ui.adsabs.harvard.edu/abs/2021MNRAS.508.4738M} {508, 4738}

\bibitem[\protect\citeauthoryear{{Maschmann}, {Melchior}, {Mamon},
  {Chilingarian}  \& {Katkov}}{{Maschmann} et~al.}{2020}]{maschmann_2020}
{Maschmann} D.,  {Melchior} A.-L.,  {Mamon} G.~A.,  {Chilingarian} I.~V.,
  {Katkov} I.~Y.,  2020, \mn@doi [\aap] {10.1051/0004-6361/202037868}, \href
  {https://ui.adsabs.harvard.edu/abs/2020A&A...641A.171M} {641, A171}

\bibitem[\protect\citeauthoryear{{Meenakshi}, {Mukherjee}, {Wagner},
  {Nesvadba}, {Morganti}, {Janssen}  \& {Bicknell}}{{Meenakshi}
  et~al.}{2022}]{Meenakshi_2022a}
{Meenakshi} M.,  {Mukherjee} D.,  {Wagner} A.~Y.,  {Nesvadba} N. P.~H.,
  {Morganti} R.,  {Janssen} R. M.~J.,   {Bicknell} G.~V.,  2022, \mn@doi
  [\mnras] {10.1093/mnras/stac167}, \href
  {https://ui.adsabs.harvard.edu/abs/2022MNRAS.511.1622M} {511, 1622}

\bibitem[\protect\citeauthoryear{{Mignone}, {Zanni}, {Tzeferacos}, {van
  Straalen}, {Colella}  \& {Bodo}}{{Mignone} et~al.}{2012}]{mignone_2012}
{Mignone} A.,  {Zanni} C.,  {Tzeferacos} P.,  {van Straalen} B.,  {Colella} P.,
    {Bodo} G.,  2012, \mn@doi [\apjs] {10.1088/0067-0049/198/1/7}, \href
  {https://ui.adsabs.harvard.edu/abs/2012ApJS..198....7M} {198, 7}

\bibitem[\protect\citeauthoryear{{Mooney} et~al.,}{{Mooney}
  et~al.}{2021}]{mooney_2021}
{Mooney} S.,  et~al., 2021, \mn@doi [\apjs] {10.3847/1538-4365/ac1c0b}, \href
  {https://ui.adsabs.harvard.edu/abs/2021ApJS..257...30M} {257, 30}

\bibitem[\protect\citeauthoryear{{Morganti}}{{Morganti}}{2017}]{morganti_2017}
{Morganti} R.,  2017, \mn@doi [Frontiers in Astronomy and Space Sciences]
  {10.3389/fspas.2017.00042}, \href
  {https://ui.adsabs.harvard.edu/abs/2017FrASS...4...42M} {4, 42}

\bibitem[\protect\citeauthoryear{{Morganti}, {Tadhunter}, {Dickson}  \&
  {Shaw}}{{Morganti} et~al.}{1997}]{morganti_1997}
{Morganti} R.,  {Tadhunter} C.~N.,  {Dickson} R.,   {Shaw} M.,  1997, \aap,
  \href {https://ui.adsabs.harvard.edu/abs/1997A&A...326..130M} {326, 130}

\bibitem[\protect\citeauthoryear{{Morganti}, {Oosterloo}, {Tadhunter}, {van
  Moorsel}  \& {Emonts}}{{Morganti} et~al.}{2005}]{morganti_2005}
{Morganti} R.,  {Oosterloo} T.~A.,  {Tadhunter} C.~N.,  {van Moorsel} G.,
  {Emonts} B.,  2005, \mn@doi [\aap] {10.1051/0004-6361:20053175}, \href
  {https://ui.adsabs.harvard.edu/abs/2005A&A...439..521M} {439, 521}

\bibitem[\protect\citeauthoryear{{Morganti}, {Holt}, {Saripalli}, {Oosterloo}
  \& {Tadhunter}}{{Morganti} et~al.}{2007}]{raffaella07}
{Morganti} R.,  {Holt} J.,  {Saripalli} L.,  {Oosterloo} T.~A.,   {Tadhunter}
  C.~N.,  2007, \mn@doi [\aap] {10.1051/0004-6361:20077888}, \href
  {https://ui.adsabs.harvard.edu/abs/2007A&A...476..735M} {476, 735}

\bibitem[\protect\citeauthoryear{{Morganti}, {Oosterloo}, {Oonk}, {Frieswijk}
  \& {Tadhunter}}{{Morganti} et~al.}{2015}]{raffaella15}
{Morganti} R.,  {Oosterloo} T.,  {Oonk} J.~B.~R.,  {Frieswijk} W.,
  {Tadhunter} C.,  2015, \mn@doi [\aap] {10.1051/0004-6361/201525860}, \href
  {https://ui.adsabs.harvard.edu/abs/2015A&A...580A...1M} {580, A1}

\bibitem[\protect\citeauthoryear{Morganti, Oosterloo, Murthy  \&
  Tadhunter}{Morganti et~al.}{2021a}]{morganti_2021a}
Morganti R.,  Oosterloo T.,  Murthy S.,   Tadhunter C.,  2021a, \mn@doi
  [Astronomische Nachrichten] {https://doi.org/10.1002/asna.20210037}, 342,
  1135

\bibitem[\protect\citeauthoryear{{Morganti}, {Oosterloo}, {Tadhunter},
  {Bernhard}  \& {Raymond Oonk}}{{Morganti} et~al.}{2021b}]{morganti_2021b}
{Morganti} R.,  {Oosterloo} T.,  {Tadhunter} C.,  {Bernhard} E.~P.,   {Raymond
  Oonk} J.~B.,  2021b, \mn@doi [\aap] {10.1051/0004-6361/202141766}, \href
  {https://ui.adsabs.harvard.edu/abs/2021A&A...656A..55M} {656, A55}

\bibitem[\protect\citeauthoryear{{Moy} \& {Rocca-Volmerange}}{{Moy} \&
  {Rocca-Volmerange}}{2002}]{moy_2002}
{Moy} E.,  {Rocca-Volmerange} B.,  2002, \mn@doi [\aap]
  {10.1051/0004-6361:20011727}, \href
  {https://ui.adsabs.harvard.edu/abs/2002A&A...383...46M} {383, 46}

\bibitem[\protect\citeauthoryear{{Mukherjee}, {Bicknell}, {Sutherland}  \&
  {Wagner}}{{Mukherjee} et~al.}{2016}]{dipanjan16}
{Mukherjee} D.,  {Bicknell} G.~V.,  {Sutherland} R.,   {Wagner} A.,  2016,
  \mn@doi [\mnras] {10.1093/mnras/stw1368}, \href
  {https://ui.adsabs.harvard.edu/abs/2016MNRAS.461..967M} {461, 967}

\bibitem[\protect\citeauthoryear{{Mukherjee}, {Wagner}, {Bicknell}, {Morganti},
  {Oosterloo}, {Nesvadba}  \& {Sutherland}}{{Mukherjee}
  et~al.}{2018a}]{dipanjan5063}
{Mukherjee} D.,  {Wagner} A.~Y.,  {Bicknell} G.~V.,  {Morganti} R.,
  {Oosterloo} T.,  {Nesvadba} N.,   {Sutherland} R.~S.,  2018a, \mn@doi
  [\mnras] {10.1093/mnras/sty067}, \href
  {https://ui.adsabs.harvard.edu/abs/2018MNRAS.476...80M} {476, 80}

\bibitem[\protect\citeauthoryear{{Mukherjee}, {Bicknell}, {Wagner},
  {Sutherland}  \& {Silk}}{{Mukherjee} et~al.}{2018b}]{dipanjan2018}
{Mukherjee} D.,  {Bicknell} G.~V.,  {Wagner} A. e.~Y.,  {Sutherland} R.~S.,
  {Silk} J.,  2018b, \mn@doi [\mnras] {10.1093/mnras/sty1776}, \href
  {https://ui.adsabs.harvard.edu/abs/2018MNRAS.479.5544M} {479, 5544}

\bibitem[\protect\citeauthoryear{{Mukherjee}, {Bicknell}  \&
  {Wagner}}{{Mukherjee} et~al.}{2021}]{mukherjee2022}
{Mukherjee} D.,  {Bicknell} G.~V.,   {Wagner} A.~Y.,  2021, \mn@doi
  [Astronomische Nachrichten] {10.1002/asna.20210061}, \href
  {https://ui.adsabs.harvard.edu/abs/2021AN....342.1140M} {342, 1140}

\bibitem[\protect\citeauthoryear{{Murthy} et~al.,}{{Murthy}
  et~al.}{2019}]{murthy_2019}
{Murthy} S.,  et~al., 2019, \mn@doi [\aap] {10.1051/0004-6361/201935931}, \href
  {https://ui.adsabs.harvard.edu/abs/2019A&A...629A..58M} {629, A58}

\bibitem[\protect\citeauthoryear{{Murthy}, {Morganti}, {Wagner}, {Oosterloo},
  {Guillard}, {Mukherjee}  \& {Bicknell}}{{Murthy} et~al.}{2022}]{suma_2022}
{Murthy} S.,  {Morganti} R.,  {Wagner} A.~Y.,  {Oosterloo} T.,  {Guillard} P.,
  {Mukherjee} D.,   {Bicknell} G.,  2022, \mn@doi [Nature Astronomy]
  {10.1038/s41550-021-01596-6}, \href
  {https://ui.adsabs.harvard.edu/abs/2022NatAs...6..488M} {6, 488}

\bibitem[\protect\citeauthoryear{{Nesvadba}, {Lehnert}, {De Breuck}, {Gilbert}
  \& {van Breugel}}{{Nesvadba} et~al.}{2007}]{nicole07}
{Nesvadba} N.~P.~H.,  {Lehnert} M.~D.,  {De Breuck} C.,  {Gilbert} A.,   {van
  Breugel} W.,  2007, \mn@doi [\aap] {10.1051/0004-6361:20078175}, \href
  {https://ui.adsabs.harvard.edu/abs/2007A&A...475..145N} {475, 145}

\bibitem[\protect\citeauthoryear{{Nesvadba}, {Lehnert}, {De Breuck}, {Gilbert}
  \& {van Breugel}}{{Nesvadba} et~al.}{2008}]{nicole08}
{Nesvadba} N.~P.~H.,  {Lehnert} M.~D.,  {De Breuck} C.,  {Gilbert} A.~M.,
  {van Breugel} W.,  2008, \mn@doi [\aap] {10.1051/0004-6361:200810346}, \href
  {https://ui.adsabs.harvard.edu/abs/2008A&A...491..407N} {491, 407}

\bibitem[\protect\citeauthoryear{{Nesvadba} et~al.,}{{Nesvadba}
  et~al.}{2010}]{nesvadba10}
{Nesvadba} N.~P.~H.,  et~al., 2010, \mn@doi [\aap]
  {10.1051/0004-6361/200913333}, \href
  {https://ui.adsabs.harvard.edu/abs/2010A&A...521A..65N} {521, A65}

\bibitem[\protect\citeauthoryear{{Nesvadba}, {De Breuck}, {Lehnert}, {Best}  \&
  {Collet}}{{Nesvadba} et~al.}{2017}]{nesvadba_2017}
{Nesvadba} N.~P.~H.,  {De Breuck} C.,  {Lehnert} M.~D.,  {Best} P.~N.,
  {Collet} C.,  2017, \mn@doi [\aap] {10.1051/0004-6361/201528040}, \href
  {https://ui.adsabs.harvard.edu/abs/2017A&A...599A.123N} {599, A123}

\bibitem[\protect\citeauthoryear{{Nevin}, {Comerford},
  {M{\"u}ller-S{\'a}nchez}, {Barrows}  \& {Cooper}}{{Nevin}
  et~al.}{2018}]{nevin_2018}
{Nevin} R.,  {Comerford} J.~M.,  {M{\"u}ller-S{\'a}nchez} F.,  {Barrows} R.,
  {Cooper} M.~C.,  2018, \mn@doi [\mnras] {10.1093/mnras/stx2433}, \href
  {https://ui.adsabs.harvard.edu/abs/2018MNRAS.473.2160N} {473, 2160}

\bibitem[\protect\citeauthoryear{{O'Dea}}{{O'Dea}}{1998}]{chris_1998}
{O'Dea} C.~P.,  1998, \mn@doi [\pasp] {10.1086/316162}, \href
  {https://ui.adsabs.harvard.edu/abs/1998PASP..110..493O} {110, 493}

\bibitem[\protect\citeauthoryear{{Oosterloo}, {Morganti}, {Tzioumis},
  {Reynolds}, {King}, {McCulloch}  \& {Tsvetanov}}{{Oosterloo}
  et~al.}{2000}]{oosterloo}
{Oosterloo} T.~A.,  {Morganti} R.,  {Tzioumis} A.,  {Reynolds} J.,  {King} E.,
  {McCulloch} P.,   {Tsvetanov} Z.,  2000, \mn@doi [\aj] {10.1086/301358},
  \href {https://ui.adsabs.harvard.edu/abs/2000AJ....119.2085O} {119, 2085}

\bibitem[\protect\citeauthoryear{{Oosterloo}, {Raymond Oonk}, {Morganti},
  {Combes}, {Dasyra}, {Salom{\'e}}, {Vlahakis}  \& {Tadhunter}}{{Oosterloo}
  et~al.}{2017}]{oosterloo_2017}
{Oosterloo} T.,  {Raymond Oonk} J.~B.,  {Morganti} R.,  {Combes} F.,  {Dasyra}
  K.,  {Salom{\'e}} P.,  {Vlahakis} N.,   {Tadhunter} C.,  2017, \mn@doi [\aap]
  {10.1051/0004-6361/201731781}, \href
  {https://ui.adsabs.harvard.edu/abs/2017A&A...608A..38O} {608, A38}

\bibitem[\protect\citeauthoryear{{Oosterloo}, {Morganti}, {Tadhunter}, {Raymond
  Oonk}, {Bignall}, {Tzioumis}  \& {Reynolds}}{{Oosterloo}
  et~al.}{2019}]{oosterloo_2019}
{Oosterloo} T.,  {Morganti} R.,  {Tadhunter} C.,  {Raymond Oonk} J.~B.,
  {Bignall} H.~E.,  {Tzioumis} T.,   {Reynolds} C.,  2019, \mn@doi [\aap]
  {10.1051/0004-6361/201936248}, \href
  {https://ui.adsabs.harvard.edu/abs/2019A&A...632A..66O} {632, A66}

\bibitem[\protect\citeauthoryear{{Padmanabhan}}{{Padmanabhan}}{2006}]{paddy_2006}
{Padmanabhan} T.,  2006, {An Invitation To Astrophysics}, An Invitation To
  Astrophysics. Series: World Scientific Series in Astronomy and Astrophysics,
  \mn@doi{10.1142/6010}

\bibitem[\protect\citeauthoryear{{Rakshit} \& {Woo}}{{Rakshit} \&
  {Woo}}{2018}]{rakshit_2018}
{Rakshit} S.,  {Woo} J.-H.,  2018, \mn@doi [\apj] {10.3847/1538-4357/aad9f8},
  \href {https://ui.adsabs.harvard.edu/abs/2018ApJ...865....5R} {865, 5}

\bibitem[\protect\citeauthoryear{{Rubinur}, {Das}  \& {Kharb}}{{Rubinur}
  et~al.}{2019}]{rubinur_2019}
{Rubinur} K.,  {Das} M.,   {Kharb} P.,  2019, \mn@doi [\mnras]
  {10.1093/mnras/stz334}, \href
  {https://ui.adsabs.harvard.edu/abs/2019MNRAS.484.4933R} {484, 4933}

\bibitem[\protect\citeauthoryear{{Ruffa} et~al.,}{{Ruffa}
  et~al.}{2019}]{ruffa_2019b}
{Ruffa} I.,  et~al., 2019, \mn@doi [\mnras] {10.1093/mnras/stz2368}, \href
  {https://ui.adsabs.harvard.edu/abs/2019MNRAS.489.3739R} {489, 3739}

\bibitem[\protect\citeauthoryear{{Ruffa}, {Laing}, {Prandoni}, {Paladino},
  {Parma}, {Davis}  \& {Bureau}}{{Ruffa} et~al.}{2020}]{ruffa_2020a}
{Ruffa} I.,  {Laing} R.~A.,  {Prandoni} I.,  {Paladino} R.,  {Parma} P.,
  {Davis} T.~A.,   {Bureau} M.,  2020, \mn@doi [\mnras]
  {10.1093/mnras/staa3166}, \href
  {https://ui.adsabs.harvard.edu/abs/2020MNRAS.499.5719R} {499, 5719}

\bibitem[\protect\citeauthoryear{{Sabater} et~al.,}{{Sabater}
  et~al.}{2019}]{sabater_2019}
{Sabater} J.,  et~al., 2019, \mn@doi [\aap] {10.1051/0004-6361/201833883},
  \href {https://ui.adsabs.harvard.edu/abs/2019A&A...622A..17S} {622, A17}

\bibitem[\protect\citeauthoryear{{Santoro}, {Rose}, {Morganti}, {Tadhunter},
  {Oosterloo}  \& {Holt}}{{Santoro} et~al.}{2018}]{santoro_2018}
{Santoro} F.,  {Rose} M.,  {Morganti} R.,  {Tadhunter} C.,  {Oosterloo} T.~A.,
   {Holt} J.,  2018, \mn@doi [\aap] {10.1051/0004-6361/201833248}, \href
  {https://ui.adsabs.harvard.edu/abs/2018A&A...617A.139S} {617, A139}

\bibitem[\protect\citeauthoryear{{Santoro}, {Tadhunter}, {Baron}, {Morganti}
  \& {Holt}}{{Santoro} et~al.}{2020}]{santoro_2020}
{Santoro} F.,  {Tadhunter} C.,  {Baron} D.,  {Morganti} R.,   {Holt} J.,  2020,
  \mn@doi [\aap] {10.1051/0004-6361/202039077}, \href
  {https://ui.adsabs.harvard.edu/abs/2020A&A...644A..54S} {644, A54}

\bibitem[\protect\citeauthoryear{Saxton, Bicknell, Sutherland  \&
  Midgley}{Saxton et~al.}{2005}]{Saxton2005}
Saxton C.~J.,  Bicknell G.~V.,  Sutherland R.~S.,   Midgley S.,  2005, \mn@doi
  [Mon. Not. R. Astron. Soc.] {10.1111/j.1365-2966.2005.08962.x}, 359, 781

\bibitem[\protect\citeauthoryear{{Schulz}, {Morganti}, {Nyland}, {Paragi},
  {Mahony}  \& {Oosterloo}}{{Schulz} et~al.}{2021}]{schulz_2021}
{Schulz} R.,  {Morganti} R.,  {Nyland} K.,  {Paragi} Z.,  {Mahony} E.~K.,
  {Oosterloo} T.,  2021, \mn@doi [\aap] {10.1051/0004-6361/202037677}, \href
  {https://ui.adsabs.harvard.edu/abs/2021A&A...647A..63S} {647, A63}

\bibitem[\protect\citeauthoryear{{Shih}, {Stockton}  \& {Kewley}}{{Shih}
  et~al.}{2013}]{shih_2013}
{Shih} H.-Y.,  {Stockton} A.,   {Kewley} L.,  2013, \mn@doi [\apj]
  {10.1088/0004-637X/772/2/138}, \href
  {https://ui.adsabs.harvard.edu/abs/2013ApJ...772..138S} {772, 138}

\bibitem[\protect\citeauthoryear{{Sol{\'o}rzano-I{\~n}arrea} \&
  {Tadhunter}}{{Sol{\'o}rzano-I{\~n}arrea} \& {Tadhunter}}{2003}]{solorzano03}
{Sol{\'o}rzano-I{\~n}arrea} C.,  {Tadhunter} C.~N.,  2003, \mn@doi [\mnras]
  {10.1046/j.1365-8711.2003.06376.x}, \href
  {https://ui.adsabs.harvard.edu/abs/2003MNRAS.340..705S} {340, 705}

\bibitem[\protect\citeauthoryear{{Speranza} et~al.,}{{Speranza}
  et~al.}{2021}]{speranza_2021}
{Speranza} G.,  et~al., 2021, arXiv e-prints, \href
  {https://ui.adsabs.harvard.edu/abs/2021arXiv210609743S} {p. arXiv:2106.09743}

\bibitem[\protect\citeauthoryear{{Sutherland} \& {Bicknell}}{{Sutherland} \&
  {Bicknell}}{2007}]{Sutherland07}
{Sutherland} R.~S.,  {Bicknell} G.~V.,  2007, \mn@doi [\apjs] {10.1086/520640},
  \href {https://ui.adsabs.harvard.edu/abs/2007ApJS..173...37S} {173, 37}

\bibitem[\protect\citeauthoryear{{Sutherland}, {Dopita}, {Binette}  \&
  {Groves}}{{Sutherland} et~al.}{2018}]{mappings_2018}
{Sutherland} R.,  {Dopita} M.,  {Binette} L.,   {Groves} B.,  2018, {MAPPINGS
  V: Astrophysical plasma modeling code} (\mn@eprint {ascl} {1807.005})

\bibitem[\protect\citeauthoryear{{Tadhunter}}{{Tadhunter}}{2002}]{tadhunter_2002}
{Tadhunter} C.~N.,  2002, in Revista Mexicana de Astronomia y Astrofisica
  Conference Series. pp 213--221 (\mn@eprint {arXiv} {astro-ph/0201315})

\bibitem[\protect\citeauthoryear{{Tadhunter}, {Villar-Martin}, {Morganti},
  {Bland-Hawthorn}  \& {Axon}}{{Tadhunter} et~al.}{2000}]{tadhunter_2000}
{Tadhunter} C.~N.,  {Villar-Martin} M.,  {Morganti} R.,  {Bland-Hawthorn} J.,
  {Axon} D.,  2000, \mn@doi [\mnras] {10.1046/j.1365-8711.2000.03416.x}, \href
  {https://ui.adsabs.harvard.edu/abs/2000MNRAS.314..849T} {314, 849}

\bibitem[\protect\citeauthoryear{{Tadhunter}, {Morganti}, {Rose}, {Oonk}  \&
  {Oosterloo}}{{Tadhunter} et~al.}{2014}]{tadhunter_2014}
{Tadhunter} C.,  {Morganti} R.,  {Rose} M.,  {Oonk} J.~B.~R.,   {Oosterloo} T.,
   2014, \mn@doi [\nat] {10.1038/nature13520}, \href
  {https://ui.adsabs.harvard.edu/abs/2014Natur.511..440T} {511, 440}

\bibitem[\protect\citeauthoryear{{Tadhunter} et~al.,}{{Tadhunter}
  et~al.}{2018}]{clive18}
{Tadhunter} C.,  et~al., 2018, \mn@doi [\mnras] {10.1093/mnras/sty1064}, \href
  {https://ui.adsabs.harvard.edu/abs/2018MNRAS.478.1558T} {478, 1558}

\bibitem[\protect\citeauthoryear{{Travascio}, {Fabbiano}, {Paggi}, {Elvis},
  {Maksym}, {Morganti}, {Oosterloo}  \& {Fiore}}{{Travascio}
  et~al.}{2021}]{travascio_2021}
{Travascio} A.,  {Fabbiano} G.,  {Paggi} A.,  {Elvis} M.,  {Maksym} W.~P.,
  {Morganti} R.,  {Oosterloo} T.,   {Fiore} F.,  2021, \mn@doi [\apj]
  {10.3847/1538-4357/ac18c7}, \href
  {https://ui.adsabs.harvard.edu/abs/2021ApJ...921..129T} {921, 129}

\bibitem[\protect\citeauthoryear{{Vayner}, {Zakamska}, {Wright}, {Armus},
  {Murray}  \& {Walth}}{{Vayner} et~al.}{2021}]{Vayner_2021b}
{Vayner} A.,  {Zakamska} N.,  {Wright} S.~A.,  {Armus} L.,  {Murray} N.,
  {Walth} G.,  2021, \mn@doi [\apj] {10.3847/1538-4357/ac2b9e}, \href
  {https://ui.adsabs.harvard.edu/abs/2021ApJ...923...59V} {923, 59}

\bibitem[\protect\citeauthoryear{{Venturi} et~al.,}{{Venturi}
  et~al.}{2021}]{Venturi21}
{Venturi} G.,  et~al., 2021, \mn@doi [\aap] {10.1051/0004-6361/202039869},
  \href {https://ui.adsabs.harvard.edu/abs/2021A&A...648A..17V} {648, A17}

\bibitem[\protect\citeauthoryear{{Villar-Martin}, {Tadhunter}  \&
  {Clark}}{{Villar-Martin} et~al.}{1997}]{montse_1997}
{Villar-Martin} M.,  {Tadhunter} C.,   {Clark} N.,  1997, \aap, \href
  {https://ui.adsabs.harvard.edu/abs/1997A&A...323...21V} {323, 21}

\bibitem[\protect\citeauthoryear{{Villar-Mart{\'\i}n}, {Tadhunter}, {Morganti},
  {Axon}  \& {Koekemoer}}{{Villar-Mart{\'\i}n} et~al.}{1999}]{montse_1999}
{Villar-Mart{\'\i}n} M.,  {Tadhunter} C.,  {Morganti} R.,  {Axon} D.,
  {Koekemoer} A.,  1999, \mn@doi [\mnras] {10.1046/j.1365-8711.1999.02603.x},
  \href {https://ui.adsabs.harvard.edu/abs/1999MNRAS.307...24V} {307, 24}

\bibitem[\protect\citeauthoryear{{Villar-Mart{\'\i}n}
  et~al.,}{{Villar-Mart{\'\i}n} et~al.}{2017}]{montse_2017}
{Villar-Mart{\'\i}n} M.,  et~al., 2017, \mn@doi [\mnras]
  {10.1093/mnras/stx2209}, \href
  {https://ui.adsabs.harvard.edu/abs/2017MNRAS.472.4659V} {472, 4659}

\bibitem[\protect\citeauthoryear{{Wagner} \& {Bicknell}}{{Wagner} \&
  {Bicknell}}{2011}]{wagner11}
{Wagner} A.~Y.,  {Bicknell} G.~V.,  2011, \mn@doi [\apj]
  {10.1088/0004-637X/738/1/117}, \href
  {https://ui.adsabs.harvard.edu/abs/2011ApJ...738..117W} {738, 117}

\bibitem[\protect\citeauthoryear{{Wagner}, {Bicknell}  \& {Umemura}}{{Wagner}
  et~al.}{2012}]{wagner12}
{Wagner} A.~Y.,  {Bicknell} G.~V.,   {Umemura} M.,  2012, \mn@doi [\apj]
  {10.1088/0004-637X/757/2/136}, \href
  {https://ui.adsabs.harvard.edu/abs/2012ApJ...757..136W} {757, 136}

\bibitem[\protect\citeauthoryear{{Wang} et~al.,}{{Wang}
  et~al.}{2011}]{wang_2011}
{Wang} J.,  et~al., 2011, \mn@doi [\apj] {10.1088/0004-637X/729/1/75}, \href
  {https://ui.adsabs.harvard.edu/abs/2011ApJ...729...75W} {729, 75}

\bibitem[\protect\citeauthoryear{{Whittle}}{{Whittle}}{1985}]{whittle_1985}
{Whittle} M.,  1985, \mn@doi [\mnras] {10.1093/mnras/213.1.1}, \href
  {https://ui.adsabs.harvard.edu/abs/1985MNRAS.213....1W} {213, 1}

\bibitem[\protect\citeauthoryear{{Zakamska} et~al.,}{{Zakamska}
  et~al.}{2016}]{nadia}
{Zakamska} N.~L.,  et~al., 2016, \mn@doi [\mnras] {10.1093/mnras/stw718}, \href
  {https://ui.adsabs.harvard.edu/abs/2016MNRAS.459.3144Z} {459, 3144}

\bibitem[\protect\citeauthoryear{{Zovaro}, {Sharp}, {Nesvadba}, {Bicknell},
  {Mukherjee}, {Wagner}, {Groves}  \& {Krishna}}{{Zovaro}
  et~al.}{2019a}]{zovaro_2019}
{Zovaro} H. R.~M.,  {Sharp} R.,  {Nesvadba} N. P.~H.,  {Bicknell} G.~V.,
  {Mukherjee} D.,  {Wagner} A.~Y.,  {Groves} B.,   {Krishna} S.,  2019a,
  \mn@doi [\mnras] {10.1093/mnras/stz233}, \href
  {https://ui.adsabs.harvard.edu/abs/2019MNRAS.484.3393Z} {484, 3393}

\bibitem[\protect\citeauthoryear{{Zovaro}, {Nesvadba}, {Sharp}, {Bicknell},
  {Groves}, {Mukherjee}  \& {Wagner}}{{Zovaro} et~al.}{2019b}]{zovaro_2019b}
{Zovaro} H. R.~M.,  {Nesvadba} N. P.~H.,  {Sharp} R.,  {Bicknell} G.~V.,
  {Groves} B.,  {Mukherjee} D.,   {Wagner} A.~Y.,  2019b, \mn@doi [\mnras]
  {10.1093/mnras/stz2459}, \href
  {https://ui.adsabs.harvard.edu/abs/2019MNRAS.489.4944Z} {489, 4944}

\makeatother
\end{thebibliography}


\appendix

\section{Radiative shock and impact of resolution}
\label{cooling-length}

\begin{figure}
    \centering
    \includegraphics[width=0.8\linewidth]{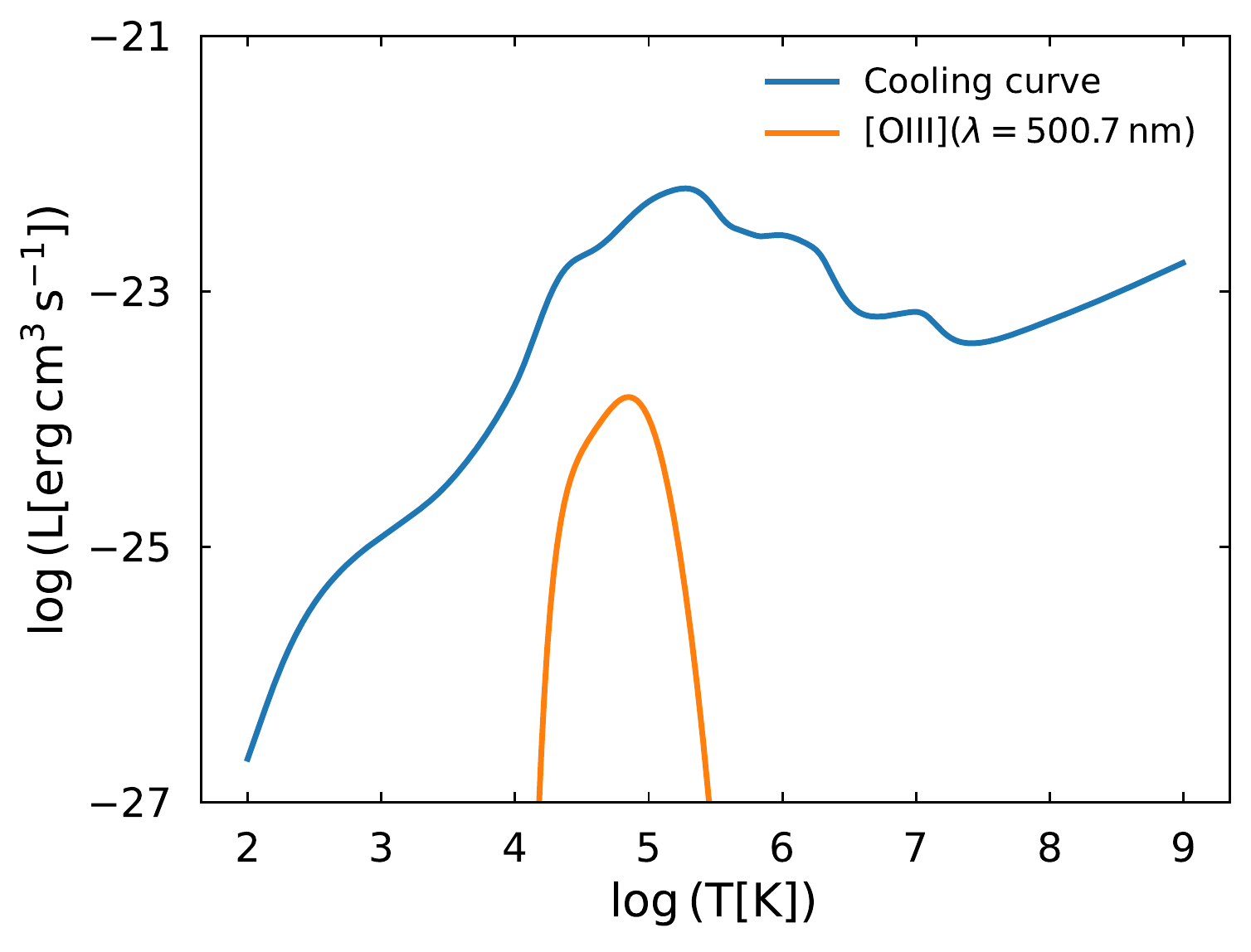}
    \caption{Cooling curves from \textsc{\textsc{mappings v}} \citep{mappings_2018} as a function of temperature for non-equilibrium cooling of shocked gas. The blue curve shows the total cooling function, and the orange represents the contribution from [\ion{O}{III}] emission.}
    \label{fig:cool-curve}
\end{figure}

\begin{figure*}
 \centerline{
\def\arraystretch{1.0}
\setlength{\tabcolsep}{0.0pt}
\begin{tabular}{lcr}
     \hspace{-1cm}
      \includegraphics[width=0.35\linewidth]{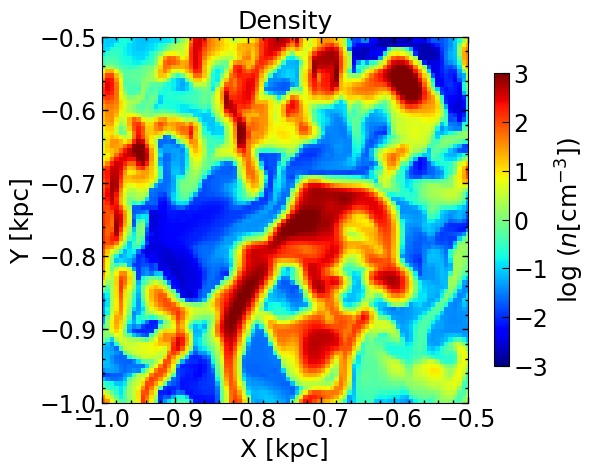} &
    \includegraphics[width=0.34\linewidth]{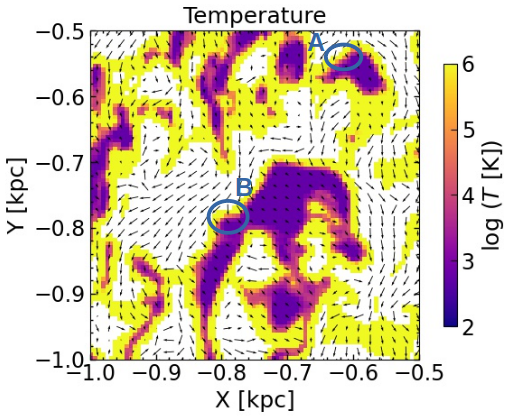} &
    \includegraphics[width=0.357\linewidth]{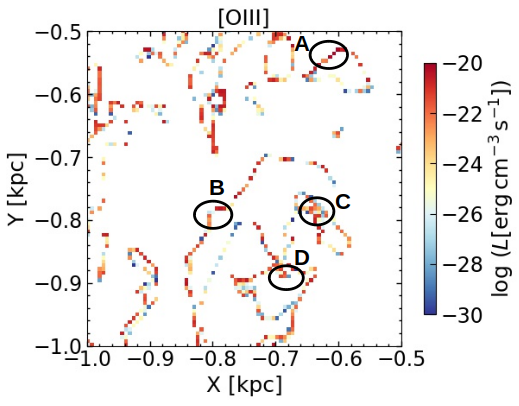}
     \end{tabular}}
 \caption{Zoomed-in view of logarithmic density, temperature over-plotted with velocities ($V_x,V_y$) and [\ion{O}{III}] emissivity in the central regions of Z=0 plane of Sim.~D at 0.98~Myr. The length of the velocity vectors are proportional to their magnitude, up to a maximum of $\sim 800 \kms$. The encircled regions in the middle and right panel show different emission zones, i.e. shocked regions (A and B) and mixing regions of hot external gas with cold gas in the clouds i.e. C and D (see Appendix~\ref{cooling-length} for detail).}
  \label{fig:temp_oiii}
\end{figure*}
In this section, we examine the radiative shocks in the simulations used for our study as well as the spatial variation of density, temperature and [\ion{O}{III}] emission zones, for a more detailed picture of the jet-gas interaction. We closely follow the approach used in \textsc{pluto} \citep[see Appendix in][]{mignone_2012} to identify the shocks in the simulations used for our study. A shocked cell is identified if the following conditions are satisfied:\\
(1)  \textbf{$\nabla \cdot V<0$}, where $V \equiv V_i=(V_x,V_y,V_z)_i$ is the velocity of the shocked cell, and \\
(2) $\frac{\Delta P}{P_{\mathrm{min}}}>5/3$, where $P$ denotes the pressure, and $\Delta P= P_{i+1}-P_{i-1}$, and $P_{\mathrm{min}}=\mathrm{min}(P_{i+1},P_{i-1})$.

The calculations are performed for 1-dimension and therefore we use a threshold of 5/3 for the pressure condition. We first verify that the temperature, density, and pressure of the shocked cell are greater than in the next neighbouring cell, and then perform the  analysis below. The physical quantities in the pre-shock and post-shock cells are denoted by the subscripts $i-1$ and $i$, respectively. 
After identifying a cell containing shocked gas, the cooling rate ($\Lambda[\mathrm{erg\,cm^{3}\,s^{-1}}]$) is obtained from the cell temperature using the cooling function pre-calculated from \textsc{\textsc{mappings v}} \citep{mappings_2018}, shown by the blue curve in Fig.~\ref{fig:cool-curve}. The figure also shows the cooling rate via the [\ion{O}{III}] line in orange.
The cooling time of the radiative shock is calculated from $t_{\mathrm{cool}}=\frac{P_{i}}{(\gamma-1)}\frac{1}{\Lambda n_{i-1}^2}$, where $n$ is the number density.
The pressure and density of the post-shock cell are used to calculate the shock velocity, which for the strong shock limit is $V_s=\sqrt{\frac{16 P_i}{3 \rho_i}}$, where $\rho$ denotes the mass density of the post-shock cell. The product of shock velocity and cooling time gives the cooling length of the radiative shock. We find that the cooling lengths for the simulations used in our study mostly lies between 0.014 and 1~pc. This is smaller than our cell resolution, which implies that the cooling layer of radiative shocks propagating into the clouds is generally not resolved in our simulations.

However, one must note that the above estimates of one dimensional shocks applied to the shocked cells in the simulations give only an approximate estimate of the cooling length. Due to the broad distribution of cloud morphologies, the shocks in the simulations have a complex 3D structure. This is highlighted in Fig.~\ref{fig:temp_oiii} where we present the cross-section of some clouds over a $500\times500$ pc region in the $X-Y$ plane of simulation D. By comparing the variation of the physical quantities across such shock boundaries, we find that the hydrodynamic shocks, which are indicated by a discontinuity in temperature/pressure in the outer layers of dense regions, are resolved by 3 to 4 computational cells in the direction perpendicular to the cloud's surface (see for e.g. encircled regions A and B in the middle panel of Fig.~\ref{fig:temp_oiii}). As indicated by the velocity vectors in the middle panel, these shocked regions are hit by the jet percolating through the complex cloud distribution in its flood-channel phase \citep[see e.g.][]{wagner12}, resulting in high [\ion{O}{III}] emission at the shock front (see encircled regions A and B in the right panel). 
Such [\ion{O}{III}] emitting regions in the post-shock cooling zones of shocks penetrating the outer layers of the clouds are mostly resolved by 1 to 2 cells. Although further spatially resolving such regions would yield more accurate estimates of the [\ion{O}{III}] luminosities, our current method correctly identifies the approximate locations of such shocked zones on the cloud's outer surfaces, and thus also correctly traces the bulk kinematics of shocked cloud that will contribute to the broadening of the simulated spectral lines.

Moreover, we note that the [\ion{O}{III}] emission also occurs due to mixing of the hot external gas with the cold gas in the clouds, as shown by the encircled regions C and D in Fig.~\ref{fig:temp_oiii}). Such regions are not directly affected by the shocks induced by the jet, but represent gas heated due to mixing with external hot medium at the boundaries of clouds, which subsequently cools due to radiative decay and contributes to the total emission from the system. The [\ion{O}{III}] emitting regions identified in our simulations thus represent an ensemble of shocked regions and hot cooling plasma mixed with the clouds. Thus, although the shocks are not well-resolved in our study, and the total luminosities are likely approximate upper limits, the integrated sum of the [\ion{O}{III}] emitting regions on the image plane, is an acceptable tracer of the distribution of the collisionally ionised gas and its dynamics.


\section{Comparing different methods for estimating velocity widths}

\label{widths-compare}

\begin{figure*}
 \centerline{
\def\arraystretch{1.0}
\setlength{\tabcolsep}{0.0pt}
\begin{tabular}{lcr}
    \includegraphics[width=0.34\linewidth]{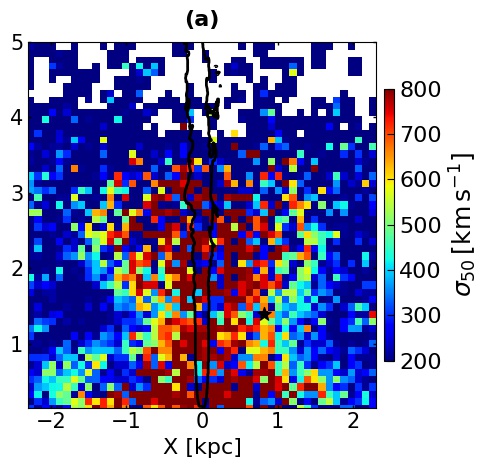}&
    \includegraphics[width=0.355\linewidth]{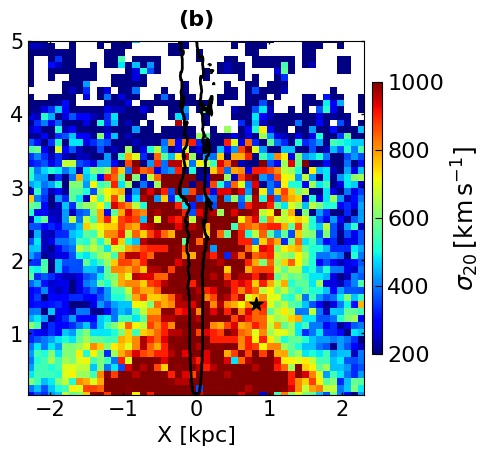}&
    \includegraphics[width=0.345\linewidth]{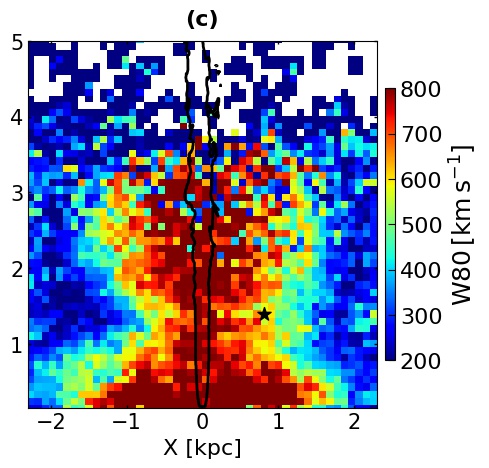}
     \end{tabular}}
     
      \centerline{
\def\arraystretch{1.0}
\setlength{\tabcolsep}{0.0pt}
\begin{tabular}{lcr}
\hspace{-1cm}
    \includegraphics[width=0.33\linewidth]{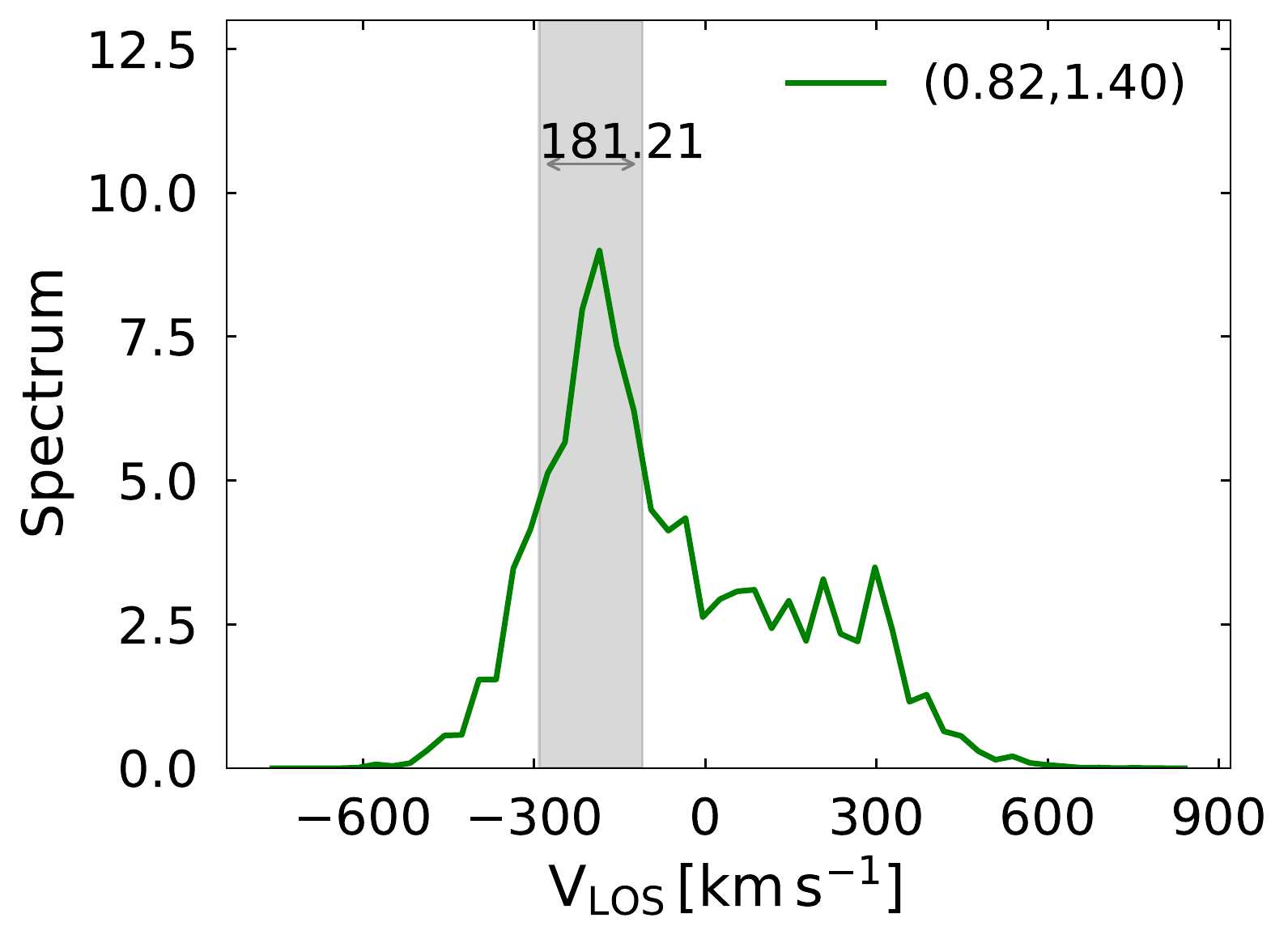}&
    \includegraphics[width=0.33\linewidth]{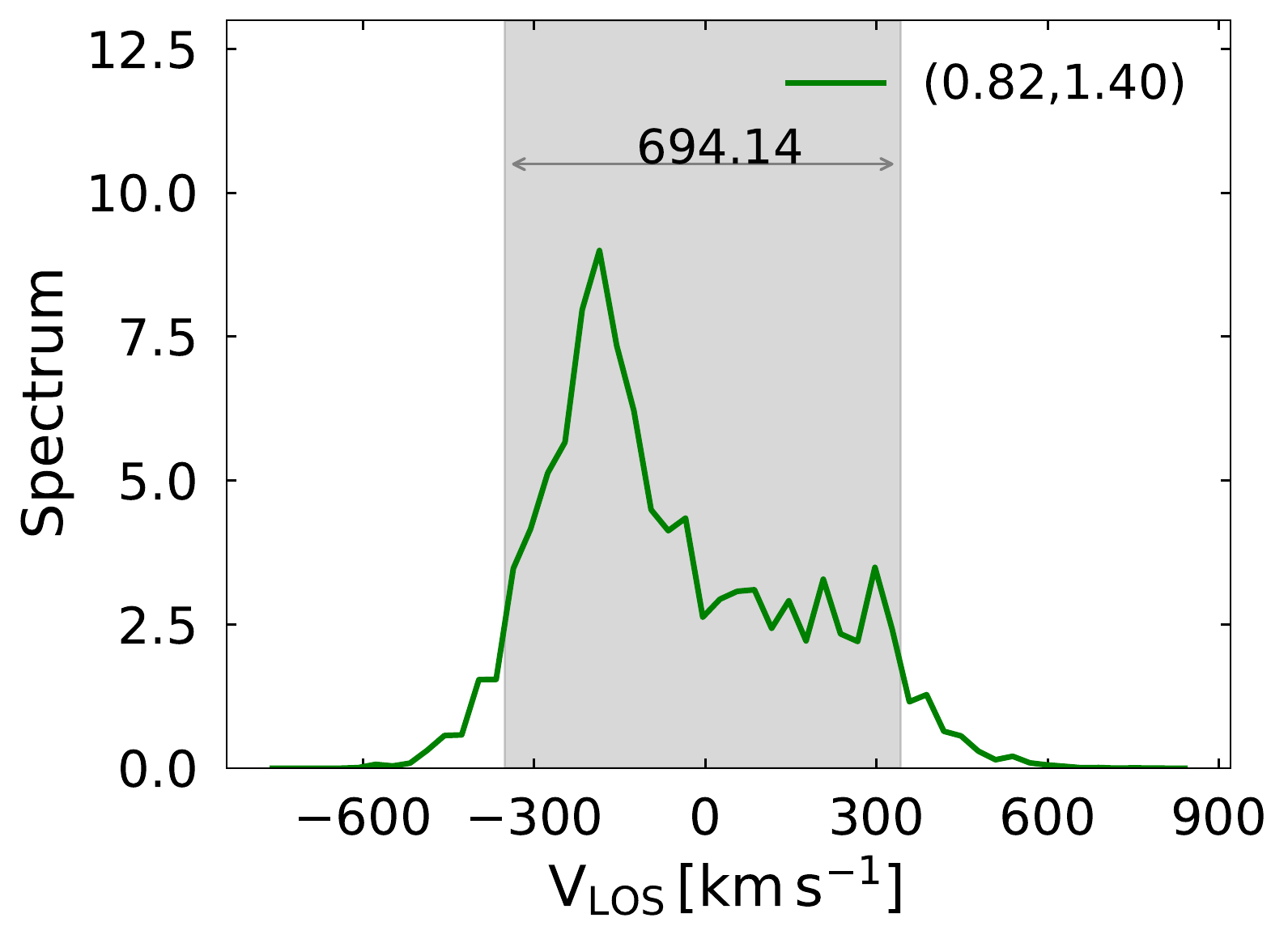}&
    \includegraphics[width=0.33\linewidth]{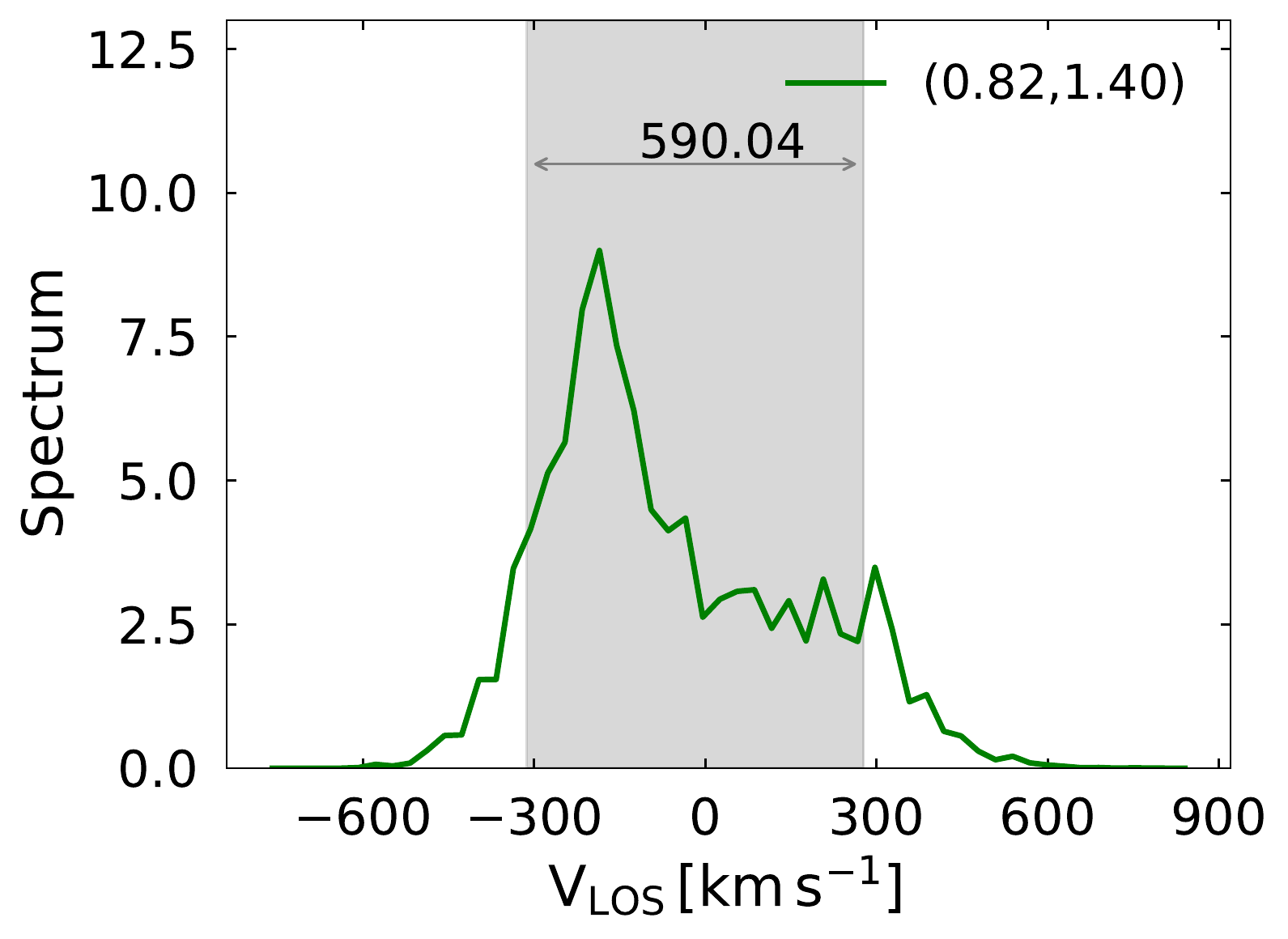}
     \end{tabular}}
      \caption{\textbf{Top:} $\sigma_{50}$ (a) and $\sigma_{20}$ widths (b) estimated at $50\%$ and $20\%$ peak intensity values, as shown in \citet{heckman_1981}, and W80 widths (c) calculated using the approach from \citet{whittle_1985} for the edge-on view ($\theta_\mathrm{I}=90^\circ,\phi_{\mathrm{I}}=270^\circ$) of Sim.~Bs ($\mathrm{P_J}=10^{45}\ergs$) at 3.1~Myr. The black contour shows the projected jet tracer at a value of 0.5 (maximum value is 1). \textbf{Bottom:} The green curve shows the spectrum for the domain marked with $\star$ (black) in the top panel. The gray shaded regions represents the width estimates and the coordinates mark the centre of the domain.}
\label{fig:w80-compare}
\end{figure*}

The estimate of velocity widths and their spatial variation are used to examine the effect of jet on the gas kinematics of the host's ISM. Several studies use the approach given in \citet{whittle_1985}, such as \citet{harrison15,nadia,coatman,Venturi21}, which relates the velocity widths to the integrated intensity of the spectrum. Some studies also follow the approach used in \citet{heckman_1981}, for e.g. \citet{nesvadba_2017,ferguilo20}, where the widths are associated to the peak intensity of the spectrum. In our study, we have followed the integrated intensity approach to estimate the W80 widths by removing the 20$\%$ of the intensity from the wings \cite[see][]{whittle_1985}, which is explained in Sec.~\ref{method_emission}. In Fig.~\ref{fig:w80-compare}, we compare the widths measured using different approaches for Sim.~Bs at 3.1~Myr. The $\sigma_{50}$ and $\sigma_{20}$ widths are estimated using approach given in \citet{heckman_1981}, where these widths corresponds to the velocity range covering $50\%$ and $20\%$ of the peak intensity in the individual spectra. The $\sigma_{50}$ estimates gives lower estimates at some regions due to the sharp cut-off of $50\%$ which excludes significant velocity space for some spectra, as is also shown in the example in Fig.~\ref{fig:w80-compare}. On the other hand, the $\sigma_{20}$ measure gives higher width measures, since it covers wider velocity range than $\sigma_{50}$ measures. However, as one can see that the overall morphology of widths on the image plane is similar for all three measures, different methods are unlikely to affect the qualitative results of our study.

\section{Modelling effect of outflows on the disc velocity field}
\label{vel_map}
\begin{figure*}
 \centerline{
\def\arraystretch{1.0}
\setlength{\tabcolsep}{0.0pt}
\begin{tabular}{lcr}
\hspace{-1cm}
    \includegraphics[width=0.35\linewidth]{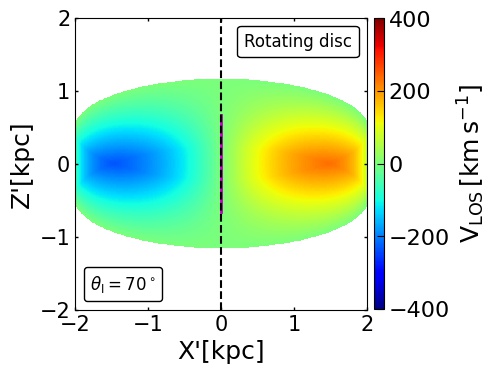}&
    \includegraphics[width=0.35\linewidth]{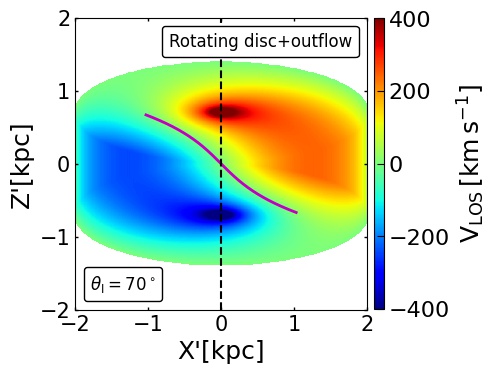}&
    \includegraphics[width=0.35\linewidth]{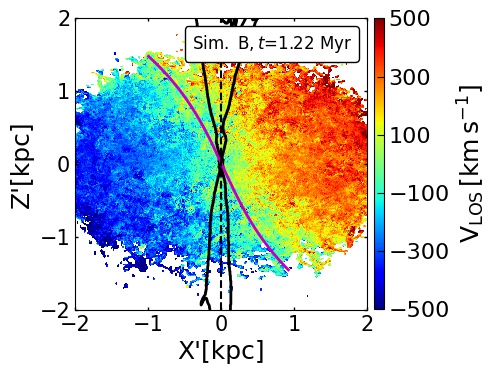}
     \end{tabular}}
      \caption{\textbf{Left and middle:} Analytical prediction for the projected velocity of the rotating disc and radial outflows superimposed on the rotating disc at $\theta_\mathrm{I}=70^\circ$ image plane. \textbf{Right:} Velocity map for Sim.~B ($\theta_\mathrm{J}=0^\circ,P_\mathrm{J}=10^{45}\ergs$) at the same disc orientation. The magenta curve follows the kinematic centers (zero-velocity regions) fitted using 3rd-degree polynomial.}
\label{fig:disc_rot}
\end{figure*}
In this appendix, we present a simple analytical toy model for predicting how the outflows from the AGN can shape the projected velocity map of the disc. We prepare a disc model of radius $\sim2~\kpc$, and height $1.5~\kpc$ (Z=$\pm0.75~\kpc$) which is rotating with a uniform angular velocity. For this, we use the disc rotation profile used in the simulations from M2018b, and the velocity map of the disc at $\theta_\mathrm{I}=70^\circ, \phi_\mathrm{I}=270^\circ$ is shown in Fig.~\ref{fig:disc_rot}. To mimic the outflows from the jet-driven bubble, a radially outwards velocity field of $v_r = 500/r~\kms$, where $r$ is the cylindrical radius, is overlaid on the rotating disc. 

As one can see in the left panel of Fig.~\ref{fig:disc_rot}, the initial disc velocity field appears symmetric in the blue and red-shifted motion. However, the presence of outflows disturbs the original velocity field, and the two counter movements of the gas are no longer symmetrically separated by the projected minor axis (i.e X'=0) of the disc. Moreover, the region separating the blue and red-shifted motion, which closely follows the zero-velocity regions (kinematic centers) appears curved. Such realignments can also be seen in the Sim.~B and Sim.~D (see Fig.~\ref{fig:vel_70_deg}) at the same disc orientation. We fit the mean of the kinematic centers ($V_\mathrm{{LOS}}<\pm10~\kms$) in Fig.~\ref{fig:disc_rot} using a third-degree polynomial, which is shown in magenta curve. Also, as compared to the unperturbed disc in the left panel in Fig.~\ref{fig:disc_rot}, the presence of outflows enhance the observed LOS velocity of the disc (see middle panel), similar to what we observe in the Fig.~\ref{fig:rot_curve} and~\ref{fig:vel_70_deg}.

\section{How would noise shape the observed velocity dispersion?}
\label{noise}
\begin{figure}
 \centerline{
\def\arraystretch{1.0}
\setlength{\tabcolsep}{0.0pt}
\begin{tabular}{lcr}
  \hspace{-0.5cm}
      {\includegraphics[width=\linewidth]{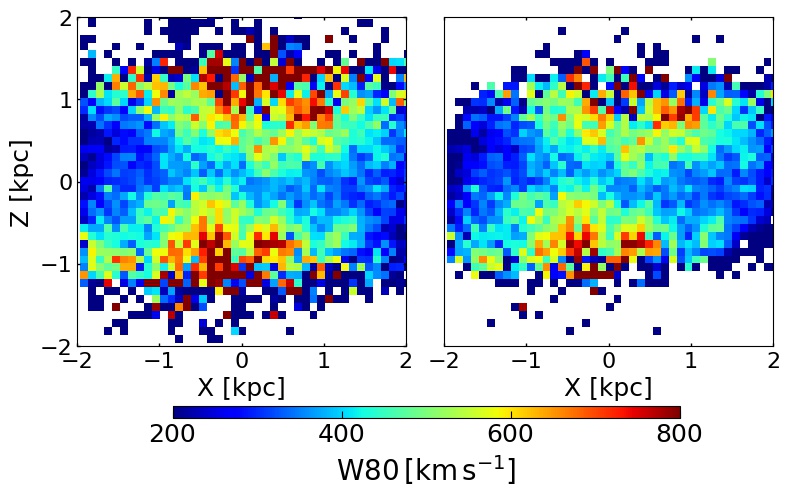}}
     \end{tabular}}
\caption{W80 widths for Sim.~D ($\theta_\mathrm{J}=45^\circ$) edge-on view at 2.31~Myr for a detector of sensitivity threshold up to 3-dex (\textbf{Left}) and 2-dex (\textbf{Right}) from the maximum cumulative flux.}
  \label{fig:w80_cut}
\end{figure}
In this study, we have used the data from hydro-dynamic  simulations to estimate the observed quantities, such as flux and shocked gas velocity dispersion. However, the real-time observation of the distant galaxies is significantly affected by the random noise in the data and the detector sensitivity. Only highly luminous emission can be sensed by the detector, which consequently shape the observed kinematics of the system. As can be seen in the Fig.~\ref{fig:OIII-simD} (edge-on at 2.31~Myr, for example), the high W80 regions at the upper and lower part of the disc coincide with low flux emitting zones. Thus, if there had been noise in the data, a detector with a weak sensitivity to low flux might not capture the dynamics of these regions. Also, the low-flux emitting wings in the spectra (see for e.g. Fig.~\ref{fig:w80_Sim_Bs}) may get submerged in the noise, and reduce the value of observed widths. To see how the observed kinematics will vary with the inclusion of random noise, we perform the following analysis on the W80 measures for the edge-on view of Sim.~D at 2.31~Myr. We use different flux thresholds, namely three and two-dex less than the maximum cumulative flux in the bins of the [\ion{O}{III}] spectra ($\sim10^{37}\ergs(\kms)^{-1}$), and put the flux to zero in the bins with values lower than these thresholds. This implies that the bins below the threshold flux are submerged with the noise and are not detected. One can see in Fig.~\ref{fig:w80_cut} that using a 3-dex cut does not alter much the morphology of widths on the image plane from Fig.~\ref{fig:OIII-simD}, however, some regions in the upper and lower regions indicate lower widths than earlier, or are not detected at all. However, setting a two-dex cut in flux reduce the extent of high-width biconical structures, which are the low-flux emitting hot-gas less-dense fragments (see column density map in Fig.\ref{fig:OIII-simD}). Thus, including noise in the data, and incorporating detector sensitivity, may reduce the strength of observed complex kinematics in the systems, but the overall morphology and the implications of jet-disc interaction are not significantly altered. 
In our analysis, we have kept the same spatial resolution in the bins, however, one can also perform the Voronoi binning of the real data i.e adaptive binning to obtain a given S/N values in the low-flux emitting regions \citep[see][]{michele_2003}. This can cause some degradation of the spatial resolution at large bins, but the dynamical information is not completely lost.

\bsp	
\label{lastpage}
\end{document}